\documentclass[twoside,12pt]{article}
\usepackage{graphicx}
\usepackage{amssymb,amsmath,cite}

\def\ket#1{|#1\rangle}

\def\scal#1#2{\langle#1|#2\rangle}
\def\matr#1#2#3{\langle#1|#2|#3\rangle}
\def\matri#1#2#3{\langle#1||#2||#3\rangle}

\def\ave#1{\langle #1\rangle}

\def\Tr{{\rm Tr}}

\begin{document}

\pdfoutput=1


\title{\vspace{1cm}Quantum phase transitions in the interacting boson model}


\author{Pavel Cejnar$^1$ and Jan Jolie$^2$}
\maketitle

\begin{center}
{\it
$^1$\,Institute of Particle and Nuclear Physics, Faculty of Mathematics and Physics, 
Charles University, V Hole{\v s}ovi{\v c}k{\'a}ch 2, 180\,00 Prague, Czech Republic
\\
$^2$\,Institute of Nuclear Physics, University of Cologne,
Z\"ulpicher Strasse 77, 50937 Cologne, Germany
}
\end{center}

\begin{abstract}
This review is focused on various properties of quantum phase transitions (QPTs) in the Interacting Boson Model (IBM) of nuclear structure.
The model describes collective modes of motions in atomic nuclei at low energies in terms of a finite number $N$ of mutually interacting $s$ and $d$ bosons.
Closely related approaches are applied in molecular physics.
In the $N\to\infty$ limit, the ground state is a boson condensate that exhibits shape-phase transitions between spherical (I), deformed prolate (II), and deformed oblate (III) forms when the interaction strengths are varied.
Finite-$N$ precursors of such behavior are verified by robust variations of nuclear properties (nuclear masses, excitation energies, transition probabilities for low lying levels) across the chart of nuclides.
Simultaneously, the model serves as a theoretical laboratory for studying diverse general features of QPTs in interacting many-body systems, which differ in many respects from lattice models of solid-state physics.
We outline the most important fields of the present interest:
(a) The coexistence of first- and second-order phase transitions supports studies related to the microscopic origin of the QPT phenomena.
(b) The competing quantum phases are characterized by specific dynamical symmetries and novel symmetry related approaches are developed to describe also the transitional dynamical domains.
(c) In some parameter regions, the QPT-like behavior can be ascribed also to individual excited states, which is linked to the thermodynamical and classical descriptions of the system.
(d) The model and its phase structure can be extended in many directions: by separating proton and neutron excitations, considering odd-fermion degrees of freedom or different particle-hole configurations, by including other types of bosons, higher order interactions, and by imposing external rotation.
All these aspects of IBM phase transitions are relevant in the interpretation of experimental data and important for a fundamental understanding of the QPT phenomenon.
\\

\\


\end{abstract}



\newpage

\tableofcontents

\newpage

\section{Introduction}
\label{se:int}

During the last decades, the classical notion of thermodynamical phase transitions \cite{Reichl} has been extended in several directions. 
One of these extensions is connected with {\em finite systems}, counting from thousands down to just few tens of particle constituents, see e.g. Refs.~\cite{Fisher82,Lee90,Borrmann99,Gross01}. 
The corresponding objects are Bose-Einstein condensates, atomic clusters, quantum dots and other mesoscopic systems.
As the temperature or other external parameters are varied, different \lq\lq phases\rq\rq, i.e. specific structural configurations of these systems, suddenly emerge. 
Finiteness of the system unavoidably results in smoothening of all relevant phase transitional observables, but it makes sense to treat these effects as precursors of true phase transitions that would take place in an asymptotic regime.
In particular, phase transitions are related to the scaling of some essential properties with the increasing size.

Another extension of standard phase transitions is to systems of interacting quantum objects at {\em zero temperature}. 
It turns out that when changing the interaction strength in some systems of this sort, one crosses a certain critical point of a nonanalytic change (in the infinite size limit) between ordered and disordered \lq\lq phases\rq\rq, which are represented by two distinct types of the ground state wave function. 
Since at zero temperature thermal fluctuations disappear, the only internal motion that can be responsible for the onset of disorder are quantum fluctuations.
This situation is therefore referred to as the quantum phase transition (QPT).
Quantum phase transitions were quickly recognized very relevant in a wide range of systems in condensed matter and many body physics, see e.g. Refs.~\cite{Hertz76,Gilmore78,Gilmore79,Dieperink80,Gilmore81,Sachdev99,Vojta03}.

Atomic nuclei are at an intersection of both above-outlined extensions of phase transitions.
\lq\lq Classical\rq\rq\ phase transitions in nuclei are driven by intensive thermodynamical variables---temperature and/or rotational frequency---and take several well-known forms: 
(a) Transition from Fermi liquid to an ideal gas of nucleons is observed in multifragmentation phenomena induced by heavy-ion collisions \cite{Siemens83,Pochodzalla95,Elliot02}. 
(b) Experimental and theoretical studies (see e.g. \cite{Schiller01,Liu01}) support the idea that nuclear matter exhibits a phase transition from the paired (superfluid) to an unpaired phase \cite{Mottelson60,Goodman81}.
(c) An analogous and maybe related effect is observed in the lowest rotational band of many nuclei as a sudden growth of the moment of inertia \cite{Nilsson95}.
(d) Transitions between different quadrupole deformed shapes are anticipated to appear in a wide range of hot rotating nuclei \cite{Alhassid86}.

Manifestations of \lq\lq quantum\rq\rq\ phase transitions in nuclei at zero temperature can be mostly observed as abrupt changes of nuclear shapes when crossing certain borders in the plane of neutron versus proton number, ${\cal N}\times{\cal Z}$, see e.g. \cite{Dieperink80,Casten93,Casten07,Casten08}.
Such transitions can be studied theoretically within some phenomenological models of nuclear structure, in which the variation of effective interaction strengths defines the relevant domain of application on the chart of nuclides, see e.g. Refs.~\cite{Dieperink80,Thouless61,Federman79,Ring80,Davis86}.

This review is aimed at quantum phase transitions in the interacting boson model (IBM).
Since its discovery by Arima and Iachello in 1975 \cite{Arima76,Arima78,Arima79}, this model has played an important role in modeling collective motions of atomic nuclei \cite{Elliot85,IA87,Bo88,II91,Ta93,FI94}.
The success of the IBM is based on a simple and elegant algebraic formulation, and on a wide span of relevant dynamical regimes. 
Closely related algebraic approaches are applied in molecular physics \cite{FI94,IL95}, hadronic physics \cite{Bijker94,Bijker00}, and other areas \cite{Iachello92,Iachello02}.

The IBM in its $sd$-boson \lq\lq incarnation\rq\rq\ exhibits both first- and second-order phase transitions between spherical, deformed-prolate and deformed-oblate shapes of the ground state \cite{Dieperink80,Dieperink80b,Feng81,Roosmalen82}.
Although the nonanalytic nature of these transitions can be verified only in the unrealistic limit of the infinite boson number, $N\to\infty$ (while in application to nuclei, $N$ coincides with the number of valence particle or hole pairs, so $N<20$ in all realistic cases), some finite-size precursors turn out to be very neat already at low boson numbers \cite{Casten07}.
The shape transitional predictions of the IBM can be applied to low-energy spectroscopic data for numerous isotopic/isotonic chains.

Apart from describing actual data, the IBM can also be considered as a valuable theoretical tool for studying some general features 
of many-body quantum and mesoscopic systems.
More than 30 years of the model history gradually disclosed that seemingly simple IBM hamiltonians encode a rich variety of complex physical phenomena, such as ground- and excited-state quantum phase transitions, effects of coexisting regular and chaotic motions, manifestations of various types of exact and approximate symmetries etc.
The model offers a deeper insight into the roots of these phenomena.
In this sense, it can be compared to the Hubbard model of solid-state physics, which was designed as a model describing rather specific systems, but in the course of time turned into a general testing ground for application of diverse theoretical concepts.

In attempts to understand the origin and consequences of quantum phase transitions in the IBM, considerable effort has been spent and a large number of results obtained---see the impressive (but certainly still incomplete) list of references [\citen{Dieperink80b}--\citen{Caprio08}] below.
Although these efforts are probably not yet completed, this might be about the right time to make a current summary.
 
In this review, the focus is set on both experimental and theoretical applications of the interacting boson model.
With regard to the recent reviews of mostly experimental aspects of shape-phase transitions in nuclei \cite{Casten07,Casten08}, more emphasis is put to the theoretical side of the problem. 
We start by describing some general features of quantum phase transitions in finite bosonic models and by pointing out some differences from the lattice models.
A considerable part of the review deals with phase-transitional features of the simplest version of the model, the so-called IBM-1.
It is presented as a solid reference frame for experimental studies, which at the same time serves as a theoretical workshop for analyzing the QPT underlying mechanisms.
Applications in some of the more advanced versions of the IBM are illustrated afterwards.
At the end, we will discuss recent generalization of the QPT type of behavior to excited states.



\section{Quantum phase transitions in finite bosonic systems}
\label{se:qpt}

Effective models of nuclear structure contain free parameters (interaction strengths) whose variations naturally induce changes of the ground-state wave function.
In a finite system such changes are always of the \lq\lq crossover\rq\rq\ type, i.e., smooth in all observables.
However, in some situations rather sharp transitions are encountered, which may signal true quantum phase transitions in the \lq\lq thermodynamical limit\rq\rq.
To decide whether this is the case or not, one needs to identify a parameter, let us denote it $\aleph$, that measures the size of the system.
Through this paper, the role of $\aleph$ is played by the total number of bosons, $N$.
A necessary requirement for calling a given abrupt structural change of the ground state a \lq\lq quantum phase transition\rq\rq\ (more precisely, a finite-$\aleph$ {\em precursor\/} of QPT) is that some of the related observables become discontinuous in the limit $\aleph\to\infty$.
Let us use this feature as a working definition of the quantum phase transition in the type of finite many-body systems that we study here.

Since in this review the focus is set on sharp transitions of nuclear shapes, the collective degrees of freedom are of primary interest. 
These can be represented by {\em bosonic\/} types of excitations, with individual bosons occupying a suitable {\em finite-}dimensional single-particle Hilbert space.
The theoretical framework for this type of description is provided by the family of interacting boson models (Sec.~\ref{se:iba}), whose quantum phase transitions will be studied in the following sections.
It turns out, however, that in relation to QPTs all finite bosonic models have some general properties in common.
We therefore start our discussion by summarizing these properties and showing related examples in simpler IBM-like models.

\subsection{Infinite-size limit of finite bosonic models}
\label{se:fin}

Consider a bosonic many-body hamiltonian with one- and two-body (and possibly higher) interaction terms conserving the total number of particles: 
\begin{equation}
H=E_0+\sum_{k}\epsilon_{k}\, b^{\dagger}_k b_k+
\sum_{k,l,m,n}\nu_{klmn}\, b^{\dagger}_k b^{\dagger}_l b_m b_n+\dots
\label{H}
\end{equation}
Here $b^{\dagger}_k$ and $b_k$, respectively, create and annihilate a boson of the $k$th type, while $\epsilon_{k}$, and $\nu_{klmn}$ represent one- and two-body interaction strengths satisfying the condition
$$
\nu_{klmn}=\nu_{lkmn}=\nu_{klnm}=\nu_{lknm}=
\nu_{nmlk}=\nu_{nmkl}=\nu_{mnlk}=\nu_{mnkl}\,,
$$
which follows from the exchange symmetry and hermicity.
$E_0$ is just an arbitrary energy shift. 
Let us note that (a part of) the subscript values may also be understood as denoting different components of the same boson species (e.g. different angular-momentum projections).
Afterwards it will turn important that one of the bosons can be treated separately from the others, and for this boson we reserve the subscript $k=0$.
We therefore choose the numbering $k=0,1,2,\dots,f$, where $f$ will be shown to coincide with the number of classical degrees of freedom associated with the system.

The total number of bosons, $N=\sum_k b^{\dagger}_k b_k$, is conserved by the hamiltonian (\ref{H}). Therefore, we can consider realizations of the many-body system with different values of $N$. However, because of the finiteness of the single-particle Hilbert space (its dimension is $f+1$), the increasing number of interacting particles makes the total energy grow faster than linearly with $N$. Namely, the two-body part will grow proportional to $N(N-1)$, the three-body part (if any) as $N(N-1)(N-2)$ etc. Thus the energy is not an {\em extensive\/} quantity. To solve this problem, we have to attenuate individual terms by the respective factors, which means that we switch to a modified (extensive) hamiltonian $H$ with interaction strengths decreasing with $N$. Expressing then the energy per particle, ${\cal H}\equiv\tfrac{1}{N}H$ (an {\em intensive\/} quantity), one has: 
\begin{equation}
{\cal H}=\varepsilon_0+\tfrac{1}{N}\sum_{k}\epsilon_{k}\, b^{\dagger}_k b_k+
\tfrac{1}{N(N-1)}\sum_{k,l,m,n}\nu_{klmn}\, b^{\dagger}_k b^{\dagger}_l b_m b_n+\dots
\label{h}
\end{equation}
(we keep the adjustable energy shift, now denoted as $\varepsilon_0$).
Strictly speaking, this hamiltonian represents a different system than the original hamiltonian (\ref{H}), although via an appropriate readjusting of the interaction strengths in Eq.~(\ref{h}) we can return to the system (\ref{H}) for each individual value of $N$.

Now, we can introduce self-adjoint coordinate and momentum operators
\begin{equation}
q_k=\tfrac{1}{\sqrt{N}}\left(\alpha_k b_k^{\dagger}+\alpha_k^* b_k\right)
\,,\quad
p_k=\tfrac{i}{\sqrt{N}}\left(\beta_k b_k^{\dagger}-\beta_k^* b_k\right)
\,,
\label{qp}
\end{equation}
or, equivalently,
\begin{equation}
b_k^{\dagger}=\tfrac{\sqrt{N}}{\Gamma}\left(\beta_k^* q_k-i\alpha_k^* p_k\right)
\,,\quad
b_k=\tfrac{\sqrt{N}}{\Gamma}\left(\beta_k q_k+i\alpha_k p_k\right)
\,,
\label{bb}
\end{equation}
with $\Gamma=\alpha_k\beta^*_k+\alpha^*_k\beta_k$
The commutator of $q_k$ and $p_l$ reads as:
\begin{equation}
[q_k,p_l]=i\,\tfrac{\Gamma}{N}\,\delta_{kl}
\,.
\label{komu}
\end{equation}
We want the commutator to be universal, independent of $k$, so we take $\alpha_k$ and $\beta_k$ such that $\Gamma$ represents a constant. For the sake of simplicity we may choose $\alpha_k=\alpha^*_k=\beta_k=\beta^*_k=\tfrac{1}{\sqrt{2}}$, hence $\Gamma=1$.

The value $\tfrac{\Gamma}{N}$ in the commutator (\ref{komu}) plays the role of the Planck constant $\hbar$. We therefore conclude that in finite bosonic systems, in which the scaling $H\to{\cal H}$ from Eqs.~(\ref{H}) and (\ref{h}) makes the coordinate-momentum representation (\ref{qp}) suitable, the $N\to\infty$ limit represents just the {\em classical limit}. This is rather important finding since in the QPT systems with an infinite single-particle Hilbert space, where the scaling (\ref{h}) is not employed, the $N\to\infty$ limit represents the transition to the quantum field theory.

It is obvious that if Eq.~(\ref{bb}) is substituted to Eq.~(\ref{h}), while replacing $N(N-1)$ by $N^2$ for large $N$, we obtain a coordinate-momentum representation of ${\cal H}$ with no explicit dependence on $N$.
Using the commutation rule (\ref{komu}) and neglecting the ${\cal O}(N^{-1})$ contraction terms, the hamiltonian can be written in the following simple form:
\begin{eqnarray}
{\cal H}_{\rm cl}({\bf q},{\bf p})=\varepsilon_0+\tfrac{1}{2}\sum_k\epsilon_k\left(p_k^2+q_k^2\right)
\!\!\!&+&\!\!\!\tfrac{1}{4}\sum_{k,l,m,n}\nu_{klmn}\left(p_kp_lp_mp_n+q_kq_lq_mq_n\right)
\label{hamc}\\
\!\!\!&+&\!\!\!\tfrac{1}{2}\sum_{k,l,m,n}\nu_{klmn}\left(p_lp_nq_kp_m+p_lp_mq_kq_n-p_mp_nq_kp_l\right)
\,.
\nonumber
\end{eqnarray}
Note that here we have used the above special choice of coefficients $\alpha$ and $\beta$.
In fact, as the contractions were skipped, Eq.~(\ref{hamc}) represents the classical (i.e. $N\to\infty$) limit of hamiltonian (\ref{h}).

The classical hamiltonian (\ref{hamc}) can be equivalently obtained using {\em condensate states\/} \cite{Gilmore79,Dieperink80,Dieperink80b,Feng81,Bohr80,Ginocchio80,Ginocchio80b,Klein81,Isacker81,Klein82}
\begin{equation}
\ket{N,{\bf c}}\propto\biggl[\sum_k c_k b_k^{\dag}\biggr]^N\ket{0}
\,,\qquad
c_k=\tfrac{1}{\sqrt{2}}(q_k-ip_k)
\label{conden}
\end{equation}
where $\ket{0}$ is the boson vacuum, and taking 
\begin{equation}
{\cal H}_{\rm cl}({\bf q},{\bf p})\equiv\frac{\matr{N,{\bf c}}{{\cal H}}{N,{\bf c}}}{\scal{N,{\bf c}}{N,{\bf c}}}
=\ave{{\cal H}}_{\ket{N,{\bf c}}}
\label{klas}
\end{equation}
in the limit $N\to\infty$.
The condensate states (\ref{conden}) respect the conservation of the total boson number.
Alternatively, one may use the {\em Glauber coherent states\/} \cite{Hatch82}
\begin{equation}
\ket{\ave{N},{\bf d}}\propto\exp\biggl[\sum_k d_k b_k^{\dag}\biggr]\ket{0}
\,,
\label{glaub}
\end{equation}
which do not have a fixed $N$, but allow to vary the average $\ave{N}=\sum_k|d_k|^2$.
For $N,\ave{N}\to\infty$, both Glauber and condensate states yield the same results if taking $c_k\propto\ave{N}^{-1/2}d_k$.

It is easy to show that quantum fluctuations measured by the dispersion of the energy per particle in states (\ref{conden}) vanish  for asymptotic $N$:
\begin{equation} 
\ave{{\cal H}^2}_{\ket{N,{\bf c}}}-\ave{{\cal H}}_{\ket{N,{\bf c}}}^2
={\cal O}(N^{-1})
\,.
\label{dis0}
\end{equation}
This is a consequence of the attenuation of interaction terms in Eq.~(\ref{h}) with an increasing number of bosons. 
As both condensate and coherent states form overcomplete bases in the bosonic Hilbert space, the above result illustrates the fact that the asymptotic number of bosons indeed represents the classical limit of the present class of systems.
In contrast, for an infinite lattice system with a finite range of interactions (not hindered by $\propto N^{-1}$) a quantity analogous to (\ref{dis0}) behaves as ${\cal O}(1)$.

There is an additional condition following from the conservation of the total boson number, namely $\sum_k\left(p_k^2+q_k^2\right)=2$, which is equivalent to the normalization $\scal{N,{\bf c}}{N,{\bf c}}=1$.
This constraint can be used to eliminate one of the degrees of freedom, for instance that connected with the $b_0$ boson.
Indeed, the absolute value $|c_0|$ can be calculated from the other values $|c_k|$ with $k>0$ through the normalization condition, while the phase $\phi_0$ of $c_0$ can be chosen arbitrary, e.g. $\phi_0=0\Rightarrow p_0=0$.
We therefore obtain a system with $f$ degrees of freedom restricted by the condition $\sum_{k>0}(p_k^2+q_k^2)\leq 2$.
Furthermore, all coordinates $q_k$ with $k>0$ can be expressed relative to $q_0$ using the transformation $q_k\mapsto{\tilde q_k}=\sqrt{2}(q_k/q_0)$, which is (for $p_0=0$) equivalent to setting $c_0=1$ in Eqs.~(\ref{conden}) and (\ref{glaub}).
The phase space of the new coordinates $({\tilde q}_1,\dots,{\tilde q}_k)$ and the corresponding momenta $({\tilde p}_1,\dots,{\tilde p}_k)$ is unlimited.
This procedure, however, implies the appearance of characteristic \lq\lq form factors\rq\rq\ in ${\cal H}_{\rm cl}$, namely the factors $[1+\sum_{k>0}(p_k^2+q_k^2)]^{-m}$ in the terms with $m$-body interactions.

The above results can be viewed as a consequence of a more sophisticated group theoretical procedure, based on so-called coset spaces \cite{Gilmore79,Zhang90}.
The procedure starts by the identification of a certain subalgebra (called maximum stability subalgebra) of the spectrum generating algebra U($f$+1) associated with the bosonic system under study (the spectrum generating algebra is formed of bilinear products $b^{\dagger}_kb_l$).
In the present case, the maximum stability subalgebra is taken as U($f$)$\otimes$U(1), where the single separated degree of freedom is the one connected with the $b_0$ boson.
The {\em algebraic coherent states\/} \cite{Zhang90} associated with the factor algebra U($f$+1)/[U(f)$\otimes$U(1)] read as
\begin{equation}
\ket{N,{\bf z}}\propto\exp\biggl[\sum_{k>0} \left(z_k b^{\dag}_kb_0+z^*_k b^{\dag}_0b_k\right)\biggr]\left[b_0^{\dag}\right]^N\ket{0}
\,.
\label{alcoh}
\end{equation}
Using the Baker-Campbell-Hausdorf formula, one can prove that Eq.~(\ref{alcoh}) transforms into the form (\ref{conden}) with $c_0=\cosh|{\bf z}|$ and $c_k=\tfrac{z_k}{|{\bf z}|}\sinh|{\bf z}|$, where $|{\bf z}|=\sqrt{\sum_{k>0}|z_k|^2}$.

In the following, we implicitly use the above reduced set of $f$ normalized coordinates and the associated momenta, but skip tildes from the notation.
We utilize shorthand symbols ${\bf q}\equiv(q_1,\dots,q_f)$ and ${\bf p}\equiv(p_1,\dots,p_f)$.
An important quantity for the analysis of the ground-state phase transitions is the potential energy.
It is obtained from the hamiltonian by setting all momenta to zero, hence $V({\bf q})\equiv{\cal H}_{\rm cl}|_{{\bf p}=0}$.
Although the original hamiltonian (\ref{hamc}) with $f+1$ degrees of freedom contains only quadratic and quartic terms in its potential $V({\bf q})$, the elimination of $q_0$ creates also cubic terms.
Moreover, the factors $(1+\sum_{k>0} q_k^2)^{-m}$ make the Taylor expansion of $V({\bf q})$ infinite.
An explicit formula for the potential will be given below in relation to the interacting boson model (Subsec.~\ref{se:geo}).
Here, the whole procedure was described just to elucidate the general method for obtaining $V({\bf q})$ that will be needed in the following.


\subsection{Nonanalytic evolutions with control parameters}
\label{se:co/di}

In the classical limit, $N\to\infty$, the ground state of hamiltonian (\ref{H}) can be identified with the global minimum of the potential energy $V({\bf q})\equiv{\cal H}_{\rm cl}|_{{\bf p}=0}$.
Coordinates of the global minimum are ${\bf q}_{\rm m}$ and the ground-state energy reads as $E_0=V({\bf q}_{\rm m})$.
Under \lq\lq normal\rq\rq\ conditions, $V$ and ${\bf q}_{\rm m}$ depend on the hamiltonian parameters $\epsilon_k$ and $\nu_{klmn}$ in a smooth, analytic way. 
However, in some cases---and these are of interest in this review---the changes of both ${\bf q}_{\rm m}$ and $E_0$ are nonanalytic for some values of the hamiltonian parameters.
Such situations are of course well documented in the literature.
There exist two standard approaches which make it possible to treat nonanalytic evolutions in a systematic way: the catastrophe theory and the Landau theory.

\subsubsection{Catastrophe theory}
\label{se:cat}

The catastrophe theory, initiated by Thom in the mid 1970's, has been broadly advertised among all kinds of scientists and technicians.
In short, the theory deals with systems in which \lq\lq continuous causes\rq\rq\ can lead to \lq\lq discontinuous effects\rq\rq.
Its application in quantum physics was pioneered by Gilmore \cite{Gilmore81}, who showed that the catastrophe theory describes and classifies nonanalytic evolutions of the ground state properties of some many-body systems.
The application of the catastrophe theory in the interacting boson model was first outlined by Feng, Gilmore, and Deans \cite{Feng81} and later elaborated by L{\'o}pez-Moreno and Casta{\~n}os \cite{Moreno96}.

The key idea opening the whole field is the concept of structural instability \cite{Stewart82}.
It can be easily explained on the potential of the quartic form $V(x)=x^4$, which will turn to be very relevant in the forthcoming analyses.
But consider at first the harmonic-oscillator potential $V(x)=x^2$.
Add a small perturbation to it, $V'(x)=x^2+\varepsilon f(x)$, with $\varepsilon$ being an infinitesimally small number and $f(x)$ a smooth function with all derivatives locally (around $x\approx 0$) restricted by a common bound.
Although the perturbation distorts the harmonic behavior close to the potential minimum, it is not difficult to see that there is no way how it could change the local topology of the problem, i.e., the number and ordering of the minima and maxima (if any) around $x\approx 0$.
The harmonic potential is structurally stable.
The same holds, less trivially, e.g. for a double-well potential $V(x)=x^4-x^2$, with the boundedness of $f(x)$ now being imposed within a broader interval including all local extremes of $V(x)$.

On the other hand, the pure quartic oscillator, $V(x)=x^4$, is structurally unstable, since a small perturbation $\varepsilon x^2$ with $\varepsilon<0$ {\em does\/} change the topology of the problem: the local minimum at $x=0$ becomes a maximum and emits two minima located symmetrically at $x=\pm\sqrt{\tfrac{1}{2}|\varepsilon|}$.
If we consider a family of potentials
\begin{equation}
V(\eta; x)=(2\eta-1)x^2+(1-\eta)x^4
\label{excusp}
\end{equation}
depending on parameter $\eta\in[0,1]$, a structurally unstable pure quartic potential $\propto x^4$ is trapped at $\eta=\tfrac{1}{2}$ in between structurally stable potentials on both sides $\eta<\tfrac{1}{2}$ and $\eta>\tfrac{1}{2}$.

Let us consider more general families of potentials with variables $x,y,z,\dots$ (denote their number by $n$) and adjustable parameters $a,b,c,\dots$ (there are $r$ of them).
It turns out that for $r\leq 5$ all possible forms of such families can be smoothly mapped (the transformation affecting the parameters as well as the variables) onto 13 \lq\lq canonical\rq\rq\ forms \cite{Stewart82}.
These forms classify all possible catastrophes in low dimension ($r\leq 5$).
Example (\ref{excusp}) belongs to the equivalence class called the {\em cusp}.
Since ground-state phase transitions between spherical and deformed shapes in nuclei and other many-body systems are closely related to the cusp, we will introduce this special type of catastrophe.

\begin{figure}[t]
\centering
\includegraphics[width=13.9cm]{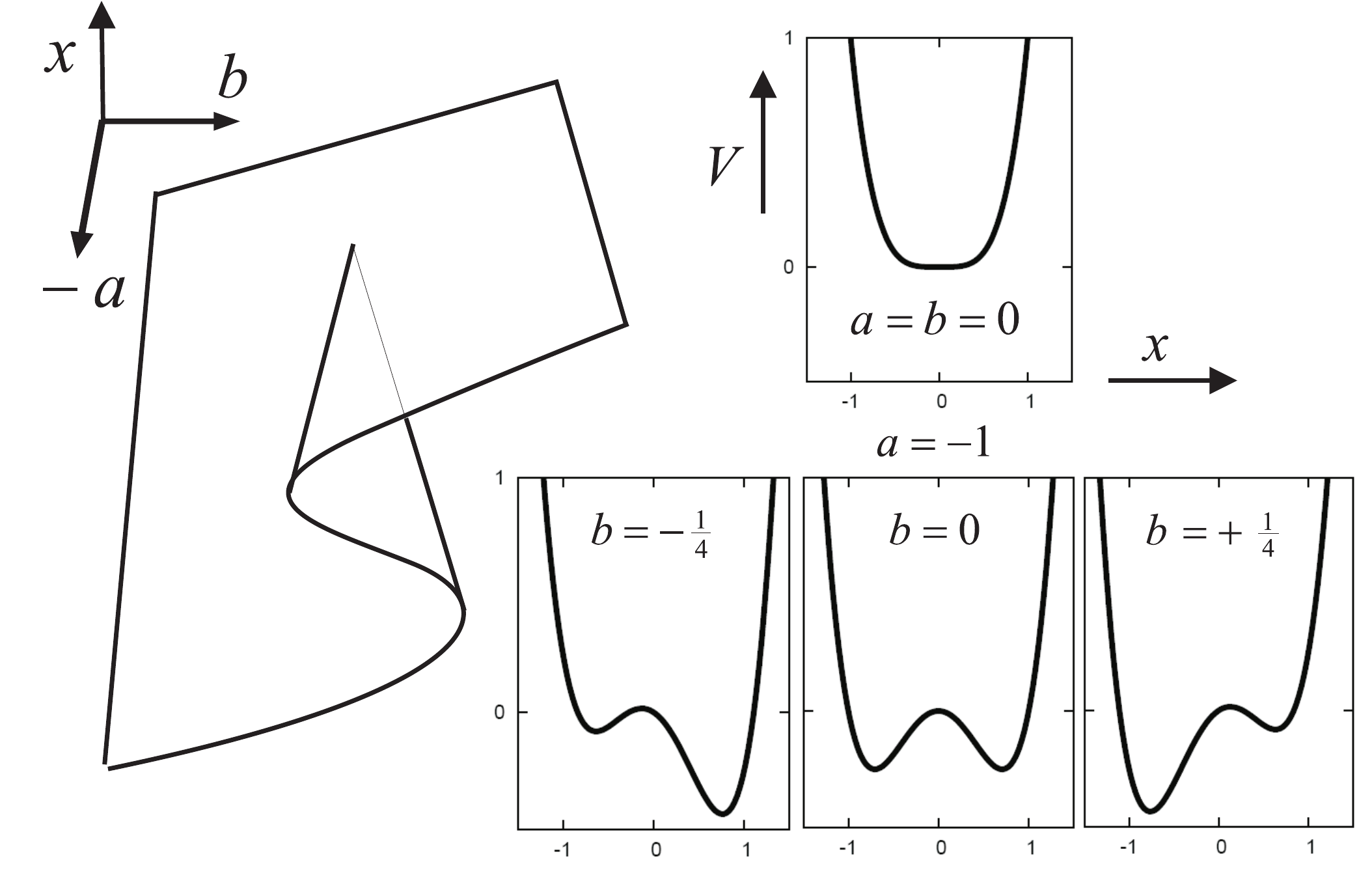}
\caption{\protect{\small The topology of the cusp catastrophe (left). The surface schematically indicates local equilibrium points (stable and unstable) of potential (\ref{cusp}) depending on control parameters $a$ and $b$. The bent part of the surface demarcates region (\ref{cusp2}) of bistable potentials. Sample potentials from this region are shown on the right-hand side.}}
\label{fi:cus}
\end{figure}

The general cusp potential reads as
\begin{equation}
V_{\rm cusp}(a,b;x)=x^4+ax^2+bx
\,,
\label{cusp}
\end{equation}
so it has $r=2$ and $n=1$.
In the parameter plane $a\times b$ the potential has a single minimum except in the cusp-like region
\begin{equation}
a<0\,,\qquad |b|\leq\tfrac{4}{3\sqrt{6}}\sqrt{(-a)^3}\,,
\label{cusp2}
\end{equation}
where two local minima exist and one local maximum in between (a bistable form), see Fig.~\ref{fi:cus}.
Both minima are degenerate at $b=0$, so that if $b$ changes from negative to positive values the minima swap and the system described by potential (\ref{cusp}) undergoes a phase transition.
In the region of parameter $b$ demarcated by condition (\ref{cusp2}) the two coexisting minima indicate a phase coexistence interval typical for first-order phase transitions.
The limiting parameter values for the bistable form of the potential are called {\em spinodal\/} and {\em antispinodal\/} points.

We assume that the system driven through the phase coexistence region is ideally equilibrated, i.e., dwells at the bottom of the lowest potential minimum.
Crossing the critical point $b=0$ (for $a<0$) then implies that the rate of change of the system's energy
flips from the value characteristic for one minimum to the value characteristic for the other minimum.
The evolution of the ground-state energy has its first derivative discontinuous, which is indeed the defining condition of a {\em first-order phase transition}.

Consider now constant $b=0$ and $a$ decreasing from positive to negative values.
In this case, the single potential minimum present at $a>0$ bifurcates at $a=0$, forming two degenerate branches characterizing the situation at $a<0$.
The system can choose any of these branches as they are fully equivalent.
If doing so, it turns out that the first derivative of the lowest energy varies in a continuous way this time, but the second derivative jumps.
This is so-called {\em second-order phase transition}.
Hence the cusp catastrophe accommodates both basic archetypes of phase transitional evolution in the systems which are of interest in this review.

\subsubsection{Landau theory}
\label{se:lan}

Landau theory of phase transitions was formulated in the late 1930's \cite{Landau37} as an attempt to develop a general method of analysis for various types of phase transitions in condensed matter physics (especially in crystals). It relies on two basic conditions, namely on (a) the assumption that the free energy is an analytic function of a quantity called {\em order parameter\/}, and on (b) the fact that the expression for the free energy must obey the symmetries of the system. Condition (a) is further strengthened by expressing the free energy as a Taylor series in the order parameter. It is now known that the Landau theory---being essentially the mean-field theory---fails in many systems. However, in the class of models we look at here it holds. The reason is the above-explained coincidence of the $N\to\infty$ and $\hbar\to 0$ limits in models with finite single-particle Hilbert spaces and properly scaled strengths of interactions. 

Hamiltonian (\ref{H}) has a number of external parameters.
We will now assume a one-dimensional smooth path in the multidimensional parameter space, i.e., consider the hamiltonian parameters depending on a single real parameter $\eta$.
A phase-transitional evolution of the ground state (if any) shows up as a nonanalytic change of the ground-state energy $E_0(\eta)=V(\eta;{\bf q}_{\rm m}(\eta))$ at a certain critical point $\eta=\eta_{\rm c}$.
Since the dependence of the potential $V(\eta;{\bf q})$ on $\eta$ is smooth (as the path is smooth), the nonanalytic evolution of $E_0(\eta)$ is always connected to a nonanalyticity in the trajectory ${\bf q}_{\rm m}(\eta)$ of the potential minimum.
We can write
$$
{\rm discontinuity\ of\ } \tfrac{d^k}{d\eta^k}E_0
\quad\Leftrightarrow\quad
{\rm discontinuity\ of\ } \tfrac{d^{k-1}}{d\eta^{k-1}}{\bf q}_{\rm m}
\quad\Leftrightarrow\quad
k{\rm th\,order\ phase\ transition\,.}
$$
This results from the following sequence of expressions 
\begin{equation}
\tfrac{d}{d\eta}E_0=
\left(\tfrac{\partial}{\partial\eta}V\right)_{{\bf q}_{\rm m}}\!\!+
\underbrace{({\bf\nabla}V)_{{\bf q}_{\rm m}}}_0\cdot\tfrac{d}{d\eta}{\bf q}_{\rm m}
\,,\quad\ \ 
\tfrac{d^2}{d\eta^2}E_0=\left(\tfrac{\partial^2}{\partial\eta^2}V\right)_{{\bf q}_{\rm m}}\!\!+
(\tfrac{\partial}{\partial\eta}{\bf\nabla}V)_{{\bf q}_{\rm m}}\cdot\tfrac{d}{d\eta}{\bf q}_{\rm m}
%
\quad\dots\,
\end{equation}
(where the dot represents the scalar product), which can be continued to an arbitrary order.
Thus if ${\bf q}_{\rm m}$ jumps, so does $\tfrac{d}{d\eta}E_0$ (first-order transition), the jump of $\tfrac{d}{d\eta}{\bf q}_{\rm m}$ implies the same for $\tfrac{d^2}{d\eta^2}E_0$ (second-order transition) etc.
The transition orders introduced here are consistent with the {\em Ehrenfest classification\/} of phase transitions.
This classification is not applicable in general (since in real systems some derivatives may diverge), but it holds within the Landau theory.

Now we come to the essence of Landau theory \cite{Landau37,Landau01}.
At zero temperature, the equilibrium free energy coincides with the ground-state energy and the role of thermodynamical variables is taken by the hamiltonian external parameters.
The phase of the system can be characterized by a suitably chosen order parameter $\xi$, which in the present context is a certain function of coordinates, $\xi\equiv\xi({\bf q})$.
The free energy can be expressed as a function of $\xi$, we denote it $V_{\rm L}(\eta;\xi)$, and its equilibrium value, obtained by minimization, coincides with the ground-state energy: $V_{\rm L}(\eta;\xi_{\rm m})=E_0(\eta)=V(\eta;{\bf q_{\rm m}})$.
Therefore, the above-derived properties of phase transitions of different orders hold also after the replacement $V(\eta;{\bf q})\mapsto V_{\rm L}(\eta;\xi)$.

As will turn out later, the order parameter relevant in our case is given by  
\begin{equation}
\xi=\pm\sqrt{\sum_{k>0}q_k^2}
\,.
\label{order}
\end{equation}
It represents a radius in the $f$-dimensional configuration space (the $q_0$ coordinate was eliminated in the way described in Subsec.~\ref{se:fin}) with the sign $\pm 1$. 
Let us note that the existence of phases characterized by $\xi_{\rm m}>0$ and $\xi_{\rm m}<0$ is an important ingredient of the general Landau theory.
The concrete meaning of the sign in Eq.~(\ref{order}) will be discussed in Subsec.~\ref{se:geo}.
The above definition implies that for $\xi_{\rm m}=0$ the system's ground state is a condensate of $b_0$ bosons, see Eq.~(\ref{conden}), while for $\xi_{\rm m}\neq 0$ the ground state is represented by a more complicated mixture of more types of bosons.

We will be mostly interested in transitions between the $\xi_{\rm m}=0$ and $\xi_{\rm m}\neq 0$ phases.
If there is a second-order transition of this kind, in its vicinity the order parameter takes arbitrarily small values.
Therefore, the free energy can be expressed as a power expansion in $\xi$,
\begin{equation}
V_{\rm L}(\eta;\xi)=V_0(\eta)+A(\eta)\,\xi^2+B(\eta)\,\xi^3+C(\eta)\,\xi^4+\dots
\,,
\label{free}
\end{equation}
where the linear term was omitted because of symmetry constraints that apply in our case (see Subsec.~\ref{se:geo}) as well as in Landau's original context.
The condition for the second-order phase transition between $\xi_{\rm m}=0$ and $\xi_{\rm m}\neq 0$ phases at $\eta=\eta_{\rm c}$ reads as
\begin{equation}
A(\eta_{\rm c})=0
\,,\quad
B(\eta_{\rm c})=0
\,,
\label{cond2nd}
\end{equation}
with $C(\eta_{\rm c})>0$.
The $\xi_{\rm m}=0$ phase is located on the $A>0$ side, while the $\xi_{\rm m}\neq 0$ phase on the $A<0$ side of $\eta_{\rm c}$.
The evolution of $\xi_{\rm m}$ on the $A<0$ side close to $A=0$ is $\xi_{\rm m}\propto\sqrt{|A|}$, hence for a linear dependence on $\eta$ one has $\xi_{\rm m}\propto\sqrt{|\eta-\eta_{\rm c}|}$. 
This implies the {\em critical exponent\/} for the order parameter\footnote
{
General definition \cite{Reichl} of the critical exponent $\lambda$ for a quantity $A(\epsilon)$, where $\epsilon\equiv\frac{\eta-\eta_{\rm c}}{\eta_{\rm c}}$, is: $\lambda=\lim_{\epsilon\to 0}\tfrac{\ln A(\epsilon)}{\ln\epsilon}$.
}
equal to $\tfrac{1}{2}$.
Note that this particular value is specific for the mean-field description of phase transitions of the above type \cite{Reichl} and therefore applies in the finite bosonic models studied here.

On the other hand, the first-order phase transition between $\xi_{\rm m}=0$ and $\xi_{\rm m}\neq 0$ phases takes place if conditions 
\begin{equation}
V_{\rm L}(\eta_{\rm c};\xi)=V_{\rm L}(\eta_{\rm c};0)
\,,\qquad
\tfrac{d}{d\xi}V_{\rm L}(\eta_{\rm c};\xi)=0
\label{cond1straw}
\end{equation}
are fulfilled for a certain value $\xi=\xi_{\rm m}\neq 0$.
Close to the second-order phase transition, where the terms of $V_{\rm L}$ with $\xi^5$ and higher can be neglected and $C>0$, one finds that a simultaneous solution of conditions (\ref{cond1straw}) exists if
\begin{equation}
B^2(\eta_{\rm c})=4A(\eta_{\rm c})C(\eta_{\rm c})\,.
\label{cond1st}
\end{equation}

If higher than quartic terms in the potential (\ref{free}) are neglected, it can be show that ${\cal T}=\tfrac{B^2}{AC}$ represents the only essential parameter of the potential with $A,B\neq 0$ and $C>0$. 
The remaining three \lq\lq unimportant\rq\rq\ parameters can be associated with an energy shift and two scale factors.
Eq.~(\ref{cond1st}) can therefore be written simply as ${\cal T}_{\rm c}=4$.
The $\xi_{\rm m}=0$ phase is located on the ${\cal T}<{\cal T}_{\rm c}$ side and the $\xi_{\rm m}\neq 0$ phase on the ${\cal T}>{\cal T}_{\rm c}$ side.
The spinodal and antispinodal points, demarcating the phase-coexistence region, are given by ${\cal T}_{\rm s}=\tfrac{32}{9}=3.555...$ and ${\cal T}_{\rm a}=+\infty$.
Let us stress that for an infinite-order potential (\ref{free}) these numerical results may in general be valid only in a vicinity of the second-order phase transition (although for some infinite-order potentials, including those relevant here, they hold everywhere).

A closer inspection of Eq.~(\ref{free}) reveals that besides the first- and second-order transitions between $\xi_{\rm m}=0$ and $\xi_{\rm m}\neq 0$ phases, there exists also a first-order transition between $\xi_{\rm m}>0$ and $\xi_{\rm m}<0$ phases.
The critical point for this transition is given by
\begin{equation}
A(\eta_{\rm c})<0
\,,\quad
B(\eta_{\rm c})=0
\,;
\label{cond1st'}
\end{equation}
at this point the equilibrium order parameter apparently changes the sign.
The $\xi_{\rm m}>0$ phase exists on the $B(\eta)<0$ side and the $\xi_{\rm m}<0$ phase on the $B(\eta)>0$ side.

Therefore, the thermodynamical potential (\ref{free}) describes a system with three phases: $\xi_{\rm m}=0$ (phase~I), $\xi_{\rm m}>0$ (phase~II), and $\xi_{\rm m}<0$ (phase~III).
We see that the first-order phase transition between these phases appear on places determined by a single sharp constraint, either Eq.~(\ref{cond1st}) or (\ref{cond1st'}), while the second-order transition is limited to places determined by two constraints, Eq.~(\ref{cond2nd}). 
The second-order constraints simultaneously fulfill both first-order constraints.
Therefore, in a general parameter space of dimension $p$ the first-order phase boundaries form two hypersurfaces of dimension $(p-1)$ and the second-order phase transition lies in their $(p-2)$ dimensional intersection.
This phase structure will be illustrated by the interacting boson model.

There is a relation between the Landau potential (\ref{free}) with terms up to $\xi^4$ and the cusp potential (\ref{cusp}).
Clearly, the potential $V_{\rm L}\propto\xi^4$ at the second-order critical point coincides with the germ of the cusp catastrophe.
The transition between the two forms can be achieved by a smooth transformation which does not affect the cusp topology. 
If setting $C=1$ in the Landau potential (the scale), the transformation involves just appropriate parameter-dependent shifts $x\mapsto x-x_0$ and $V\mapsto V-V_0$ in the cusp potential.
The mapping $(A,B)\mapsto(a,b)$ between topologically equivalent forms (\ref{free}) and (\ref{cusp}) then reads as \cite{Cejnar07}:
\begin{equation}
a=A-\tfrac{3}{8}B^2
\,,\qquad
b=\tfrac{1}{2}B\left(\tfrac{1}{4}B^2-A\right)
\\.
\end{equation}
We may therefore conclude that the phase structure sketched in the previous paragraph belongs to the cusp equivalence class.

\subsection{Hamiltonians with a linear dependence on the control parameter}
\label{se:lin}

The classical analysis of the previous subsection needs to be connected with specific quantum signatures.  
To this end, consider a class of quantum models with a linear dependence on a single real control parameter $\eta$.
The hamiltonian reads as
\begin{equation}
H(\eta)=H_0+\eta\,V
\,,
\label{Hlin}
\end{equation}
where $H_0$ and $V$ are mutually incompatible terms, $[H_0,V]\neq 0$.
For the Landau analysis the linearity means that coefficients in Eq.~(\ref{free}) will be linear functions of $\eta$, which alone would not be an important achievement.
On the quantum level, however, the linearity represents a considerable advantage.
In particular, the knowledge of the whole spectrum $E_i(\eta)$ ($i=0,1,\dots,n-1$) at a {\em single\/} value of the control parameter, together with the complete set of instantaneous matrix elements $\matr{\psi_i(\eta)}{V}{\psi_j(\eta)}$, where $\ket{\psi_i(\eta)}$ is the $i$th eigenstate of $H(\eta)$, determine all spectral observables for {\em any\/} value of the control parameter.

In the many-body case, the linear ansatz is naturally satisfied if the control parameter represents a weight factor at a certain interaction term of the hamiltonian. 
Otherwise it can be justified by the Taylor expansion of a general nonlinear hamiltonian around the point of interest.
The dimensionless parameter $\eta$ may, in principle, vary within the unlimited domain $\eta\in(\-\infty,+\infty)$, but in practice we are always interested in a certain restricted interval $\eta\in[\eta_1,\eta_2]$.
Since we can redefine $H_0$ such that it coincides with $H(\eta_1)$ and absorb the value $(\eta_2-\eta_1)$ in $V$,
the form (\ref{Hlin}) in its full generality can be studied using the constraint $\eta\in[0,1]$.
An equivalent expression is then 
\begin{equation}
H(\eta)=(1-\eta)\underbrace{H(0)}_{H_0}+\eta\underbrace{H(1)}_{H_0+V}
\,.
\label{Hlin01}
\end{equation}  
Quantum phase transitions are likely to appear if the limiting hamiltonians $H(0)$ and $H(1)$ correspond to two essentially different dynamical modes of the system, e.g. such represented by distinct dynamical symmetries (examples given below).

The evolution of the hamiltonian eigenvectors can be expressed as a unitary transformation in the Hilbert space that naturally conserves traces of all operators.
This yields very simple predictions for bulk properties of the spectrum of hamiltonian (\ref{Hlin}), namely for the average energy ${\overline E}=\frac{1}{n}\sum_{i=0}^{n-1}E_i$ and the spread of the spectrum measured by the statistical dispersion $(\Delta E)^2=\frac{1}{n}\sum_{i=0}^{n-1}(E_i-{\overline E})^2$.
While the spectrum average behaves linearly with $\eta$, the dispersion is quadratic \cite{Heinze06}:
\begin{eqnarray}
n{\overline E}&=&{\rm Tr}H_0+\eta{\rm Tr}V
\,,\label{aveE}\\
n^2(\Delta E)^2&=&\left[n\Tr H_0^2-\Tr^2H_0\right]+
2\eta\left[n\Tr(H_0 V)-\Tr H_0\Tr V\right]+
\eta^2\left[n\Tr V^2-\Tr^2V\right]
\,.
\label{disp}
\end{eqnarray}
The minimum of Eq.~(\ref{disp}) is at
\begin{equation}
\eta_{\rm m}=-\frac{n\Tr(H_0 V)-\Tr H_0\Tr V}{n\Tr V^2-\Tr^2V}
\,,
\label{etam}
\end{equation}
where the proximity of energy levels induces rapid structural changes of the hamiltonian eigenfunctions $\ket{\psi_i(\eta)}$, as can be seen from basic perturbation theory applied to $H(\eta+\delta\eta)=H(\eta)+\delta\eta\,V$, e.g. from the overlap formula
\begin{equation}
|\scal{\psi_i(\eta)}{\psi_i(\eta+\delta\eta)}|^2\approx1-(\delta\eta)^2\sum_{j(\neq i)}
\frac{|\matr{\psi_i(\eta)}{V}{\psi_j(\eta)}|^2}{\left[E_i(\eta)-E_j(\eta)\right]^2}
\label{decay1}
\end{equation}
with the squared distance of levels in the denominator.
On the other hand, for $\eta$ far away from $\eta_{\rm m}$ the second term in Eq.~(\ref{Hlin}) totally prevails and the corresponding wave functions approximately coincide with eigenfunctions of $V$. 
The spectrum just linearly blows up.

The most interesting physics happens around the minimum of dispersion (\ref{disp}).
This applies also to ground-state quantum phase transitions driven by $\eta$. 
If such a transition exists in the given model, it most likely appears at a critical point $\eta=\eta_{\rm c}$ that lies somewhere close to $\eta_{\rm m}$.
We will assume that our choice of $H_0$ and $V$ in Eq.~(\ref{Hlin}) was made so that both $\eta_{\rm m}$ and $\eta_{\rm c}$ (if any) are contained in the interval $\eta\in[0,1]$.

Let us have a closer look on situations when quantum phase transitions related to the structure of the ground state can typically take place.
First, it is clear that since matrix elements of the hamiltonian in an arbitrary fixed basis vary with $\eta$ in a smooth (linear) way, any nonanalyticity of the eigenvalue and eigenvector evolutions can only occur if the Hilbert space dimension increases asymptotically, $n\to\infty$ (which for the bosonic models considered here means $N\to\infty$).

Elementary calculation yields the following expressions for the derivatives of the ground-state energy:
\begin{equation}
\tfrac{d}{d\eta}E_0(\eta)=\matr{\psi_0(\eta)}{V}{\psi_0(\eta)}
\,,\quad
\tfrac{d^2}{d\eta^2}E_0(\eta)=-2\sum_{i>0}\frac{|\matr{\psi_i(\eta)}{V}{\psi_0(\eta)}|^2}{E_i(\eta)-E_0(\eta)}
\,.
\label{derivatives}
\end{equation}
Since the second derivative in Eq.~(\ref{derivatives}) cannot be positive ($E_i>E_0\ \forall\,i>0$), the ground-state average $\ave{V}_0\equiv\matr{\psi_0}{V}{\psi_0}$ never increases. 
A common situation for the ground-state quantum phase transition to appear is when $V$ is a semi-positively definite operator or when some constrains do not allow the average $\ave{V}_0$ to become negative. Then it is likely that the continuously decreasing average $\ave{V}_0$ incidentally reaches zero at a certain point $\eta_{\rm c}$. If so, $\eta_{\rm c}$ represents a critical point of the ground-state evolution. Indeed, at this point the first derivative of the ground-state energy in Eq.~(\ref{derivatives}) gets fixed, $\tfrac{d}{d\eta}E_0=0$ for $\eta\geq\eta_{\rm c}$, and the second derivative jumps to zero, $\tfrac{d^2}{d\eta^2}E_0=0$ for $\eta>\eta_{\rm c}$. From Eq.~(\ref{derivatives}) we know that everywhere on the right of the critical point there is $\matr{\psi_0}{V}{\psi_i}\matr{\psi_i}{V}{\psi_0}=0$ for all $i$ (including $i=0$), and this yields
$\ave{V^2}_0\equiv\matr{\psi_0}{V^2}{\psi_0}=0$ for $\eta>\eta_{\rm c}$. In other words, the ground state becomes an eigenstate of $V$ with zero eigenvalue.

The situation described above constitutes a second-order ground-state phase transition from a \lq\lq less symmetric\rq\rq, $\ave{V}_0>0$, to a \lq\lq more symmetric\rq\rq, $\ave{V}_0=\ave{V^2}_0=0$, phase. 
The notion of symmetry is invoked here in relation to the spontaneous symmetry breaking: the \lq\lq less symmetric\rq\rq\ form of the ground state usually brakes a certain symmetry that the hamiltonian itself maintains.
The ground-state average $\ave{V}_0$ can be used as a quantum order parameter and related to the classical (mean-field) order parameter introduced in Subsec.~\ref{se:lan}.

As will be discussed below, in case of the second-order QPT a nonanalytic (for $n\to\infty$) change of this parameter is connected with a singular growth of the level density at $\eta\to\eta_{\rm c}$ and $E\to E_0$.
Indeed, as follows from the Pechukas-Yukawa theory \cite{Pechukas83,Yukawa85}, the evolution of levels with $\eta$ for a linear hamiltonian (\ref{Hlin}) is analogous to one-dimensional dynamics of a 2D Coulomb gas (however, with the product charges also subject to specific variations).
To make any of the trajectories nonanalytic, one needs to produce an infinite local growth of \lq\lq charge\rq\rq\ at the corresponding place.
The Pechukas-Yukawa approach and its implications for quantum phase transitions will be discussed in Subsec.~\ref{se:dy}.

For the second-order QPT the order parameter changes continuously, but with a discontinuous derivative.
In contrast, the first-order QPT involves a discontinuous (for $n\to\infty$) jump of the order parameter itself, i.e., the discontinuity of already the first derivative in Eq.~(\ref{derivatives}) at $\eta=\eta_{\rm c}$.
In terms of the Coulomb gas analogy, this needs locally an infinite \lq\lq force\rq\rq, which may be caused by a crossing of a pair of levels (or their sharp anticrossing, indistinguishable from a real crossing).
Although such effects are usually accompanied by an infinite growth of the level density, similar to the second-order transition, this is not necessarily the case.
Therefore, the mechanisms underlying the first- and second-order transitions (a sharp crossing or anticrossing of two levels and a local singularity of the level density, respectively) may be interrelated, but in general are different. 


\subsection{Simple example: Lipkin model}
\label{se:lip}

We would like to sketch here QPT properties of the model introduced by Lipkin, Meshkov and Glick in 1965 \cite{Lipkin65,Meshkov65,Glick65}, which is often referred to as the Lipkin model. 
This will take us very close to the case of the interacting boson model that will be opened in the next section. 
One can find extensive literature investigating various properties of the Lipkin model, including its phase transitional behavior, see e.g. Refs.~\cite{Gilmore78,Gilmore81,Ring80,Heiss88,Dusuel04,Vidal04,Heiss05,Leyvraz05}. 

The model is formulated in terms of pseudospin operators $\{J_z,J_+,J_-\}$ that form the SU(2) algebra. 
A simplified hamiltonian may be taken for example as follows,
\begin{equation}
H'(\zeta)=J_z-\zeta\tfrac{1}{N}(J_++J_-)^2
\,,
\label{LipHam}
\end{equation}
where $\zeta$ is a control parameter that should not yet be identified with $\eta$ from Eq.~(\ref{Hlin01}), see below, and $N$ is related to the spin quantum number $j=\tfrac{1}{2}N$ denoting the chosen SU(2) representation. 
There exist various modifications of Eq.~(\ref{LipHam}) with other quadratic combinations of the pseudospin operators in the form $V=\tfrac{1}{N}[J_-^2+J_+^2+w(J_-J_++J_+J_-)]$ containing an additional parameter $w$.

The usual interpretation of the SU(2) algebra involved in Eq.~(\ref{LipHam}) is in terms of fermionic operators:
\begin{equation}
J_z=-\tfrac{1}{2}\sum_{i=1}^{\Omega}a_{i-}^{\dagger}a_{i-}+\tfrac{1}{2}\sum_{i=1}^{\Omega}a_{i+}^{\dagger}a_{i+}
\,,\quad
J_{\pm}=\sum_{i=1}^{\Omega}a_{i\pm}^{\dagger}a_{i\mp}
\,.
\label{pseudof}
\end{equation}
Here, $a_{i-}^{\dagger}$ and $a_{i-}$ create and annihilate fermions on the lower single-particle level, while $a_{i+}^{\dagger}$ and $a_{i+}$ do the same for the upper level. Both levels have capacity $\Omega$ and the total number of fermions, $N=\sum_i(a_{i-}^{\dagger}a_{i-}+a_{i+}^{\dagger}a_{i+})=N_-+N_+$, must satisfy the condition $N\leq\Omega$. 
The hamiltonian (\ref{LipHam}) and its generalized forms conserve the parity
\begin{equation}
\Pi=(-)^{N_+}=(-)^{J_z+j}
\,.
\label{LipPar}
\end{equation}
Alternatively, assuming $J_{\bullet}=\frac{1}{2}\sum_{i}\sigma_{\bullet}^{(i)}$, the Lipkin hamiltonian can be interpreted as describing interactions in an infinite array of spin-$\tfrac{1}{2}$ particles. 
As the strength of interactions between fermions on both levels---or between individual spins---increases with $\zeta$, the ground state is tempted to switch from the form with $(N_-,N_+)=(N,0)$ (all fermions on the lower level, or all spins down) to a \lq\lq diamagnetic\rq\rq\ form with both average occupation numbers $\ave{N_+}$ and $\ave{N_-}$ nonzero.
Indeed, this happens at a certain critical point $\zeta_{\rm c}$, which will be determined below for a slightly modified hamiltonian by the methods outlined in the previous subsections.

Using the Holstein-Primakoff mapping,
\begin{equation}
J_z=b^{\dagger}b-j
\,,\quad
J_+=b^{\dagger}\sqrt{2j-b^{\dagger}b}
\,,\quad
J_-=\sqrt{2j-b^{\dagger}b}\ b
\,,
\label{LipHP}
\end{equation}
one can translate the hamiltonian into the bosonic form.
However, this type of bosonic representation is not convenient for the present purposes since the total number of bosons $b^{\dagger}b$ is not conserved and since the boson interactions are of unlimited order (because of the square root).
A simpler alternative is to employ the Swinger mapping,
\begin{equation}
J_z=\tfrac{1}{2}(t^{\dagger}t-s^{\dagger}s)
\,,\quad
J_+=t^{\dagger}s
\,,\quad
J_-=s^{\dagger}t
\,,
\label{pseudob}
\end{equation}
where $s^{\dagger},s$ and $t^{\dagger},t$ create and annihilate two types of bosons. 
Since the parity (\ref{LipPar}) can be expressed as $\Pi=(-)^{n_t}$, where $n_t=t^{\dagger}t$ is the $t$-boson number operator, it is natural to consider $s$ to be a scalar and $t$ a pseudoscalar boson. 
The original pseudospin algebra of the model is now expressed in terms of the spectrum generating algebra U(2) of the system of $s$ and $t$ bosons with $N=s^{\dagger}s+t^{\dagger}t=2j$. We write hamiltonian (\ref{LipHam}) in a slightly modified form,
\begin{equation}
H(\eta)=(1-\eta)\left[-\tfrac{1}{N}(t^{\dagger}s+s^{\dagger}t)(t^{\dagger}s+s^{\dagger}t)\right]+\eta n_t=H'(\zeta)-\zeta J_z+(1-\zeta)\tfrac{N}{2}
\,,
\label{LipHam2}
\end{equation}
where the control parameter $\zeta$ was replaced by $1-\eta\in[0,1]$. 
The limits $\eta=0$ and $\eta=1$ are characterized by O(2) and U(1) dynamical symmetries, respectively, since the corresponding hamiltonians coincide with the Casimir invariants of the O(2) or U(1) subalgebras of U(2). 

The U(1) case ($\eta=1$) represents a system of noninteracting bosons, with the $s$- and $t$-boson energies set to $\epsilon_s=0$ and $\epsilon_t=1$, respectively. The term with $(1-\eta)$ introduces interactions between bosons of both types. If these interactions are too weak, the ground state for infinite boson numbers will be a pure $s$-boson condensate. At a certain critical interaction strength, however, the ground state wave function flips into a mixed condensate of $s$ and $t$ bosons. The critical strength can be obtained from the variational analysis with trial states $\ket{N,q}\propto(s^{\dagger}+qt^{\dagger})^N\ket{0}$, as described in Subsec.~\ref{se:fin}, which leads to the following expression for the classical potential energy:
\begin{equation}
V(\eta;q)=\frac{(5\eta-4)q^2+\eta q^4}{(1+q^2)^2}
\,.
\label{LipPot}
\end{equation}
Taking into account the Taylor expansion $(1+q^2)^{-2}=1-2q^2+3q^4-4q^6+5q^8-\dots$, we see that the potential (\ref{LipPot}) has the general Landau-like form (\ref{free}) with the cubic term missing. Consequently, the nonanalytic evolution at $\eta_{\rm c}=\tfrac{4}{5}$ (where the quadratic term in the numerator changes its sign) represents a second-order ground state phase transition.
For $\eta\geq\eta_{\rm c}$, the potential in Eq.~(\ref{LipPot}) has a minimum at $q_{\rm m}=0$, hence the ground state is indeed the pure $s$ condensate, as anticipated. For $\eta=(\eta_{\rm c}-\varepsilon)<\eta_{\rm c}$, there exist two degenerate minima at 
\begin{equation}
q_{\rm m}=\pm\sqrt{\frac{5\varepsilon}{8-5\varepsilon}}
\,,
\label{LipMin}
\end{equation}
describing the mixed condensates. For $\varepsilon\to 0$, the minima converge to 0 as $q_{\rm m}\approx\pm\sqrt{\tfrac{5}{8}\varepsilon}$. The critical exponent for the \lq\lq order parameter\rq\rq\ $q_{\rm m}$ is therefore equal to $\tfrac{1}{2}$.

The Lipkin hamiltonian can be generalized to get also the first-order phase transitions \cite{Feng81}. 
The only way to do this (while preserving the two-body character of the model) is to sacrifice the parity conservation.
Indeed, as proposed in Ref.~\cite{Vidal06}, modifying hamiltonian (\ref{LipHam2}) to the form
\begin{equation}
H^{\chi}(\eta)=(1-\eta)\left[-\tfrac{1}{N}(t^{\dagger}s+s^{\dagger}t+\chi t^{\dagger}t)(t^{\dagger}s+s^{\dagger}t+\chi t^{\dagger}t)\right]+\eta n_t
\,,
\label{LipHam3}
\end{equation}
which contains parity-violating terms such as $t^{\dagger}t^{\dagger}ts$ etc., one obtains the potential
\begin{equation}
V^{\chi}(\eta;q)=\frac{(5\eta-4)q^2-4\chi(1-\eta)q^3+(\eta+\eta\chi^2-\chi^2)q^4}{(1+q^2)^2}
\,.
\label{LipPot2}
\end{equation}
Here, the cubic term is already present for $\chi\neq 0$ and the ground state first-order phase transition takes place. It turns out that for potential (\ref{LipPot2}) the condition (\ref{cond1st}) holds exactly regardless of the distance from the second-order critical point. The first-order phase transition therefore appears at
\begin{equation}
\eta^{\chi}_{\rm c}=\frac{4+\chi^2}{5+\chi^2}
\,.
\label{LipCr}
\end{equation}
In the following section we will see that these results are very close to those obtained within the interacting boson model.

\subsection{Historical and terminological remarks}
\label{se:hister}

As mentioned in Sec.~\ref{se:int}, the term {\em quantum phase transition\/} comes from physics of infinite lattice systems of spin-like objects interacting via finite-range interactions.
The order-disorder phase transitions in such systems at zero temperature are driven by external control parameters and can be related to zero point motions, i.e. purely quantum fluctuations, of the lattice constituents.
The advent of this kind of physics was marked by a pioneering work of Hertz \cite{Hertz76} in 1976.
At present, the QPT field belongs to one of the most rapidly growing branches of condensed matter physics \cite{Sachdev99,Vojta03}.

On the other hand, the use of the QPT term in the context of models presently studied \cite{Gilmore78,Gilmore79,Dieperink80,Gilmore81} might seem slightly confusing, since---as we saw---the infinite-$N$ limit of such models is just the limit of classical physics.
However, some of the models that belong to the category \lq\lq finite\rq\rq\ are indeed very close to those studied in condensed matter physics.
This is most evident for the Lipkin model \cite{Lipkin65,Meshkov65,Glick65}, which can be cast as a hamiltonian describing an infinite chain of spin-$\tfrac{1}{2}$ particles interacting by infinite-range interactions.
A seemingly minor difference from the other lattice models---the infinite range of interactions---creates the necessity to damp the interaction constant with increasing $N$ and therefore leads to all consequences following from the convergence to the mean-field description with $N\to\infty$.
Various forms of the interacting boson model \cite{IA87}, including the one with $s$ and $d$ bosons, is then just a natural extension of the Lipkin model from U(2) to higher spectrum generating algebras.
As will be shown in the forthcoming sections, this extension results in a considerable enrichment of the phase structure, allowing also the first-order phase transitions to appear \cite{Gilmore79,Dieperink80,Dieperink80b,Feng81}.

The history of {\em interaction-driven phase transitions}, as one may alternatively call such phenomena, is however much longer (some historical remarks can be found in Ref.~\cite{Rosensteel05}).
To our knowledge the first author who used the term \lq\lq phase transition\rq\rq\ in this context was Thouless in 1961 \cite{Thouless61}.
It was in connection with what is now known as a \lq\lq collapse of the random phase approximation\rq\rq\ (RPA) in nuclei: at some critical value of the hamiltonian control parameter the RPA phonon frequency (determining the elementary vibrational mode of the system) drops to zero and becomes imaginary beyond this point.
This can be considered as a microscopic signature of a sudden structural change of the nuclear ground state.

The Lipkin \cite{Lipkin65,Meshkov65,Glick65} and related pseudospin models created a new wave of interest in similar phenomena, see e.g. \cite{Ring80}.
The unified language for their description (based on coherent states and the catastrophe theory) was developed by Gilmore in the late 1970's \cite{Gilmore78,Gilmore79}.
Note that Gilmore proposed the term \lq\lq ground-state energy phase transitions\rq\rq, or simply {\em ground-state phase transitions}, which we sometimes adopt.
The field continued growing in the 1980's with the discovery of phase transitions in the $sd$-IBM by Dieperink, Scholten, and Iachello \cite{Dieperink80,Dieperink80b,Feng81} and also in herefrom inspired so--called fermion dynamical symmetry model (FDSM) \cite{Ginocchio80c,Wu86,Zhang87,Zhang88,Zhang88b}.
Since the phases in these models are defined by the equilibrium shape of a nucleus in its ground state, the related QPT-like phenomena are often called {\em shape-phase transitions}.

The recent increase of interest in this field comes back to the 1990's \cite{Casten93,Moreno96,Iachello98,Rowe98,Bahri98}.
The topic was reopened by several authors pursuing different goals, mainly an analysis of nuclear structure evolution \cite{Casten93,Iachello98,Casten99,Jolie99} and the application of so-called quasi dynamical symmetries \cite{Rowe98,Bahri98} and critical point symmetries \cite{Iachello00,Iachello01,Iachello03,Iachello05,Caprio07}.
Symmetry (regardless of its concrete incarnation) seems to be a unifying theme in a great majority of shape-phase transitional studies in nuclear physics.
Whereas in the infinite lattice models the quantum phase transition separates ordered and disordered phases of the lattice, in nuclear-related models the transition is usually between two specific dynamical symmetries of the system, i.e. between two different types of order.
In this respect, such transitions can be compared to {\em structural phase transitions\/} in solid-state physics.

In studies of quantum shape-phase transitions in nuclei, the interacting boson model, in its various forms, has attracted the major attention.
Apart from a comparison with actual nuclear data \cite{Dieperink80,Frank89,Iachello98}, the IBM soon became an important testing ground for theoretical investigations of general concepts and methods, see e.g. Refs.~\cite{Moreno96,Cejnar00,Cejnar01}.
Both these aspects are relevant since the IBM with its rich phase structure offers a basic example of a system whose features differ in many respects from the traditional QPT systems studied in the context of solid state physics.
Therefore, the model may provide essential hints for deeper understanding of the QPT physics in general.

 
\section{The interacting boson approximation}
\label{se:iba}

The Interacting Boson Model (IBM) was proposed in 1975 by Iachello and Arima to describe collective excitations of heavy or medium mass atomic nuclei \cite{Arima76,Arima78,Arima79}. 
This model combined ingredients of the two most successful paradigms used in nuclear physics at that time: the shell model and the geometrical (or collective) model \cite{Ta93}.  
The  shell  model considers the nucleus as an ensemble of weakly interacting fermions occupying single-particle orbits in the nuclear mean field.  
Despite a considerable truncation of the model Hilbert space achieved by activating only the nucleons on valence shells, calculations in heavy nuclei away from magic numbers were prohibitively complex.   
The geometric model attacked the nuclear many-body problem from the other side: 
Heavy nuclei can in some situations be considered as droplets of a quantum liquid, with elementary excitations identified with highly correlated collective vibrations and rotations. 
In even-even nuclei, the basic constituents of the model are quadrupole phonons which can be represented as bosons with spin and parity $l^{\pi}=2^+$. 
The collective model has been successful in describing certain classes of nuclei away from closed shells. 

The IBM is intermediate between these two complementary approaches in that it connects the bosonic behavior of the geometric model to the fermionic nature of the shell model. 
This is achieved using the pairing property of short-range residual interactions. 
Pairwise coupled nucleons (or holes, vacancies left by missing nucleons if the shell is more than half-filled) behave much like bosons.
The energetically most likely combination of two identical nucleons coupled by a short-range force is the one with zero total angular momentum.
This can be approximated by an $s$ boson, while the bifermion combination with angular momentum $2$ maps onto a $d$ boson.

The original version of the interacting boson model, nowadays abbreviated as IBM-1 \cite{IA87}, is applicable to even--even nuclei.
The IBM-1 does not separate bosons connected with proton-proton and neutron-neutron pairs (this is done in an extended version of the model, the IBM-2 \cite{IA87}, which is suitable for the description of isovector collective excitations) and does not consider bosons connected with mixed proton-neutron pairs (these bosons, increasingly relevant as approaching the ${\cal N}\approx{\cal Z}$ nuclei, are introduced in more advanced versions of the model, IBM-3 and IBM-4 \cite{Elliot85}).
The IBM-1 also does not consider single-nucleon excitations and their couplings with nucleon pairs (these are treated in the interacting boson-fermion model, the IBFM \cite{II91}, which is applicable in odd nuclei).
Some modified versions of the IBM also include bosons with other spins and parities, such as $g$ ($4^+$), $p$ ($1^-$), and $f$ ($3^-$) bosons \cite{IA87}.
Various IBM extensions and their quantum phase transitions will be discussed in Sec.~\ref{se:ext}, while here and in the following two sections we will focus on properties of the IBM-1.

\subsection{Foundations and the algebraic structure}
\label{se:alg}

The interacting boson model, including its simplest version IBM-1, benefits from its transparent algebraic formulation.
As indicated above, the model building blocks are $s$ and $d$ bosons that represent phenomenological images of $0^+$ and $2^+$ pairs of valence nucleons or holes and are also closely related to the basic quanta of nuclear collective excitations.
Unitary transformations among the six states $s^{\dag}\ket{0}$ and $d^\dag_m\ket{0}$, with $m=0,\pm1,\pm 2$, generate the Lie group U(6), which is identified with the {\em spectrum generating (dynamical) group\/} of the model.
The 36 generators of the associated algebra can be written in the form $b_{lm}^\dag b_{l'm'}$, where $b_{00}^\dag\equiv s^{\dag}$ and $b_{2m}^\dag\equiv d_m^{\dag}$.

Although the separate boson numbers $n_s$ and $n_d$ are apparently not conserved by the dynamical U(6) algebra, the sum $n_s+n_d=N$ is. 
It means that collective states of an even-even nucleus with $N_{\rm F}$ valence nucleons (or valence holes) are mapped to the IBM-1 Hilbert space of $N=\tfrac{1}{2}N_{\rm F}$ bosons.
This is in contrast to the quadrupole phonon model, a bosonized version of the geometric model, where bosons directly represent quanta of collective excitations, so that their number varies within one nucleus \cite{Janssen74}. 
For instance, in the latter model the ground state of a spherical nucleus is treated as the vacuum, a state with $N=0$, while in the IBM the same state coincides with a condensate of $s$ bosons with given $N>0$.
In spite of these differences, both models can be connected and rooted in an underlying microscopic treatment \cite{Klein82,Blaizot78,Blaizot78b,Marshalek06}.

An $sd$-boson hamiltonian with one- and two-body interactions that conserves the total boson number and the total angular momentum has the following general form:
\begin{equation}
H=E_0+\epsilon_d\,n_d+\sum_{l_1l_2l'_1l'_2l}v^{(l)}_{l_1l_2l'_1l'_2}\left[[b^\dag_{l_1}\times b^\dag_{l_2}]^{(l)}\times[\tilde b_{l'_1}\times\tilde b_{l'_2}]^{(l)}\right]^{(0)}_0.
\label{3_ibmham}
\end{equation}
The first term is a constant, which may be included to quantify the nuclear binding energy of the core. 
The second term represents the relevant one-body contributions (the $s$-boson part can be eliminated using the relation $n_s=N-n_d$). 
The third part corresponds to {\em two-body interaction}, the coefficients $v^{(l)}_{l_1l_2l'_1l'_2}$ being related to the interaction reduced matrix elements between normalized two-boson states with total angular momentum $l$.
We use the standard notation ${\tilde b}_{lm}\equiv(-)^{l-m}b_{l(-m)}$, it is ${\tilde s}=s$ and ${\tilde d}_{m}=(-1)^{m}d_{-m}$, and $[A^{(l_1)}\times B^{(l_2)}]^{(l)}_M\equiv\sum_{m_1m_2}(l_1m_1l_2m_2|lm)A^{(l_1)}_{m_1}B^{(l_2)}_{m_2}$, where $A^{(l_1)}$ and $B^{(l_2)}$, respectively, are rank $l_1$ and $l_2$ spherical tensors.

Taking into account the hermicity of the hamiltonian and its required symmetry under the time reversal (reality of coefficients), one finds out that only six real interaction coefficients determine the properties of the spectrum (up to the constant shift $E_0$).
The hamiltonian can then be rewritten in the following multipole form \cite{IA87}:
\begin{eqnarray}
H&=&E´_0+\epsilon_d\,n_d+c_1(L\cdot L)+c_2(Q^\chi\cdot Q^\chi)+c_3(T^{(3)}\cdot T^{(3)})+c_4(T^{(4)}\cdot T^{(4)})
\,, \label{2_ibmham}\\
&&L_m=\sqrt{10}[d^\dag\times{\tilde d}]^{(1)}_m
\,,\label{L}\\
&&Q_m^\chi=s^{\dagger}{\tilde d}_m+d^{\dagger}_m s+\chi[d^{\dagger}\times{\tilde d}]^{(2)}_m
\,,\label{Q}\\
&&T_m^{(k)} =[d^{\dag}\times{\tilde d}]^{(k)}_m
\,,\label{T}
\end{eqnarray}
where we introduced the scalar product $(A^{(l)}\cdot B^{(l)})\equiv\sqrt{2l+1}[A^{(l)}\times B^{(l)}]^{(0)}_0$.
The free parameters now read as $\{E_0,\epsilon_d,\chi,c_1,\dots,c_4\}$.
Three operators $L_m$ ($m=0,\pm 1$) in Eq.~(\ref{L}) define spherical components of the {\em angular momentum\/} and generate the physical rotational group O(3) of the model (the hamiltonian is a scalar with respect to this group).\footnote{In this paper, we denote the angular momentum operators and the quantum number associated with the squared angular momentum by the same symbol. The rotational algebra and the other orthogonal algebras are referred to as O($n$), irrespective of whether the determinant is constrained to +1 or not.}
Similarly, $Q_m^{\chi}$ ($m=0,\pm 1,\pm 2$) in Eq.~(\ref{Q}) represent spherical components of the {\em quadrupole operator}, determining, e.g. the E2 transition rates.
Parameter $\chi$, which can be chosen within the interval $|\chi|\in[0,\tfrac{\sqrt{7}}{2}]$, is assumed to have the same value in the hamiltonian and in the E2 transition operator (so called consistent-$Q$ formalism \cite{Warner82}).

Numerical procedures exist to obtain the eigenvalues and eigenvectors of the IBM hamiltonian in the general case, but the problem can be solved analytically for some particular choices of parameters.
These special cases correspond to {\em dynamical symmetries\/} associated with the algebraic reductions
\begin{equation}
{\rm U}(6)\supset\left\{\begin{array}{c}{\rm U}(5)\supset{\rm O}(5)\\{\rm SU}(3)\\{\rm O}(6)\supset{\rm O}(5)\end{array}\right\}\supset{\rm O}(3)
\,.
\label{3_ibmlat}
\end{equation}
The dynamical symmetries associated with the three chains are named vibrational, U(5), rotational, SU(3), and $\gamma$-unstable, O(6). 
Each of them provides a complete basis for the numerical solution.
The algebras appearing in Eq.~(\ref{3_ibmlat}) are subalgebras of U(6) generated by operators of the type $b^{\dag}_{lm}{\tilde b}_{l'm'}$ and their linear combinations; explicit forms are listed e.g. in Ref.~\cite{IA87}.

With the subalgebras U(5), O(5), O(3), SU(3), and O(6) there are associated one linear and five quadratic Casimir operators.
Denoting by $C_n[{\rm G}]$ the $n$th-order Casimir operator of the group G, the general IBM hamiltonian with up to two-body interactions can be written in the following way,
\begin{equation}
H=E_0+a\,C_1[{\rm U(5)}]+b\,C_2[{\rm U(5)}]+c\,C_2[{\rm O(5)}]+d\,C_2[{\rm O(3)}]+e\,C_2[{\rm SU(3)}]+f\,C_2[{\rm O(6)}]
\,,\label{3_ibmhamc}
\end{equation}
where we introduced yet another parametrization with $\{E_0,a,\dots,f\}$.
Explicit expressions for Casimir operators as well as for the mapping between the alternative parameter sets are given e.g. in Ref.~\cite{IA87}. 
We point at different conventions used in the literature; below we will implicitly employ the definitions listed in Ref.~\cite{Cejnar98}.

If some of the coefficients in Eq.~(\ref{3_ibmhamc}) vanish such that the hamiltonian contains only Casimir operators of the subalgebras associated with a single reduction in Eq.~(\ref{3_ibmlat}), the system possesses the corresponding dynamical symmetry and the energy eigenvalue problem can be solved analytically.
The hamiltonian eigenfunctions are determined by the hierarchy characterizing the embedding of irreducible representations of groups in chains (\ref{3_ibmlat}).
The eigenfunctions are determined by the corresponding irrep labels (these represent conserved quantum numbers) and read as $\ket{[N],n_d,\tau,\nu_{\Delta},L}$ for the U(5) dynamical symmetry, $\ket{[N],\lambda,\mu,K_L,L}$ for the SU(3), and $\ket{[N],\sigma,\tau,\nu_{\Delta},L}$ for the O(6).
Reduction rules specifying which labels are contained in the given irrep can be found in Ref.~\cite{IA87}.
Since the eigenvectors are independent of the actual values of hamiltonian parameters, they make available clearcut predictions for observables like transition rates or transfer amplitudes, which can be used to experimentally verify whether a particular dynamical symmetry is present.
Analytic expressions for energies are as follows:
\begin{eqnarray}
H\ket{[N],n_d,\tau,\nu_{\Delta},L}=a\,n_d+b\,n_d(n_d+5)+c\,\tau(\tau+3)+d\,L(L+1)
\qquad&&{\rm U(5)}
\,,\nonumber\\
H\ket{[N],\lambda,\mu,K_L,L}=e\,[\lambda^2+\mu^2+\lambda\mu+3(\lambda+\mu)]+d\,L(L+1)
\qquad\quad\ \ \,&&{\rm SU(3)}
\,,\label{energy}\\
H\ket{[N],\sigma,\tau,\nu_{\Delta},L}=f\,\sigma(\sigma+4)+c\,\tau(\tau+3)+d\,L(L+1)
\qquad\qquad\qquad&&{\rm O(6)}
\,.\nonumber
\end{eqnarray}
Note that the energies do not depend on the so-called missing labels $\nu_{\Delta}$ and $K_L$, respectively, that characterize a nonunique embedding of O(3) irreps in O(5) and SU(3) ones.

Besides the dynamical symmetries defined above, there exist two additional ones differing from their respective counterparts in Eq.~(\ref{3_ibmlat}) by the choice of the relative phase between $s$ and $d$ bosons \cite{Isacker85,Kusnezov97,Cejnar98b,Shirokov98,Cejnar01b}.
These additional symmetries can be represented by the following chains
\begin{equation}
{\rm U}(6)\supset\left\{\begin{array}{c}\overline{{\rm SU}(3)}\\\overline{{\rm O}(6)}\supset{\rm O}(5)
\end{array}\right\}\supset{\rm O}(3)
\,,
\label{3_ibmlatadd}
\end{equation}
where $\overline{{\rm SU}(3)}$ and $\overline{{\rm O}(6)}$ can be obtained from \lq\lq standard\rq\rq\ SU(3) and O(6) via the gauge transformation $s^{\dag}\mapsto e^{i\phi}s^{\dag}$ with $\phi=\pi$ and $\pm\tfrac{\pi}{2}$, respectively
[other phases are avoided on the basis of the required reality of the interaction coefficients in Eq.~(\ref{3_ibmham}), i.e. the time-reversal invariance].
Note that since the U(5) chain a priori separates $s$ and $d$ bosons, it has no counterpart in Eq.~(\ref{3_ibmlatadd}).
Although the additional dynamical symmetries are closely related to the original ones (for instance, they yield identical spectra), the corresponding hamiltonians written in terms of the original Casimir operators look like if having no dynamical symmetry.
The ratios of coefficients in expansion (\ref{3_ibmhamc}) needed to fabricate dynamical symmetries (\ref{3_ibmlatadd}) are as follows:
\begin{eqnarray}
a\,:\,b\,:\,c\,:\,e\,:\,f&=&2\,:\,2\,:\,-6\,:\,-1\,:\,4 
\qquad\ \,\overline{{\rm SU}(3)}
\nonumber\,,\\
a\,:\,b\,:\,e\,:\,f&=&4N\!\!+\!8\,:\,-4\,:\,0\,:\,-1
\qquad\overline{{\rm O}(6)}
\,.
\label{parrat}
\end{eqnarray}
The existence of additional (\lq\lq hidden\rq\rq\cite{Kusnezov97}) dynamical symmetries (\ref{3_ibmlatadd}) results from a more general property of the IBM-1 hamiltonians, namely the existence of discrete transformations in the model parameter space that leave the spectrum unchanged (so called parameter symmetries \cite{Shirokov98}).

To conclude the basic overview of the IBM-1 properties, we note that all hamiltonians (\ref{3_ibmhamc}) with $e=0$, which are transitional between U(5) and O(6) dynamical symmetries, conserve the O(5) quantum number $\tau$, called seniority (a bosonic analog of the fermionic seniority \cite{Ta93}).
As a consequence, these hamiltonians are still integrable---although the U(5) and O(6) invariants are no more integrals of motions in the transitional regime, their role can be taken by the hamiltonian itself. 
There is no general analytical energy formula valid along the [U(5)--O(6)]$\supset$O(5) path but for these hamiltonians the original diagonalization problem has been translated to an equivalent problem of solving a certain set of algebraic equations (a semi-analytic result) \cite{Pan98,Dukelsky01,Dukelsky04}.
The structure of the spectrum and the form of the wave functions exhibit some special features along this transition.
In particular, subsets of levels with different seniority quantum numbers do not interact with each other and therefore cross with no repulsion as the model control parameters vary.

\subsection{Geometrical analysis and the shape-phase structure}
\label{se:geo}

A general method for the determination of the classical limit for hamiltonians of the type (\ref{H}) was described in Subsec.~\ref{se:fin}.
This method can be adapted in the IBM case, yielding a coordinate representation of the bosonic many-body hamiltonians (\ref{3_ibmham}), and simultaneously a geometrical interpretation of the equivalent algebraic hamiltonians (\ref{3_ibmhamc}).
The classical limit can be obtained from condensate states of the type (\ref{conden}),
\begin{equation}
\ket{N,{\bf\alpha}}=\biggl[N!\bigl(1+\sum_m|\alpha_m|^2\bigr)^{N}\biggr]^{-\frac{1}{2}}\biggl[s^{\dagger}+\sum_m\alpha_m d_m^{\dag}\biggr]^N\ket{0}
\label{condenIBM}
\end{equation}
(here with the normalization coefficients included), equivalent to the algebraic coherent state (\ref{alcoh}), or from the corresponding Glauber coherent states (\ref{glaub}).
The classical hamiltonian is constructed in analogy to Eq.~(\ref{klas}).

Coefficients $\alpha_m$ in Eq.~(\ref{condenIBM}) with $m=0,\pm 1,\pm 2$ transform as $l=2$ spherical tensors and are, in general, complex.
To extract coordinates $q_m$ and the associated momenta $p_m$, the following prescriptions can be used \cite{Hatch82}:
\begin{eqnarray}
\alpha_m=\tfrac{1}{\sqrt{2}}\left[q_m^*+ip_m\right]
\,,&&
\alpha_m^*=\tfrac{1}{\sqrt{2}}\left[q_m-ip_m^*\right]
\,,\\
q_m=\tfrac{1}{\sqrt{2}}\left[(-)^m\alpha_{-m}+\alpha_m^*\right]
\,,&&
p_m=\tfrac{i}{\sqrt{2}}\left[(-)^m\alpha^*_{-m}-\alpha_m\right]
\end{eqnarray}
Both $q_m$ and $p_m$ are again the $l=2$ spherical tensors, still complex, but since they satisfy $q_m^*=(-)^m q_{-m}$ and $p_m^*=(-)^m p_{-m}$, there are only 5 independent real values for each of them.
These define 5 degrees of freedom associated with the model.

By definition, the classical hamiltonian can only contain scalar combinations of coordinates and momenta.
There are only two independent scalar combinations of the $q$'s, namely \cite{Noack68}
\begin{equation}
[q\times q]^{(0)}=\tfrac{1}{\sqrt{5}}\beta^2
\,,\quad
\left[[q\times q]^{(2)}\times q\right]^{(0)}=-\sqrt{\tfrac{2}{35}}\beta^3\cos 3\gamma
\,,
\label{scalcomb}
\end{equation}
where we included the parametrization through the well known Bohr quadrupole shape variables $\beta$ and $\gamma$ \cite{BM75}. 
As seen, $\beta$ represents a radius in the 5-dimensional configuration space, while $\gamma$ can be associated with one of the hyperspherical angles.
The remaining 3 angles are related to Euler angles describing a rotation to the frame where both tensors $q_m$ and $p_m$ become diagonal (the principal axis system). 
The momenta associated with Euler angles appear in the kinetic energy, but the potential energy depends only on $\beta$ and $\gamma$.
It has the following general form \cite{Dieperink80,Dieperink80b,Feng81,Ginocchio80,Ginocchio80b}
\begin{equation}
V(\beta,\gamma)=\frac{A\beta^2+B\beta^3\cos 3\gamma+C\beta^4}{(1+\beta^2)^2}
\label{Vfunc}
\,,
\end{equation}
where the denominator follows from elimination of the degrees of freedom connected with the $s$ boson (see Sec.~\ref{se:fin}). 
Coefficients $A$, $B$, and $C$ are determined from the general IBM hamiltonian parameters, e.g. from $a,b,\dots,f$ in Eq.~(\ref{3_ibmhamc}).
Such expressions can be found in the literature \cite{IA87} and we do not list them here.

Potential (\ref{Vfunc}) has the general Landau-like form (\ref{free}) except it has two variables, $\beta$ and $\gamma$.
However, we see that since $\gamma$ enters the potential (\ref{Vfunc}) only through the $\cos 3\gamma$ dependence in the cubic term, the minimization in this variable can be performed separately. 
The global minimum is either at $\gamma_{\rm m}=0$ (equivalently at $\tfrac{2\pi}{3}$ or $\tfrac{4\pi}{3}$) for $B<0$, or at $\gamma_{\rm m}=\tfrac{\pi}{3}$ (equivalently at $\pi$ or $\frac{5\pi}{3}$) for $B>0$.
The second possibility can be expressed via changing the sign of the corresponding $\beta_{\rm m}$ and simultaneously setting $\gamma_{\rm m}=0$.
Therefore, the global minimum of the potential (\ref{Vfunc}) can be found using a simplified form with $\gamma=0$ and $\beta\in(-\infty,+\infty)$, which is nothing but the minimization of the Landau potential (\ref{free}) with $\xi\equiv\beta\,{\rm sign}(\cos 3\gamma)$.
We shall stress, however, that this equivalence is justified only if searching for the {\em global\/} minimum of $V(\beta,\gamma)$; any dynamical conclusions based on the above procedure (e.g. identification of secondary minima with $\beta<0$ etc.) would be strongly misleading since the dependence on $\gamma$ is in general relevant.

\begin{figure}[t]
\centering
\includegraphics[width=13.9cm]{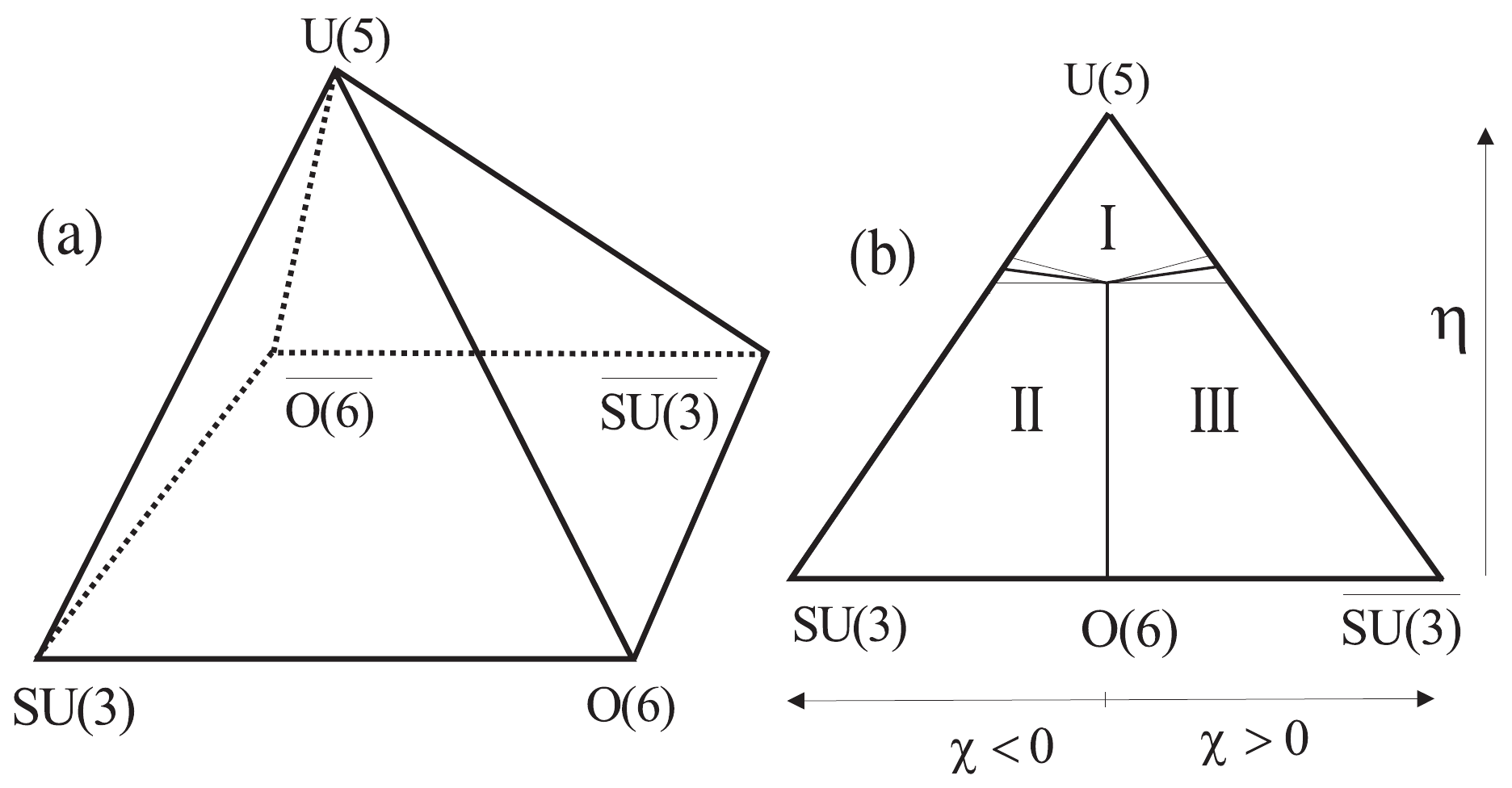}
\caption{\protect{\small (a) A symbolic representation of the parameter space of the interacting boson model with its dynamical symmetries. The rear side of the pyramid is equivalent to the front side and any triangle is related to all the others by a similarity transformation. (b) The extended Casten triangle of hamiltonian (\ref{ising}) with phases I (spherical), II (prolate), and III (oblate). The extended triangle maps the front side of the pyramid in panel (a). The I--II and I--III first-order phase transitions are connected with narrow coexistence regions while the \lq\lq triple point\rq\rq\ at the I-II-III intersection represents a second-order phase transition.}}
\label{fi:tri}
\end{figure}

To separate the \lq\lq phases\rq\rq\ determined by the potential (\ref{Vfunc}), the conditions derived in Subsec.~\ref{se:lan} can be applied \cite{Jolie02}. 
Let us stress that in the present case the phase separatrices given by Eqs.~(\ref{cond1st}) and (\ref{cond1st'}) are valid exactly.
The IBM phases can be described as follows \cite{Jolie01}:
\begin{description} 
\item{I.} Phase with $\beta_{\rm m}=0$ is the \lq\lq more symmetric\rq\rq\ phase. It appears for $A>\tfrac{B^2}{4|C|}$. This phase is interpreted as {\em spherical}, since $\beta_{\rm m}=0$ implies that the ground state is formed by a condensate of only the $s$ bosons, $\ket{\psi_0}\propto(s^{\dag})^N\ket{0}$ (up to the leading order terms in $N$), thus $\tfrac{1}{N}\matr{\psi_0}{n_d}{\psi_0}=0$. Hamiltonians with dynamical symmetry U(5) belong to this phase (we assume $\epsilon_s<\epsilon_d$).
\item{II.} Phase with $\beta_{\rm m}>0,\gamma_{\rm m}=0$ is \lq\lq less symmetric\rq\rq and appears for $A<\tfrac{B^2}{4|C|}$, $B<0$. It is deformed since the ground state is a condensate $\ket{\psi_0}\propto(s^{\dag}+\beta_{\rm m} d_0^{\dag})^N\ket{0}$, hence $\tfrac{1}{N}\matr{\psi_0}{n_d}{\psi_0}>0$. Prominent representatives of this phase are hamiltonians with the SU(3) dynamical symmetry. The deformation in this phase is conventionally considered {\em prolate}.
\item{III.} Phase with $\beta_{\rm m}>0,\gamma_{\rm m}=\tfrac{\pi}{3}$, that applies for $A<\tfrac{B^2}{4|C|}$ and $B>0$, is also deformed, but interpreted as {\em oblate}, with $\ket{\psi_0}\propto(s^{\dag}-\beta_{\rm m} d_0^{\dag})^N\ket{0}$ and $\tfrac{1}{N}\matr{\psi_0}{n_d}{\psi_0}>0$. This phase, which is just a mirror conjugate of phase~II, can be represented by hamiltonians with dynamical symmetry ${\overline{\rm SU(3)}}$.
\end{description}
Note that the dynamical symmetries O(6) and ${\overline{\rm O(6)}}$ do not represent separate phases, but lie on prolate-oblate and oblate-prolate phase separatrices between phases~II and III.
The IBM-1 parameter space and its phase structure are schematically depicted in Fig.~\ref{fi:tri}.

A brief comment is needed here on the transition between phases II and III.
According to the Ehrenfest classification, this is the first-order phase transition (the first derivative of the ground-state energy jumps) \cite{Jolie01}.
Nevertheless, it is not of the generic type, since it does not generate the zone of coexisting prolate and oblate phases.
This difference from the original Landau situation (Subsec.~\ref{se:lan}) is obviously connected with the $\gamma$ degree of freedom, whose role in the Landau analysis was explained above.
Indeed, as approaching the $B=0$ line from either side, the secondary \lq\lq minimum\rq\rq\ of the potential (\ref{Vfunc}) at $\gamma=0$ with ${\rm sign}\beta=-{\rm sign}\beta_{\rm m}$ is only a saddle point.
At the critical point, the potential becomes totally flat in $\gamma$.
Another peculiarity is that hamiltonians in parts II and III of the IBM phase diagram can be connected by a similarity transformation, which preserves the energy spectrum \cite{Shirokov98,Cejnar01b}.
This means that, in a certain sense, the phases II and III are just mirror images of each other.

\subsection{Simplified hamiltonian}
\label{se:sim}

In the previous subsection we saw that the geometrical configuration of the ground state is fully determined by 2 effective parameters---the ratios of coefficients in Eq.~(\ref{Vfunc}).
However, the most general hamiltonians (\ref{2_ibmham}) or (\ref{3_ibmhamc}) have 7 free parameters.
It is therefore clear that the phase structure of the IBM-1 can be studied---without the loss of generality---using a restricted, suitably chosen two-parameter hamiltonian.
In the following, we will work with hamiltonians of the type $H=\epsilon_d\,n_d-\kappa(Q^{\chi}\cdot Q^{\chi})$, see Ref.~\cite{Warner82}. 
The above form still contains 3 independent parameters ($\epsilon_d$, $\kappa$, and $\chi$), but can be easily reduced to a two-parameter form by choosing a specific energy scale (e.g. in units of $\epsilon_d$).
While $\epsilon_d$ represents a single-particle energy of $d$ bosons (with respect to the single-particle energy of $s$ bosons, which is set to 0), the value of $\kappa$ measures the overall strength of interactions involved in the scalar product $(Q^{\chi}\cdot Q^{\chi})$ with given value of $\chi$, see Eq.~(\ref{Q}).
Taking into account the discussions in Sec.~\ref{se:qpt}, we rescale the above hamiltonian as follows
\begin{equation}
H^{\chi}(\eta)=(1-\eta)\tfrac{1}{N}(-Q^{\chi}\cdot Q^{\chi})+\eta\,n_d
\,.
\label{ising}
\end{equation}
For a fixed $\chi$ this is the form (\ref{Hlin01}) with $\eta\in[0,1]$ and
\begin{equation}
H_0=-\tfrac{1}{N}(Q^{\chi}\cdot Q^{\chi})\,,\qquad 
V=n_d+\tfrac{1}{N}(Q^{\chi}\cdot Q^{\chi})\,. 
\end{equation}
Let us note that hamiltonian (\ref{ising}) is analogous to the Lipkin model hamiltonian (\ref{LipHam3}) considered above. It will turn out that also the phase analysis is essentially the same.

The case $\eta=1$ of hamiltonian (\ref{ising}) clearly corresponds to the U(5) dynamical symmetry, a non-interacting ensemble of $s$ and $d$ bosons.
An alternative parametrization is therefore the one with $\zeta=1-\eta$.
At $\zeta=0$, the ground state is obviously formed by a condensate of $s$ bosons (the spherical phase), but as $\zeta$ increases the rising interaction strength with $d$ bosons tends to change the nature of the ground state into a more complicated mixture of $s$ and $d$ boson states (the deformed phase).
This mechanism is rather similar to that already discussed for the Lipkin model (Subsec.~\ref{se:lip}).
However, the present system yields a richer structure since the nature of transitions between the noninteracting and interacting phases can be further tuned by changing the quadrupole operator parameter $\chi$.

Comparing the simplified hamiltonian with the one in Eq.~(\ref{2_ibmham}) we see that only two terms of the general form are preserved.
However, if rewriting Eq.~(\ref{ising}) in the form (\ref{3_ibmhamc}), we get
\begin{eqnarray}
H^{\chi}(\eta)
&=&\left[\eta+\tfrac{2}{7N}(1-\eta)\chi\left(\chi+\tfrac{\sqrt{7}}{2}\right)\right] C_1[{\rm U(5)}]
+\tfrac{2}{7N}(1-\eta)\chi\left(\chi+\tfrac{\sqrt{7}}{2}\right) C_2[{\rm U(5)}]
\nonumber\\ 
&+&\tfrac{1}{N}(\eta-1)\left(1+\tfrac{3}{\sqrt{7}}\chi+\tfrac{2}{7}\chi^2\right) C_2[{\rm O(5)}]
+\tfrac{1}{14N}(1-\eta)\chi(\chi+2\sqrt{7}) C_2[{\rm O(3)}]
\nonumber\\ 
&+&\tfrac{1}{N\sqrt{7}}(\eta-1)\chi C_2[{\rm SU(3)}]
+\tfrac{1}{N}(1-\eta)\left(1+\tfrac{2}{\sqrt{7}}\chi\right) C_2[{\rm O(6)}] 
\label{QChamiltonian}
\end{eqnarray}
(this expansion is valid for definitions of the Casimir invariants given in Ref.~\cite{Cejnar98}), which contains all relevant Casimir operators.
As we already know, the case $\eta=1$ ($\chi$ arbitrary) corresponds to the U(5) dynamical symmetry.
Expansion (\ref{QChamiltonian}) can help us to classify the other dynamical symmetries of hamiltonian (\ref{ising}).
These are located on the $\eta=0$ side of the parameter range: the SU(3) dynamical symmetry at $(\eta,\chi)=(0,-\tfrac{\sqrt{7}}{2})$, the O(6) at $(\eta,\chi)=(0,0)$, and the ${\overline{\rm SU(3)}}$ at $(\eta,\chi)=(0,+\tfrac{\sqrt{7}}{2})$ [cf. Eq.~(\ref{parrat})].

With the restrictions $\eta\in[0,1]$ and $\chi\in[-\tfrac{\sqrt{7}}{2},+\tfrac{\sqrt{7}}{2}]$, the parameter range of hamiltonian (\ref{ising}) can be imaged as a rectangle.
However, since on the $\eta=1$ side $\chi$ does not appear in the hamiltonian, it is more natural to visualize the parameter range as a triangle, see Fig.~\ref{fi:tri}(b). 
This is an {\em extended Casten triangle\/} \cite{Jolie01}, which in addition to the standard dynamical symmetries U(5), SU(3), and O(6) contains also the ${\overline{\rm SU(3)}}$. 
The ${\overline{\rm O(6)}}$ symmetry is, however, not present.

Three comments on the extended triangle are in order:
(a) Although the influence of the parameter $\chi$ on the dynamics fades away with $\eta\to 1$ (which justifies the use of triangle instead of a rectangle), the E2 transition rates are determined by the quadrupole operator $Q^{\chi}$ and therefore still depend on $\chi$ even for $\eta=1$ \cite{Werner00}.
(b) It can be shown that mirror conjugate hamiltonians with $\chi=-|\chi|$ and $\chi=+|\chi|$ are always connected by a unitary transformation, following from the discrete parameter symmetry of the IBM-1 \cite{Shirokov98,Cejnar01b}.
Therefore, the $\chi<0$ and $\chi>0$ halves of the extended Casten triangle are dynamically equivalent, although, as shown below, describing rotors with different types of deformation (prolate and oblate).
(c) We notice in Eq.~(\ref{QChamiltonian}) that on the $\eta=0$ side of the triangle the hamiltonians with $\chi\neq\pm\tfrac{\sqrt{7}}{2}$ still contain a contribution proportional to the sum $C_1[{\rm U(5)}]+C_2[{\rm U(5)}]$.
That is somewhat confusing since this case should correspond to the SU(3)--O(6)--${\overline{\rm SU(3)}}$ transitional regime.
Nevertheless, taking into account the identity
\begin{equation}
C_1[{\rm U(5)}]+C_2[{\rm U(5)}]=\tfrac{1}{2}C_2[\overline{\rm SU(3)}]-2C_2[{\rm O(6)}]+
3C_2[{\rm O(5)}]+\tfrac{1}{2}C_2[{\rm SU(3)}]-C_2[{\rm O(3)}]
\,,
\label{u5}
\end{equation}
one can always eliminate the contribution of U(5) invariants on the $\eta=0$ side by including the ${\overline{\rm SU(3)}}$ Casimir operator into the expansion (\ref{QChamiltonian}).

The geometrical analysis (Subsec.~\ref{se:geo}) of hamiltonian (\ref{ising}) yields the following results \cite{Cejnar00}: The coefficients in the potential energy formula (\ref{Vfunc}) read as
\begin{equation}
A(\eta)=5\eta-4\,,\qquad
B^{\chi}(\eta)=4\sqrt{\tfrac{2}{7}}\chi(1-\eta)\,,\qquad
C^{\chi}(\eta)=\eta-\tfrac{2}{7}\chi^2(1-\eta)\,.
\label{ABC}
\end{equation}
Notice that the latter two coefficients depend on both $\eta$ and $\chi$.
The spinodal point for the deformed-to-spherical evolution appears at $\eta=\eta_{\rm s}\equiv\tfrac{4}{5}$. At this point, $A$ changes from negative to positive values and the potential (\ref{Vfunc}) develops a local (for $\chi\neq 0$) minimum at $\beta=0$. The critical point is located at $\eta=\eta_{\rm c}$,
\begin{equation}
\eta^{\chi}_{\rm c}\equiv\frac{4+\tfrac{2}{7}\chi^2}{5+\tfrac{2}{7}\chi^2}
\,.
\label{etac}
\end{equation}
Here the depths of the $\beta=0$ and $\beta\neq 0$ minima become equal. This condition is fully equivalent to Eq.~(\ref{LipCr}) already derived for the analogous Lipkin hamiltonian. The antispinodal point, where the $\beta\neq 0$ minimum disappears, follows shortly after the critical point, $\eta_{\rm a}(\chi)\geq\eta^{\chi}_{\rm c}$, the exact location being determined by a cubic equation not given here.
For $\chi\neq 0$, the interval from $\eta_{\rm s}$ to $\eta_{\rm a}$ demarcates the phase-coexistence region for a first-order phase transition between phases I and II ($\chi<0$) or I and III ($\chi>0$).

For $\chi=0$ (that is $B=0$) all the three points---spinodal, critical, and antispinodal---merge at $\eta_{\rm c}(0)=\tfrac{4}{5}$ and the QPT between spherical and deformed phases is of the second order.
The critical exponent for the order parameter $\beta$ is $\tfrac{1}{2}$.
When crossing the line $\chi=0$ in a transverse direction within the interval $\eta\in[0,\tfrac{4}{5}]$, the potential becomes $\gamma$-flat and the sign of $B$ changes.
It means that the $(\beta,\gamma)=(\beta_{\rm m},0)$ global minimum (valid at $\chi<0$) moves to $(\beta,\gamma)=(\beta_{\rm m},\tfrac{\pi}{3})$ (valid at $\chi>0$).
Consequently, the line segment $\chi=0$, $\eta\in[0,\tfrac{4}{5}]$ represents a transition from prolate ($\chi<0$) to oblate ($\chi>0$) shapes (phases II and III, respectively).
As discussed in Subsec.~\ref{se:geo}, this is a first-order phase transition, but with no separate spinodal and antispinodal points.
The point $(\eta,\chi)=(\tfrac{4}{5},0)$ can be interpreted as a \lq\lq triple point\rq\rq\ of nuclear deformations \cite{Jolie02} since three phases---spherical, prolate, and oblate---join there, see Fig.~\ref{fi:tri}.
Such a simple phase structure is in agreement with the Landau theory (see Subsec.~\ref{se:lan}).

Let us finally focus on an alternative parametrization of the O(6)--U(5) trajectory with $\chi=0$, which is often used in the literature, see e.g. \cite{Rowe04,Rowe04b}.
It reads as
\begin{equation}
H'(\eta')=\eta'\,n_d+(1-\eta')\tfrac{1}{N}P^{\dag}P
\label{ising0}
\,,
\end{equation}
where $P^{\dag}=\tfrac{1}{2}(d^{\dag}\cdot d^{\dag}-s^{\dag}s^{\dag})$ is a boson {\em pair operator}. 
In this case the critical point appears at $\eta'=\eta'_{\rm c}\equiv\tfrac{1}{2}$.
hamiltonian (\ref{ising0}) is closely related to the form (\ref{ising}) with $\chi=0$ as one can write 
\begin{equation}
H^{0}(\eta)=\underbrace{\eta\,n_d+(1-\eta)\tfrac{4}{N}P^{\dag}P}_{(4-3\eta)H'\left(\tfrac{\eta}{4-3\eta}\right)}+(1-\eta)\tfrac{1}{N}C_2[{\rm O(5)}]+(1-\eta)(N+4)
\,.
\label{relis0}
\end{equation}
Since the Casimir invariant $C_2[{\rm O(5)}]=n_d(n_d+3)-(d^{\dag}\cdot d^{\dag})({\tilde d}\cdot{\tilde d})$ is conserved along the $\chi=0$ path, it takes values $\tau(\tau+3)$, where $\tau$ is the seniority quantum number.
This means that the hamiltonians $H'_0(\eta')$ and $H^{0}(\eta)$ yield identical eigenfunctions and spectra connected just by simple transformations following from Eq.~(\ref{relis0}).

\subsection{Computational aspects}
\label{se:num}

\begin{figure}[t]
\centering
\includegraphics[width=11.1cm]{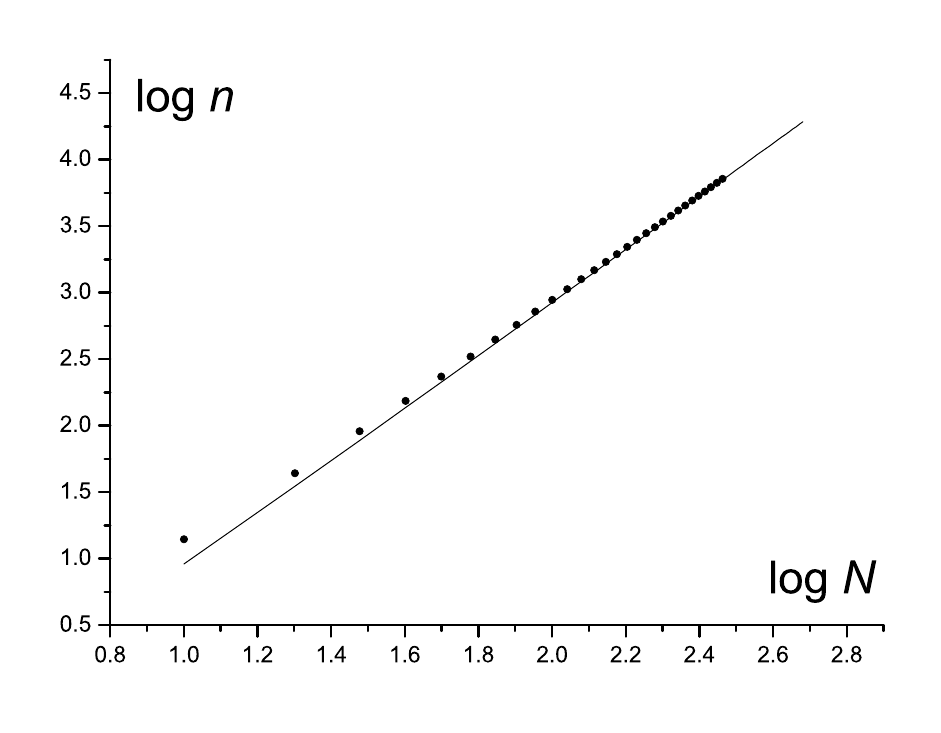}
\caption{\protect\small The dimension of the subspace of IBM states with $L=0$ as a function of the boson number. The dots correspond to actual dimensions calculated for $N=10,20,\dots,300$, the line represents an approximate formula $n\approx\tfrac{1}{12}N(N+1)$.}
\label{fi:dim}
\end{figure}

In numerical investigations of the IBM phase transitions, it is essential to calculate various spectroscopic signatures up to very large boson numbers. 
This may present a computational problem, since the dimension of the relevant Hilbert space increases.
An explicit formula for the total number of states with all angular momenta $L=0,2,3,4,\dots,2N$ reads as follows,
\begin{equation}
n_{\rm tot}=\frac{(N+5)!}{5!\,N!}\sim\frac{N^5}{120}
\,,
\label{numsta}
\end{equation}
where the approximation on the right-hand side holds for very large $N$.
Dimensions of individual $L$ subspaces can also be determined.
This can be done exactly by employing the well known reduction rules for the embeddings of various irreps corresponding to algebras in any of the IBM dynamical symmetry chains, all ending at the O(3) invariant symmetry subalgebra.
The result for $L=0$ is shown in Fig.~\ref{fi:dim}, together with an explicit asymptotic formula $n\sim\tfrac{1}{12}N^2$.
In order to determine the ground-state energy and wave function for a finite $N$, a numerical diagonalization of the IBM hamiltonian has to be carried out within this subspace.
We see that although its dimension grows only quadratically with $N$, the boson numbers $N\sim 100$ may already cause some troubles.

In fact, the most difficult part of the computation is not the diagonalization procedure itself.
Very efficient diagonalization algorithms are commonly available, especially for sparse matrices (thus applicable in the IBM case), which make it possible to go up to very high dimensions.
The part that presently seems to set the limits is the initial calculation of the hamiltonian matrix to be diagonalized.
In a majority of codes, the evaluation of matrix elements relies on the U(5) dynamical symmetry basis and makes use of the coefficients of fractional parentage \cite{Ta93}.
These are computed in a recurrent way, the complexity growing very quickly with increasing dimension.
An alternative scheme using the O(6) basis has been developed \cite{Rowe05}, but it does not solve the problem of large boson numbers.

Numerical difficulties can be overcome in the case of the transition between U(5) and O(6) dynamical symmetries along the path conserving the underlying O(5) dynamical symmetry.
As mentioned above, transitional hamiltonians of this type are fully integrable.
The Casimir invariant of the O(5) subalgebra represents an additional integral of motions, yielding the conserved seniority quantum number $\tau$.
The dimension of the $L=\tau=0$ subspace of states (including the ground state) grows as $n\sim\tfrac{1}{2}N$, which is a much slower increase than that in Fig.~\ref{fi:dim}.
An additional advantage is that the hamiltonian matrix in this subspace can be calculated directly, using explicit expressions for matrix elements of the O(6) Casimir invariant in the U(5) basis given by Arima and Iachello~\cite{Arima79}.
This makes it possible to increase the upper bound for boson numbers by several orders of magnitude (cf. Ref.~\cite{Cejnar07b}).

Some alternative, more sophisticated methods of solutions of the eigenvalue problem along the [U(5)-O(6)]$\supset$O(5) transition have been developed.
One of these methods is based on an infinite dimensional algebraic technique.
In the IBM context, the method was introduced by Pan and Draayer \cite{Pan98} and by Dukelsky and Pittel \cite{Dukelsky01}. 
The general solution resorts from an exact treatment of the pairing model invented by Richardson \cite{Richardson63a,Richardson63b} in the 1960's---see Ref.~\cite{Dukelsky04} for an overview and more references.
If the U(5)-O(6) transitional IBM hamiltonian is written in the pairing form (\ref{ising0}), the diagonalization problem for seniority $\tau$ is transformed to the problem of solving a coupled set of $n=\tfrac{1}{2}(N-\tau)$ nonlinear algebraic equations (Richardson equations).
For large boson numbers $N$ and small values of $\tau$ it is still difficult to obtain all solutions, but to determine the lowest eigenvalues is feasible.
The method was applied in a QPT analysis by Arias, Dukelsky, and Garc{\'\i}a-Ramos \cite{Arias03b}.

Another approach to the [U(5)-O(6)]$\supset$O(5) transition makes use of the so called continuous unitary transformation technique, introduced in the 1990's by G{\l}azek and Wilson \cite{Glazek93,Glazek94} and independently by Wegner \cite{Wegner94}---for a compact overview see Ref.~\cite{Dusuel05b}.
The unitary transformation $U$ that diagonalizes the given hamiltonian $H$ is found as a limit $U=\lim_{\lambda\to\infty}U(\lambda)$, where the dependence $U(\lambda)$ can be viewed as a continuous flow in the operator space from an arbitrary initial point $U(0)$.
Equations governing this flow in such a way that it converges to the solution of the eigenvalue problem have been derived and they can be iterated numerically.
In the IBM, the method was for the first time employed by Dusuel, Vidal, Arias, Dukelsky, and Garc{\'\i}a-Ramos \cite{Dusuel05b}.

A promising tool for calculations with very large boson numbers was recently presented by Ho, Rosensteel, and Rowe \cite{Ho07}. 
It is based on a generalized equation-of-motion approach put into the algebraic framework; for an explanation and original references see Refs.~\cite{Ho07,Rosensteel08}.
The method completely avoids the diagonalization step of the computation and proceeds directly to the construction of a matrix representation of the dynamical algebra of the problem.
This makes it possible to efficiently approximate the relevant observables---energies and matrix elements involving low-lying states---using  matrices whose dimensions are vastly reduced compared to the full Hilbert space dimension.
This technique has been so far applied in the second-order ($\gamma$-soft) shape phase transition within the geometric model \cite{Ho07} and in the Lipkin model \cite{Rosensteel08}, both cases being analogous to the U(5)-O(6) transition in the IBM.


\section{Quantum phase transitions in IBM-1: theoretical results}
\label{se:th}

Basic results of the mean-field analysis of ground-state shape-phase transitions in the interacting boson model-1 were already discussed in the preceding sections.
The use of coherent states for the study of IBM-1 ground-state transitions (a technique introduced by Gilmore \cite{Gilmore78,Gilmore79}) was for the first time reported by Dieperink, Scholten, and Iachello in 1980 \cite{Dieperink80}.
Interpretation of these results in terms of the catastrophe theory was given by Feng, Gilmore, and Deans in 1981 \cite{Feng81} and later, in 1996, by L{\' o}pez-Moreno and Casta{\~n}os in a detailed analysis \cite{Moreno96}.

The latter work also noticed the phase transition between prolate and oblate shapes (the transition achieved by changing the relative phase between $s$ and $d$ bosons).
This was then independently discussed in Ref.~\cite{Jolie01}, where the extended Casten triangle of Fig.~\ref{fi:tri}(b) was introduced.
The IBM-1 shape-phase structure with shape types I (spherical), II (prolate) and III (oblate) was put into the context of Landau theory by Jolie, Cejnar {\em et al.} in 2002 \cite{Jolie02,Cejnar03b}.
This kind of phase diagram, with a \lq\lq triple point\rq\rq\ in the middle, represents the simplest possibility allowed by the Landau theory with a single order parameter $\beta$ (for the discussion of differences from the original Landau context see Subsec.~\ref{se:geo}).
Atomic nuclei seem to belong to its few realizations in nature \cite{Warner02}.

From the theoretical point of view, an attractive property of the IBM is the simultaneous presence of both first- and second-order phase transitions.
This naturally allows for comparative analyses probing the general features of QPTs.
As foreseen in Ref.~\cite{Feng81} and newly proposed by Vidal {\it et al.} \cite{Vidal06}, the same structure can be reproduced also with an extended Lipkin hamiltonian containing parity violating terms (Subsec.~\ref{se:lip}).
However, if one sticks to models with two types of bosons---one scalar and one non-scalar---under the strict assumption of angular momentum or parity conservation, the IBM with $s$ and $d$ bosons is the simplest case with QPTs of both kinds \cite{Cejnar07}.

\subsection{Calculations beyond the mean field}
\label{se:bey}

So far, the phase transitional properties of the interacting boson model were discussed within the {\em mean field\/} (or Hartree-Bose) approximation, i.e. the approach based on the coherent state formalism sketched in Subsec.~\ref{se:fin}.
This approach yields correct results in the infinite-size limit, $N\to\infty$, but to predict finite-size corrections, one needs to employ more sophisticated techniques.
From the theoretical viewpoint, this is an extremely important task since the {\em scaling\/} of spectroscopic observables with $N$ plays the key role in describing the precursors of quantum phase transitions in finite systems.
Indeed, in the quantum critical points the scaling becomes singular (different from other points of the model parameter space), which may be considered as a defining feature of QPTs in finite systems.

To calculate the finite-$N$ corrections, diverse strategies are followed.
A direct possibility is to perform numerical diagonalization of the IBM hamiltonian for very large boson numbers and to analyze the convergence of the relevant quantities of interest to their asymptotic values.
In practice, this route is viable only for the transition between U(5) and O(6) dynamical symmetries that conserve (all the way) the underlying O(5) dynamical symmetry, i.e. for the second-order phase transitional path.
Indeed, as explained in Subsec.~\ref{se:alg}, the model is integrable along this path, which makes it possible to obtain results up to very large boson numbers using various more or less sophisticated techniques (Subsec.~\ref{se:num}), ranging from the Richardson equations \cite{Arias03b} to just a \lq\lq brute force\rq\rq\ diagonalization \cite{Cejnar07}.
Depending on the actual set of states included in such U(5)-O(6) calculations (i.e., in relation to the specific analysis), the upper values of the boson number can be $N=10^4$ \cite{Ramos05} or even $N=10^5$ \cite{Cejnar07}.
On the other hand, purely numerical calculations along the IBM first-order phase transitional paths can be currently performed with boson numbers not exceeding $N\sim 10^2$ \cite{Rowe04c,Rosensteel05}.

Another route to large size calculations leads through expansion techniques for various observables in powers of $N$. 
The advantage of such methods lies in their enhanced potential to determine the asymptotic scaling laws for both first- and second-order phase transitions.
An interesting approach \cite{Dusuel05,Dusuel05b,Vidal06,Arias07} is based on the Holstein-Primakoff type of mapping of $s$ and $d$ bosons onto boson $b$,
\begin{equation}
d_{\mu}^{\dagger}d_{\nu}\mapsto b_{\mu}^{\dagger}b_{\nu}\,,\qquad
s^{\dagger}s\mapsto N-n_b\,,\qquad
d_{\mu}^{\dagger}s=(s^{\dagger}d_{\mu})^{\dagger}\mapsto b_{\mu}^{\dagger}\sqrt{N-n_b}\,,
\label{Holst}
\end{equation}
where $n_b=\sum_{\mu}b_{\mu}^{\dagger}b_{\mu}$.
The following step is a shift transformation $c^{\dagger}_{\mu}=b^{\dagger}_{\mu}-\sqrt{N}\lambda^*_{\mu}$ with a complex vector $\lambda_{\mu}$ satisfying $\sum_{\mu}|\lambda_{\mu}|^2\equiv\|\lambda\|^2\in[0,1]$.
This implies $\ave{n_c}=\ave{n_b}-\|\lambda\|^2N$ for states with a fixed total boson number, and therefore allows for macroscopic average occupation numbers $\ave{n_b}$ while at the same time $\ave{n_c}\ll N$.
The resulting hamiltonian is written in terms of 5 pairs of operators $c^{\dagger}_{\mu}$, $c_{\mu}$ and the number $N$, which enables one to naturally separate terms of different orders in $N$.

The mean-field approximation represents the highest order term $\propto N^1$.
Note that here we consider an extensive IBM hamiltonian, such as the one in Eq.~(\ref{ising}). 
For hamiltonians in the intensive, i.e. energy-per-boson form, Eq.~(\ref{h}), the orders must be of course reduced by 1.
To proceed beyond the mean field, one needs to diagonalize---order by order---the respective higher hamiltonian terms.
Calculations of this type have been performed by Arias, Dukelsky, Dusuel, Garc{\'\i}a-Ramos, and Vidal for the second-order transition \cite{Dusuel05,Dusuel05b,Vidal06,Arias07} as well as for the first-order transition \cite{Vidal06,Arias07}.
The first nonvanishing quantum correction is of order $N^{0}$ and can be treated with the aid of the Bogoliubov transformation.
It is relatively straightforward in the spherical phase, while in the deformed phase one needs to proceed separately for angular momentum projections $\mu=0,\pm 1$, and $\pm 2$ (corresponding to the $\beta$, Goldstone, and $\gamma$ excitations, respectively) \cite{Arias07}.
Corrections to the resulting expressions are of order ${\cal O}(N^{-1})$.
Explicit calculations of terms $\propto N^{-k}$ with $k=1,2$ have been performed within the continuous unitary transformation formalism (briefly mentioned in Subsec.~\ref{se:num}) for the second-order transition \cite{Dusuel05,Dusuel05b}.

\begin{figure}[t]
\centering
\includegraphics[width=16.7cm]{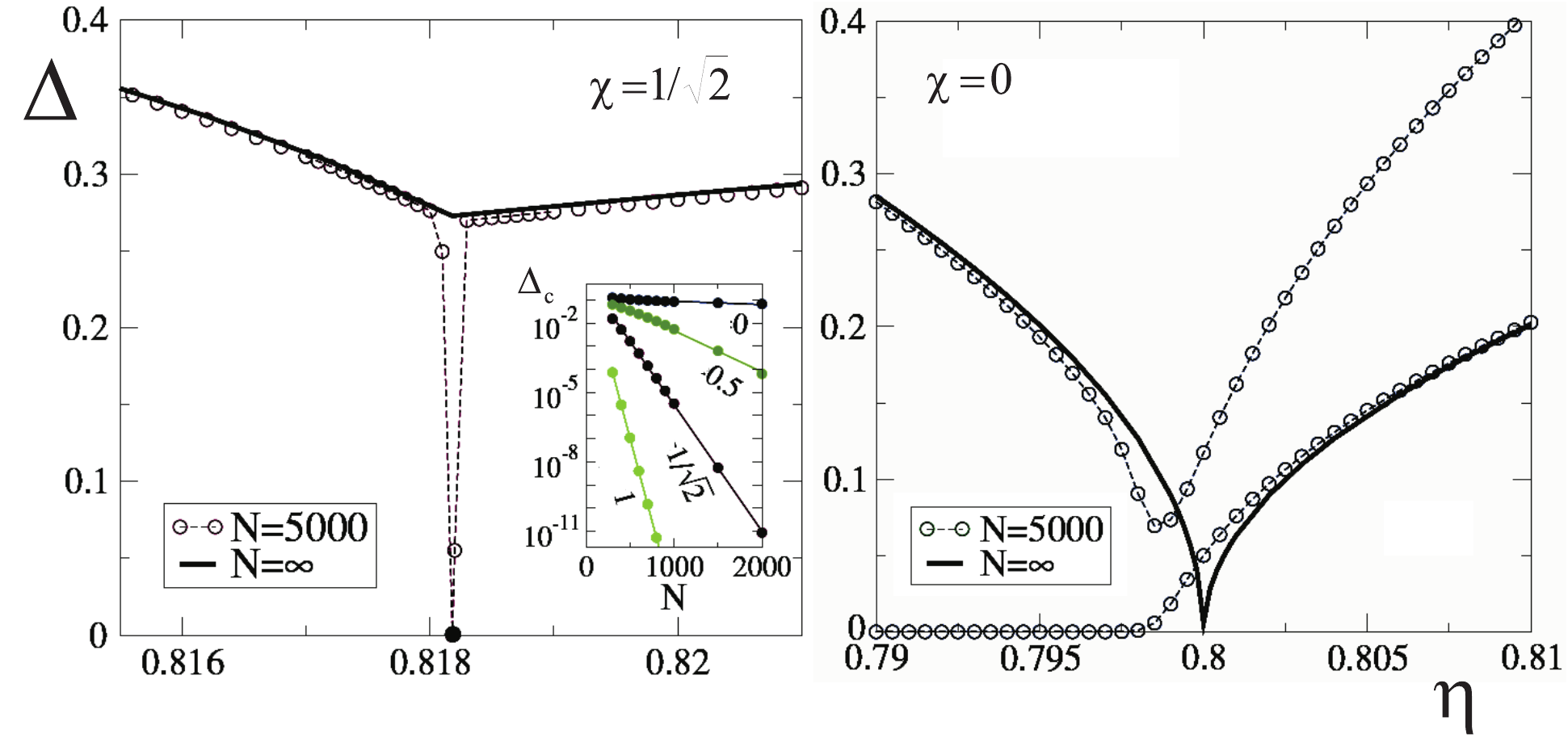}
\caption{\protect{\small The gap $\Delta$ as a function of $\eta$ around the first-order (left) and second-order (right) phase transition for the Lipkin hamiltonian (\ref{LipHam3}). The spherical phase is at $\eta>\eta_{\rm c}$, with $\eta_{\rm c}=\tfrac{9}{11}$ for $\chi=\frac{1}{\sqrt{2}}$ and $\eta_{\rm c}=\tfrac{4}{5}$ for $\chi=0$. Lines and the full dot are analytic results, open dots correspond to a numerical diagonalization. For $\chi=0$ both first and second excited states are included. The inset shows the scaling of $\Delta$ at the critical point with $N$ for various values of $\chi$. Adapted from Vidal {\it et al.} \cite{Vidal06}.}}
\label{fi:gap}
\end{figure}

Without going to technical details, we will report here results of these calculations for the gap $\Delta$ between the ground state and the first excited state.
The convergence $\Delta(\eta_{\rm c})\to 0$ with increasing $N$ at the quantum critical point $\eta_{\rm c}$ represents an essential QPT precursor.
Note that for an IBM extensive hamiltonian far away from $\eta_{\rm c}$ the average gap is expected to be independent of $N$, but in the critical region the low levels systematically reduce their spacing. 
Calculations \cite{Dusuel05,Dusuel05b,Vidal06,Arias07} are consistent with the scaling laws
\begin{equation}
\Delta(\eta_{\rm c})\propto N^{-\frac{1}{3}}\quad {\rm (second\ order\ transition)}
\,,\qquad
\Delta(\eta_{\rm c})\propto e^{-aN}\quad {\rm(first\ order\ transition)}
\,,
\label{gap}
\end{equation}
where the value of $a$ depends on the place where we cross the first-order transition, e.g. on the value of $\chi$ in the parametrization (\ref{ising}).
A simple reasoning for the scaling law in the second-order transition was given by Rowe \cite{Rowe04}.
It is based on the fact that in the second-order critical point the mean field potential is the pure quartic oscillator $V\propto N\beta^4$ (factor $N$, which was not present in the notation of Sec.~\ref{se:qpt}, is due to conversion to the extensive form).
At the same time, kinetic terms $T\propto -N\hbar^2\frac{\partial^2}{\partial\beta^2}$ are of order $N^{-1}$ (since $\hbar\propto N^{-1}$, see Subsec.~\ref{se:fin}).
Therefore, after the transformation $\beta\mapsto{\tilde\beta}\propto N^{1/3}\beta$ the hamiltonian produces an overall scaling factor $N^{-1/3}$, in agreement with Eq.~(\ref{gap}).

The behavior of the gap in both first- and second-order QPTs of the Lipkin model is illustrated in Fig.~\ref{fi:gap}, adapted from Ref.~\cite{Vidal06}.
It was obtained for hamiltonian (\ref{LipHam3}) which yields an equivalent phase diagram as the $sd$-IBM and the same scaling properties, but at the same time enables one to perform numerical calculations up to higher boson numbers ($N=5000$ in our case).
In the second-order transition ($\chi=0$), the ground state and first excited state form a nearly degenerate parity doublet for $\eta<\eta_{\rm c}$, therefore the figure shows two lowest excited states.
Scaling properties of various observables in a wider class of two-level boson models, including the $sd$-IBM, can be found in Refs.~\cite{Dusuel05,Dusuel05b,Arias07}.

\subsection{Thermodynamical analogies}
\label{se:ther}

Structural changes of the ground state in the quantum critical region were investigated by Cejnar, Jolie, Zelevinsky and Sokolov \cite{Cejnar98,Cejnar00,Cejnar01} by adopting concepts from statistical physics.
These analyses define suitable entropic quantities, the wave-function and von Neumann entropies, that make it possible to follow a parallel with thermodynamics.
The wave-function entropy \cite{Cejnar98b} is just the Shannon (information) entropy that measures the spread of an arbitrary wave function $\ket{\psi}=\sum_j\alpha^{\cal B}_{j}\ket{j^{\cal B}}$ in basis ${\cal B}\equiv\{\ket{j^{\cal B}}\}_{j=1,2,\dots}$ of the Hilbert space.
For the $i$th eigenstate of the hamiltonian $H(\eta)$ this yields:
\begin{equation}
W^{\cal B}_i(\eta)=-\sum_j|\alpha^{\cal B}_{ij}(\eta)|^2\ln|\alpha^{\cal B}_{ij}(\eta)|^2
\,.
\label{WS}
\end{equation}
Since a phase transition, being essentially a sudden restructuralization of the ground state and near excited states, can be seen as a transition between two types of basis vectors (those relevant at each side of the transition), the wave-function entropy is well suited to demonstrate basic QPT effects \cite{Cejnar00,Cejnar03b}.

On the other hand, von Neumann entropy is constructed to be independent of the basis.
It can be obtained by adding a small noisy component $\delta\eta$ to the model control parameter $\eta$, which results in turning all the hamiltonian eigenstates $\ket{\psi_i(\eta)}$ to density operators $\rho_i(\eta)$, expressing statistical character of the place in the parameter space where the eigenstates are evaluated \cite{Sokolov98}.
The $i$th state von Neumann entropy then reads as
\begin{equation}
S_i(\eta)={\rm Tr}\left[-\rho_i(\eta)\ln \rho_i(\eta)\right]
\,,
\label{SN}
\end{equation}
where $\eta$ represents the average of $\eta+\delta\eta$ under the assumption $\ave{\delta\eta}=0$.
The density operator and its entropy must depend on the size and type of the noise. 
If $\ave{\delta\eta^2}\ll\ave{\eta}^2$ the essential effects are mostly due to mutual mixing between a few closest eigenstates whose strength for a given eigenstate $i$ strongly depends on $\eta$.
The places of abrupt structural changes (quantum phase transitions) can be easily identified as pronounced maxima of von Neumann entropy \cite{Cejnar01}.

The noise-induced density operator $\rho_i(\eta)$ can also be associated with an equivalent thermally populated system (canonical density operator), which makes it possible to introduce an analog of specific heat ${\cal C}_i={\rm Tr}[\rho_i\ln^2\rho_i]-S_i^2$ defined for each state $i$ at given $\eta$ \cite{Cejnar01}.
For a linear hamiltonian (\ref{Hlin}) and small-amplitude noise one obtains an approximation
\begin{equation}
{\cal C}_i(\eta)\approx\ave{\delta\eta^2}\ln^2\ave{\delta\eta^2}\sum_{j(\neq i)}\frac{|\matr{\psi_i(\eta)}{V}{\psi_j(\eta)}|^2}{[E_i(\eta)-E_j(\eta)]^2}
\,,
\label{speheat}
\end{equation}
which, as seen from Eq.~(\ref{decay1}), represents a direct measure of the parameter-induced variation of wave functions.
Indeed, for the IBM ground state the \lq\lq specific heat\rq\rq\ introduced in this way yields very similar values as the \lq\lq specific heat\rq\rq\ defined through the first derivative of the wave-function entropy $W^{\rm U(5)}_0(\eta)$ in the U(5) basis  \cite{Cejnar03}.

As follows e.g. from Eqs.~(\ref{decay1}) and (\ref{speheat}), the effects of mixing for a given state $i$ are strongly enhanced in a vicinity of {\em avoided crossings\/} of level $i$ with neighboring levels.
In these cases, the energy denominator that appears in the perturbative expansion of the hamiltonian eigenstates makes the perturbation $V$ extremely efficient.
Recall that hamiltonian (\ref{Hlin}) can be written in the obvious self-similar form $H(\eta+\delta\eta)=H(\eta)+\delta\eta V$ for an arbitrary point $\eta$.
The crossings are avoided for levels with the same symmetry quantum numbers because such states generally yield nonzero matrix elements $\matr{\psi_i}{V}{\psi_j}$.
This prevents a statistically meaningful occurrence of exact degeneracies in spectra (corresponding to fixed symmetry quantum numbers) driven by a single (real) parameter $\eta$, although actual crossings may occur as exceptions with a vanishing statistical weight (or more frequently in integrable systems with hidden quantum numbers).
In any case, avoided or unavoided crossings of low-energy levels including the ground state seem to hold the key to understanding of both first-order and continuous QPT phenomena \cite{Cejnar00,Brentano04}.

To describe the influence of level crossings near quantum critical points in systems with increasing size, it turned out useful to extend the domain of the control parameter $\eta$ in Eq.~(\ref{Hlin}) from real to complex values \cite{Heiss88,Heiss05,Cejnar05,Cejnar07b}.
The reason for this is the fact that---unlike the real-$\eta$ case---the hamiltonian eigenvalues do cross in the complex-$\eta$ plane.
If the complex crossing occurs close to the real axis, one observes a sharp avoided crossing in the real-$\eta$ level dynamics.
Convergence of the complex crossing to real $\eta$ with an increasing size of the system implies that at the corresponding place the evolution of energies and wave functions becomes nonanalytic.

Since the complex-extended hamiltonian $H(\eta)$ is no more hermitian (its eigenvalues are complex), the points $\eta_k$ satisfying $E_i(\eta_k)=E_j(\eta_k)$ may be called {\em nonhermitian degeneracies}.
An $n$-dimensional hamiltonian of the type (\ref{Hlin}) has in total $\tfrac{1}{2}n(n-1)$ complex conjugate pairs of such points.
They are either diabolic points (a nongeneric type of crossing in complex $\eta$) \cite{Berry84} or branch, so called exceptional points (the generic type of degeneracy) \cite{Heiss88,Zirnbauer83,Rotter01}.
In a diabolic point, the two eigenvalue sheets just touch each other with a conical topology.
The branch point, on the other hand, connects the two Riemann sheets in a more tricky way, forming a singularity which can be locally described as the complex square root.
The system of $n$ Riemann sheets associated with solutions of the eigenvalue equation becomes entangled.
Only the branch points located on the Riemann sheet that for real $\eta$ corresponds to the lowest energy eigenvalue can influence the ground-state phase transitional behavior.
This, however, represents a severe problem because to numerically assign individual branch points to their respective Riemann sheets is a very difficult task even for moderate dimensions.

A method to bypass this problem has been developed and tested in the IBM phase transitions by Cejnar, Heinze, and Dobe{\v s} \cite{Cejnar05,Cejnar07b}.
Its basic idea is that branch points located on the selected Riemann sheet near the real parameter axis can be detected indirectly, using only the real energy eigenvalues $E_i(\eta)$ for ${\rm Im}\eta=0$.
To this end, we introduce a quantity
\begin{equation}
C_i(\eta)=-\tfrac{1}{n-1}\,\tfrac{d^2}{d\eta^2}\sum_{j(\neq i)}\ln\left|E_j(\eta)-E_i(\eta)\right|
\,,
\label{Cbr}
\end{equation}
where $n$ is the dimension of the relevant subspace, counting only the states with the same symmetry quantum numbers (for the ground-state considerations, the relevant subspace typically consists of states with zero angular momentum, but there may be also other conserved quantum numbers to be taken into account, e.g. seniority).
As shown in Ref.~\cite{Cejnar07b}, the quantity $C_i(\eta)$ measures the proximity of branch points located on the Riemann sheet of the $i$th state to the real parameter axis at $\eta$.
In fact, it can be derived from a two-dimensional electrostatic analogy, associating the $i$th sheet branch points with charges in the complex plane. 
$C_i(\eta)$ then represents the first derivative of the repulsive Coulomb force acting at the place $\eta$ of the real axis: $C\propto\frac{d}{d\eta}F$.
If the $i$th sheet branch points get close to the real axis, the change of force at the corresponding place is large and $C_i(\eta)$ will show a peak.
This peak becomes singular in the QPT critical points, where the distance of branch points from the real axis asymptotically vanishes.

\begin{figure}[t]
\centering
\includegraphics[width=18.5cm]{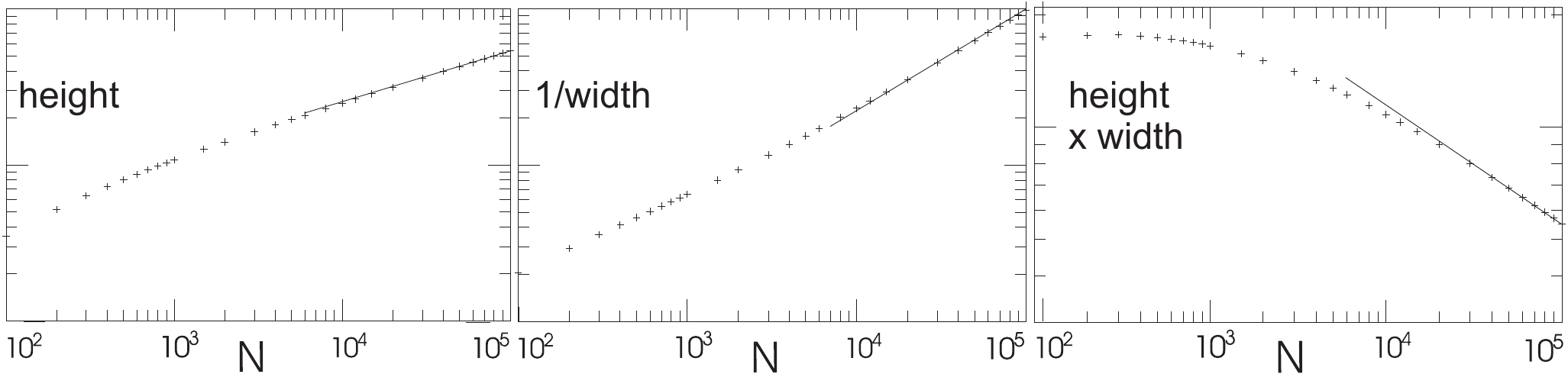}
\caption{\protect\small The evolution of the peak of the ground-state \lq\lq specific heat\rq\rq\ from Eq.~(\ref{Cbr}) at the IBM second-order QPT with increasing boson number $N$. The log-log plots show the peak height (left), the peak inverse width at half maximum (middle), and the product of the peak height and width (right). The vertical scale in all panels is irrelevant. The linear dependence in the rightmost panel represents a power-law decrease of the peak area, indicating a vanishing value of the \lq\lq latent heat\rq\rq\ $Q$. Adapted from Ref.~\cite{Cejnar07b}. }
\label{fi:speheat}
\end{figure}

At the same time, the quantity~(\ref{Cbr}) can be shown to play a similar role in quantum phase transitions as the specific heat in thermal phase transitions \cite{Cejnar05,Cejnar07b}.
In the first-order transition, the QPT peak of $C_i(\eta)$ converges to a $\delta$-function located at the critical point $\eta_{\rm c}$. 
For a continuous transition, the limiting behavior of $C_i(\eta)$, although still singular, yields a vanishing integral over an infinitesimal vicinity of the critical point: $Q=\lim_{\epsilon\to 0}\int_{\eta_{\rm c}-\epsilon}^{\eta_{\rm c}+\epsilon}C\,d\eta=0$.
This is illustrated in Fig.~\ref{fi:speheat}, which shows the boson number dependence of the peak height (panel a) and the inverse width at half maximum (panel b) of the \lq\lq specific heat\rq\rq\ $C_0(\eta)$ for the IBM ground state at the critical point of the second-order QPT between O(6) and U(5) limits. 
One observes that the peak width decreases faster than the height increases [cf. panel (c)], which leads to the estimate of $Q=0$ in the $N\to\infty$ limit.
This is in agreement with the continuous character of the transition studied.
Note that for thermal phase transitions, $Q$ would be nothing but the latent heat of the transition. 

In thermodynamics, the role of branch points is played by zeros of the partition function in  complex extended temperature.
As discussed by Yang and Lee in their 1952 seminal papers \cite{Yang52,Lee52}, complex zeros of the partition function are crucial for fundamental understanding of the thermal phase transitions.
The criteria derived for thermal phase transitions of various orders in terms of complex zeros \cite{Borrmann00} are essentially the same as those for QPTs and branch points in the complex parameter plane.
Therefore, the above outlined considerations represent a powerful analogy between thermal and quantum phase transitions.

\subsection{Signatures involving excited states}
\label{se:sig}

Although the definition of quantum phase transitions is related to the ground state, it is clear that there must also exist some QPT induced effects in the behavior of energies and transition amplitudes for {\em low-lying\/} excited states.
In IBM, such spectroscopic observables can be approached via so called quadrupole shape invariants \cite{Werner00,Werner02}, which represent expressions based on $n$th-order moments ($n\geq 2$) of the quadrupole operator $Q_{\chi}\equiv Q$ in the ground state:
\begin{equation}
q_2=\matr{0_1^+}{(Q\cdot Q)}{0_1^+}\,,\quad q_3=\matr{0_1^+}{[QQQ]^{(0)}}{0_1^+}\,,\quad q_3=\matr{0_1^+}{(Q\cdot Q)(Q\cdot Q)}{0_1^+}\,,\dots
\end{equation} 
The mean field expectation values for these invariants can be written in terms of quadrupole shape variables $\beta$ and $\gamma$, but at the same time, equivalent expressions can be derived with the aid of sums over products of reduced matrix elements $\matri{0_i^+}{Q}{2_j^+}$ for E2 transitions between individual $0^+$ and $2^+$ states \cite{Werner00}.
Since only the states with small phonon numbers contribute significantly to these sums, the shape invariants represent an important tool for identifying suitable QPT signatures in the spectra of low lying levels.

Among the other spectroscopic signatures of the IBM shape transition across the extended Casten triangle we name the following: 
(a) The ratio $R_{4/2}=E(4_1^+)/E(2_1^+)$ between excitation energies of the first $4^+$ and $2^+$ states. It takes the harmonic vibrator value $R_{4/2}\sim 2$ near the U(5) limit, the rotor value $\sim 3.33$ near the SU(3) and ${\overline{\rm SU(3)}}$ limits, and the value $\sim 2.5$ near the O(6) limit. 
(b) The quadrupole moment $Q(2_1^+)\propto\matri{2_1^+}{Q}{2_1^+}$ of the first $2^+$ state, which in the mean field approximation is proportional to the ground-state equilibrium value of $\beta$. This measure is related to the expectation value $\matr{0_1^+}{n_d}{0_1^+}\equiv\ave{n_d}_0\propto\beta^2$ of the $d$ boson number operator in the ground state [proportional to $B({\rm E2},2_1^+\to 0_1^+)$], which is however insensitive to the sign of $\beta$ and therefore not suited for the description of prolate-oblate transitions.
(c) The strength $B({\rm E2},2_2^+\to 2_1^+)$ for the transition from the second to the first $2^+$ state, which is close to zero near the SU(3) and ${\overline{\rm SU(3)}}$ limits and nonzero in the spherical and $\gamma$-unstable phases.
(d) The strength $\rho^2(E0,0_2^+\to 0_1^+)$ of the E0 transition between the first excited $0^+$ and the ground state, which is also related to the $d$-boson content of the ground state \cite{Brentano04}.
(e) The ratio $B_{4/2}=B({\rm E2},4_1^+\to 2_1^+)/B({\rm E2},2_1^+\to 0_1^+)$ of transition strengths between the lowest $L=0,2,4$ states, which exhibits a peak close to the critical point \cite{Rowe04c}.
(f) Two-particle transfer intensities for the ground-state to ground-state transition and transitions from the ground state to excited $0^+$ states, which show characteristic patterns in the critical region \cite{Fossion07}.

An effort has been spent to identify observables suitable for a direct distinction between first- and second-order QPTs in finite-$N$ numerical data.
Such observables would turn out important in numerical studies of systems with a phase transition of unknown nature.
In the IBM framework, the quantity
\begin{equation}
\nu_2=\tfrac{1}{N}\left[\matr{0_2^+}{n_d}{0_2^+}-\matr{0_1^+}{n_d}{0_1^+}\right]
\label{nu12}
\end{equation}
was proposed as such a measure by Iachello and Zamfir \cite{Iachello04}.
Indeed, while the above discussed ground state average $\tfrac{1}{N}\ave{n_d}\equiv\nu_1$ rises from $\nu_1\sim 0$ (spherical phase) to $\nu_1>0$ (deformed phase) with no particular difference between the first- and second-order transition, $\nu_2$ is much more sensitive to the transition type.
For the first-order transition it exhibits an abrupt, sign changing crossover from $\nu_2>0$ (spherical phase) to $\nu_2\sim 0$ (deformed phase), whereas for the second-order transition the change is smoother.
The key for understanding these observations is the fact that the first-order quantum phase transition can be locally interpreted as a sharp anticrossing of the two lowest energy eigenstates with the same symmetry quantum numbers \cite{Zamfir02}. 
Indeed, this is how the wiggling behavior of $\nu_2$ in the first-order transition comes about.
The quantities $\nu_1$ and $\nu_2$ can be related to the expectation values $\ave{r^2}_{\psi}$ of the squared radius (real space) in the states involved \cite{IA87}, particularly $\nu_2\propto\ave{r^2}_{0_2^+}-\ave{r^2}_{0_1^+}$ measures the isomer shift between the first two $0^+$ states.
In some cases, $\nu_2$ has to be replaced by $\nu'_2=\tfrac{1}{N}\left[\matr{2_1^+}{n_d}{2_1^+}-\matr{0_1^+}{n_d}{0_1^+}\right]$, which is proportional to the experimentally available isomer shift $\delta\ave{r^2}=\ave{r^2}_{2_1^+}-\ave{r^2}_{0_1^+}$.
Note that the isomer shifts are closely related to the expectation values of the effective E0 transition operator in the states involved, so that there is a link to the E0 strengths discussed above.

In a very recent work by Bonatsos, McCutchan, Casten, and Casperson, a new quantity was proposed to distinguish the first- and second-order QPT in the IBM framework. 
It is the energy ratio $E(6_1^+)/(0_2^+)$ between the first $6^+$ and second $0^+$ states, which takes value 1.5 in the U(5) limit, $\sim 0$ in the SU(3) limit, and 1 in the O(6) limit.
A comparison of some other spectroscopic observables for first- and second-order transitions has been presented by Pan, Draayer, Luo, and Zhang \cite{Pan03,Pan05} and by Zhang, Hou, and Liu \cite{Zhang07}.

\subsection{Symmetry related approaches}
\label{se:sy}

Essential contributions to the interpretation of low energy spectroscopic data at and around the quantum critical points in nuclei have been brought by novel generalizations of the dynamical symmetry concept.
One of these approaches attempts to ascribe specific dynamical symmetries to the first- and second-order critical points.
This might seem surprising as the critical point itself is just in between two competing symmetries, so one would expect rather disorder than a symmetry at this place.
Nevertheless, it turned out that these so called {\em critical point symmetries\/} are very successful in the description of spectroscopic data.
The first specimen of this type was discovered within the fermion dynamical symmetry model \cite{Zhang87}.
The SO(7) dynamical symmetry of that model is located at the phase transition between different ground state shapes, similarly as in the IBM the O(6) dynamical symmetry constitutes the critical point between prolate and oblate shapes when proceeding along the SU(3)--${\overline{\rm SU(3)}}$ transition \cite{Jolie01}.

A systematic, though approximate approach to the construction of the critical point symmetries was presented by Iachello in the framework of the geometric model \cite{Iachello00,Iachello01,Iachello03,Iachello05,Caprio07}.
Hereby use was made of a strongly simplified Bohr hamiltonian which for the spherical--prolate transition led to an approximate critical point symmetry denoted as X(5) \cite{Iachello01}, and for the spherical to $\gamma$-unstable transition to the symmetry called E(5) \cite{Iachello00}. 
Both these symmetries focused on the $\beta$ degree of freedom, for which an infinite square well potential was imposed, and were based on an approximate separation of $\beta$ and $\gamma$ dynamical modes.
As a result of these simplifications, virtually parameter free solutions were obtained, which prompted an extensive search for experimental examples.

Since the critical point descriptions are embedded in the framework of the geometric model, they do not reflect the finite number of nucleons in real nuclei.
Finite size effects have been treated in the IBM framework by Leviatan and Ginocchio for the E(5) case (second-order transition) \cite{Leviatan03} and by Leviatan for the X(5) case (first-order transition) \cite{Leviatan05}.
In both cases, the method employed was based on the IBM intrinsic states projected onto the appropriate symmetry quantum numbers and on the variation of the corresponding energy functional after the projection.
In the X(5) case, an additional two level mixing calculation had to be carried out as a consequence of the phase coexistence in the first-order transition.
An effective deformation that results from these calculations constitutes the structure of the excited bands at the critical points. 
However, as explicitly demonstrated by Arias and Garc{\'\i}a-Ramos {\it et al.} \cite{Arias03,Ramos05} for the second-order transition, the critical point symmetries based on the original approximation of the square well potential in $\beta$ have only a limited validity in the IBM framework even for an infinite number of bosons.
This is so because the second-order critical point potential in IBM is essentially a pure quartic oscillator, $V\propto\beta^4$.
We will therefore not review the X(5) and E(5) symmetries in detail here.
The interested reader can consult recent overviews by Caprio and Iachello \cite{Caprio07}, and by Casten and McCutchan \cite{Casten07}.

Another symmetry related approach to quantum phase transition is based on so called {\em partial dynamical symmetries}.
These symmetries express situations, when only a certain subset of properties corresponding to the full dynamical symmetry is retained by the system, whereas the remaining properties are not.
In particular, all or a part of states may be characterized by a subset of symmetry quantum numbers, or a part of states by all quantum numbers \cite{Alhassid92,Whelan93,Leviatan96,Isacker99,Leviatan02}.
Situations like this are particularly relevant in systems with coexisting regular and chaotic features, but they may also capture essential features of the phase coexistence in first-order QPTs.

The application of this concept to the first-order critical point in the IBM relies on the possibility to increase the size of potential barrier between the coexisting minima within the phase transitional region.
The two minima, whose existence is inherent to the first-order transition, had been pointed out as a potential source of nuclear isomerism \cite{Feng81} and later discussed in a similar context of coexisting spectral structures \cite{Iachello98,Jolie99}.
However, for a typical IBM hamiltonian these minima are too shallow \cite{Jolie99,Rowe04c}.
A method for constructing an IBM hamiltonian with a decent barrier separating the coexisting phases has been proposed by Leviatan \cite{Leviatan06} and recently used by the same author to demonstrate the relevance of partial dynamical symmetries at the critical point \cite{Leviatan07}.
The method makes use of the decomposition of the IBM hamiltonian into the intrinsic and collective parts \cite{Leviatan87}.
While the intrinsic part generates an energy surface with two minima and can be tuned to achieve an arbitrary barrier, the collective part is composed of kinetic terms which do not affect the shape of the energy surface. 
The dynamics at the critical point can be again described by an effective deformation determined by variation after projection and a two-level mixing calculation.

The application of this method to the IBM first-order transition led to the identification of a specific partial dynamical symmetry at the critical point \cite{Leviatan07}.
Distinct subsets of solvable states, characterized by different types of good dynamical symmetries, were shown to coexist in the system governed by the critical hamiltonian.
The two subsets of states can be constructed from the U(5) and SU(3) dynamical symmetry chains, respectively, in agreement with the limiting dynamical symmetries of the transition and with the intuition that the first-order spherical--deformed transition generates coexisting vibrational and rotational spectral structures.
In fact, also the second-order phase transition in the IBM is characterized by a partial dynamical symmetry, although of a different type than the first-order transition [all states retain a part of quantum numbers due to the underlying O(5) symmetry, as discussed above].
While the latter feature is most probably specific to the model, the applicability of the partial dynamical symmetry with two coexisting subsets of solvable states to generic first-order QPTs constitutes an interesting subject for further research.

The last symmetry-related concept we want to discuss in this subsection is based on so called {\em quasi dynamical symmetries}.
This term should not be confused with partial dynamical symmetries discussed above.
In a partial dynamical symmetry, a fixed realization of a given algebra is partly imprinted in the spectrum (in one of the above senses).
In a quasi dynamical symmetry, in turn, the symmetry is completely broken for all states, but for a part of the spectrum this happens in a highly organized manner which creates an illusion that the symmetry is preserved.
Rowe, who coined this term, describes it as follows \cite{Rowe04b}: \lq\lq Quasidynamical symmetry is an expression of the possibility that a subset of physical data may exhibit all the properties that would result if the system had a symmetry which, in fact, it does not have.\rq\rq\
The mathematical formulation of this concept relies on so called embedded representations \cite{Rowe88}, which can be associated with particular (coherent) linear combinations of equivalent irreducible representations of a certain Lie algebra that give rise to a new irreducible representation.
For systems with quantum phase transitions it often appears that two incompatible quasi dynamical symmetries can be associated with the two competing phases. 
The quasi dynamical symmetry of type I starts breaking only if the system enters the phase transitional region, from where it may emerge with a quasi dynamical symmetry of type II.

\begin{figure}[t]
\centering
\includegraphics[width=13.9cm]{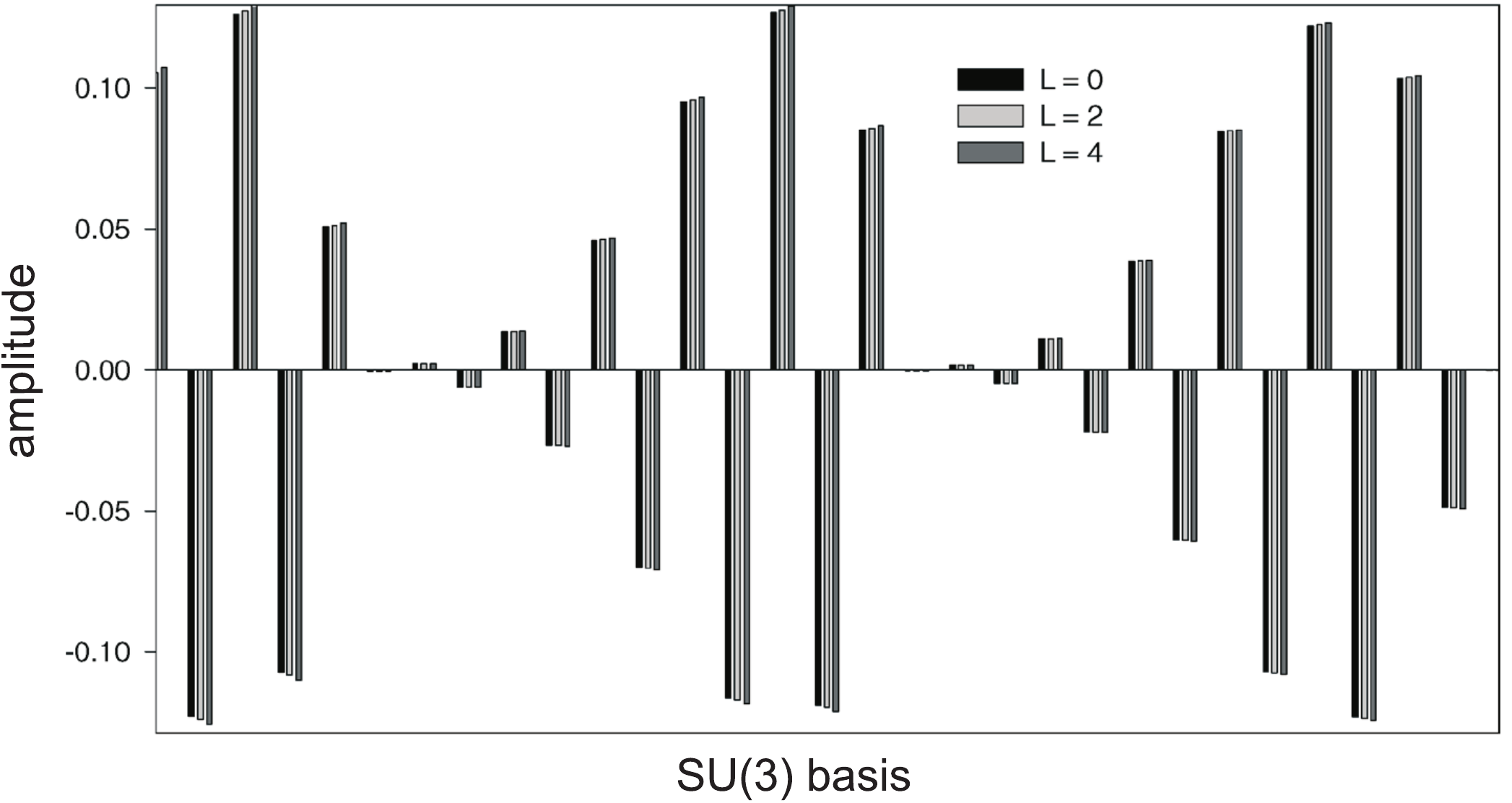}
\caption{\protect\small Amplitudes for an expansion of the lowest eigenstates with $L=0,2,4$ ($N=100$) in the SU(3) basis just before the phase transition to the spherical phase (only a part of amplitudes is shown). A highly coherent pattern of mixing illustrates the SU(3) quasi dynamical symmetry. Adapted from Rosensteel and Rowe~\cite{Rosensteel05}.}
\label{fi:su3quasi}
\end{figure}

A good example is the SU(3) quasi dynamical symmetry of the shell and related models \cite{Rochford88,Bahri00}. 
The exact Elliot SU(3) dynamical symmetry is badly broken by spin-orbit and major shell mixing interactions, which leads to a complete fragmentation of actual hamiltonian eigenstates among individual SU(3) irreps.
Nevertheless, the mixing amplitudes preserve a very high degree of coherence, so that observable quantities (transition strengths) retain the basic signatures of the unbroken SU(3).

A similar effect was shown by Rowe, Turner and Rosensteel \cite{Rowe04c,Rosensteel05} also for the first-order transitional path between the U(5) and SU(3) limits of the IBM.
An example of the decomposition of eigenstates in the SU(3) basis is shown in Fig.~\ref{fi:su3quasi}.
The three lowest yrast states shown in the figure at a value of the control parameter just before the phase transition to the spherical phase have essentially the same amplitudes in the SU(3) basis.
In the SU(3) limit, the yrast states are all classified by the same SU(3) quantum numbers $(\lambda,\mu)=(2N,0)$.
Although the SU(3) dynamical symmetry is apparently broken (the states are widely spread in the symmetry basis), the mixing has a highly coherent form.
In fact, the yrast states on the deformed side of the phase diagram can be derived from an intrinsic state of the same form as in the SU(3) limit.
The amplitudes evaluated at only a slightly different value of the control parameter---just after the phase transition---exhibit a dramatically different pattern, with the SU(3) coherence lost.
The quasi dynamical symmetry on the U(5) side of the transition can be described in terms of the random phase approximation.

Even a more comprehensive analysis guided by the idea of quasi dynamical symmetry was presented for the IBM second-order transition between U(5) and O(6) limits \cite{Rowe04,Rowe04b}, which has a direct counterpart in the geometric model \cite{Turner05}.
We will return to these results briefly in Subsec.~\ref{se:se}.


\section{Comparison with experimental data}
\label{se:com}

If the theoretical signatures of quantum phase transitions are to be compared with experimental observables in real nuclei, there appears an essential difficulty: 
The parameters that induce realistic nuclear shape transitions, i.e. proton and neutron numbers $\cal Z$ and $\cal N$, do not vary in a continuous way. 
One has to reconcile with the fact that phase transitional regions cannot be studied in arbitrary detail by fine tuning of the control parameters. 
Whether one observes an ideal transitional specimen nucleus between two distinct shapes depends on the particular discretization of the relevant section of the nuclear chart.

\begin{figure}[t]
\centering
\includegraphics[width=18.5cm]{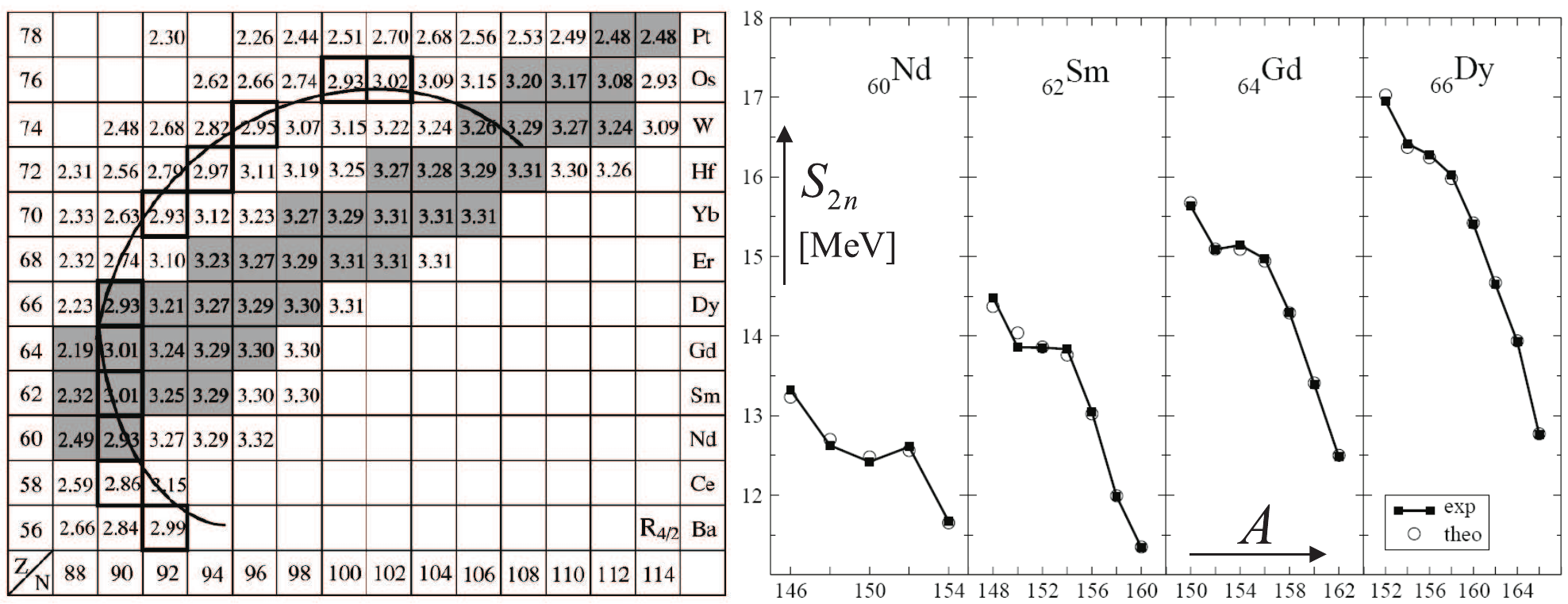}
\caption{\protect{\small Left (adopted from McCutchan {\it et al.} \cite{Libby04}): a part of the nuclear chart with values of $R_{4/2}=E(4_1^+)/E(2_1^+)$ and the $P=5$ contour, see Eq.~(\ref{promis}). The stable nuclei are shown in gray. Right (adapted from Garc{\'\i}a-Ramos {\it et al.} \cite{Ramos03}): the two-neutron separation energies of Nd, Sm, Gd and Dy isotopic chains. Experimental values (full symbols) are shown together with the IBM fit (open symbols).}}
\label{fi:chart}
\end{figure}

Essentially, shapes of atomic nuclei are determined by the changing shell structures of protons and neutrons. 
The presence of a shell closure drives the nucleus towards a dominance of the spherical shape.  
Although the detailed shell configuration represents the microscopic origin of the shape, often a simplification can be done when one deals with nuclei far away from (sub)shell closures. 
The quantities that then mostly determine the equilibrium shape of a given nucleus are the numbers $N_p$ and $N_n$ of {\em valence\/} protons and neutrons, or the corresponding holes if the nucleus is located above the midshell.
A good empirical criterion for the onset of deformation was proposed by Casten {\it et al.} \cite{Casten87,Libby04}.
It reads as 
\begin{equation}
P\equiv\frac{N_pN_n}{N_p+N_n}\approx 5\,,
\label{promis}
\end{equation}
where $P$ (derived from \lq\lq promiscuity\rq\rq) expresses an average number of interactions of each valence nucleon (or hole) with nucleons (holes) of the other type. 
For $P>5$ the nucleus can be expected to be deformed in its ground state, while for $P<5$ it is more likely spherical. 
The approximate condition~(\ref{promis}) demarcates expected regions of shape-phase transitions across the nuclear chart, as shown in the left panel of Fig.~\ref{fi:chart} \cite{Libby04}.
One notices that the regions with $P \approx 5$ close to stability are rather scarce. 
Notably, the most important transitional region is centered around ${\cal N}=90$ and at the Os and W nuclei.
It turns out that the condition (\ref{promis}) is valid close to the U(5)-SU(3) transition, but it is not sufficient when a nucleus is close to the U(5)-O(6) transition: For example, $^{184}$W has $P=5.1$ but the fit of spectroscopic observables places it near the O(6)-SU(3) leg of the symmetry triangle.

Since the focus of this review is centered more on the theory side, we present below only an outline of relevant experimental issues.
The reader interested in more detailed information can consult recent reviews \cite{Casten07,Casten08}.

\subsection{Nuclear masses}

Already in their pioneering work on shape-phase transitions in atomic nuclei, Dieperink, Scholten, and Iachello pointed out the importance of nuclear mass measurements for the study of shape transitions \cite{Dieperink80}. 
As the masses depend on binding energies, changes in the absolute energy of the ground state lead to kinks in an otherwise smooth behavior. 
For even-even nuclei these effects together with abrupt changes in nuclear radii are the only direct observables since the ground state alone has no electromagnetic moments or transitions. 

A good quantity to track changes in nuclear masses is provided by the extraction of {\em two-nucleon separation energies}, which are extracted as mass differences
\begin{equation}
S_{2n}({\cal Z},{\cal N})=M({\cal Z},{\cal N}\!-\!2)+2m_nc^2-M({\cal Z},{\cal N})\,,\quad
S_{2p}({\cal Z},{\cal N})=M({\cal Z}\!-\!2,{\cal N})+2m_pc^2-M({\cal Z},{\cal N})\,,
\label{s2n}
\end{equation}
with $M({\cal Z},{\cal N})$ the mass of a nucleus with ${\cal Z}$ protons and ${\cal N}$ neutrons, and $m_p$, $m_n$ the masses of nucleons.
Theoretically, the separation energies can be predicted from the IBM ground-state energies $E_0$ for boson numbers $N+1$ and $N$ at values $\eta+\delta\eta$ and $\eta$ of the control parameter associated with the given pair of nuclei.
Specifically,
\begin{equation}
S_{2\bullet}=E_0(N\!+\!1,\eta\!+\!\delta\eta)-E_0(N,\eta)\approx\tfrac{\partial}{\partial N}E_0(N,\eta)+\delta\eta\tfrac{\partial}{\partial\eta}E_0(N,\eta)
\,,
\label{s2n_teo}
\end{equation}
where $\bullet$ stands for $n$ or $p$.
Since $E_0$ depends on $N$ quadratically (due to one- and two-body interactions), the first term on the right-hand side yields a linear dependence on $N$ and hence a smooth contribution to the separation energy.
An irregular contribution is created by the second term:
The first-order transition, with a discontinuous $\tfrac{\partial}{\partial\eta}E_0$, leads to a discontinuity of the separation energy.
The second-order transition, with a discontinuous $\tfrac{\partial^2}{\partial\eta^2}E_0$, induces only a change of slope of the separation energy curve at the critical point.

As an example, the right-hand panel of Fig.~\ref{fi:chart} compares the experimental values of the two-neutron separation energies in the Nd, Sm, Gd and Dy isotopes to the results of IBM calculations performed in Ref.~\cite{Ramos03}. 
One clearly observes a \lq\lq discontinuity\rq\rq\ (smoothened by finite-size effects) indicating that at ${\cal N}=90$ a first-order spherical-to-deformed phase transition is occurring in each of the isotopic chains.

A systematic study of phase transitions and their effect on separation energies and masses was performed by Garc{\'\i}a-Ramos, De Coster, Fossion, and Heyde in Ref.~\cite{Ramos01}. 
These authors find a very high sensitivity to the IBM parameters around the second-order phase transition, which may indicate that the description of energies and electromagnetic transition rates is not a sufficient criterion in this region.
Compared to the first-order transition, the effect of a second-order phase transition on separation energies is by about an order of magnitude smaller (a change in the slope instead of a discontinuity).
It is therefore very important that nowadays very accurate mass measurements with $\Delta M/M\sim 10^{-7}$ become possible on stable as well as exotic nuclei \cite{Blaum06}. 
The access to exotic nuclei is essential to verify phase transitional predictions in the $P\approx 5$ regions far from the valley of stability.

\subsection{Low-energy collective states}

\begin{figure}[t]
\centering
\includegraphics[width=13cm]{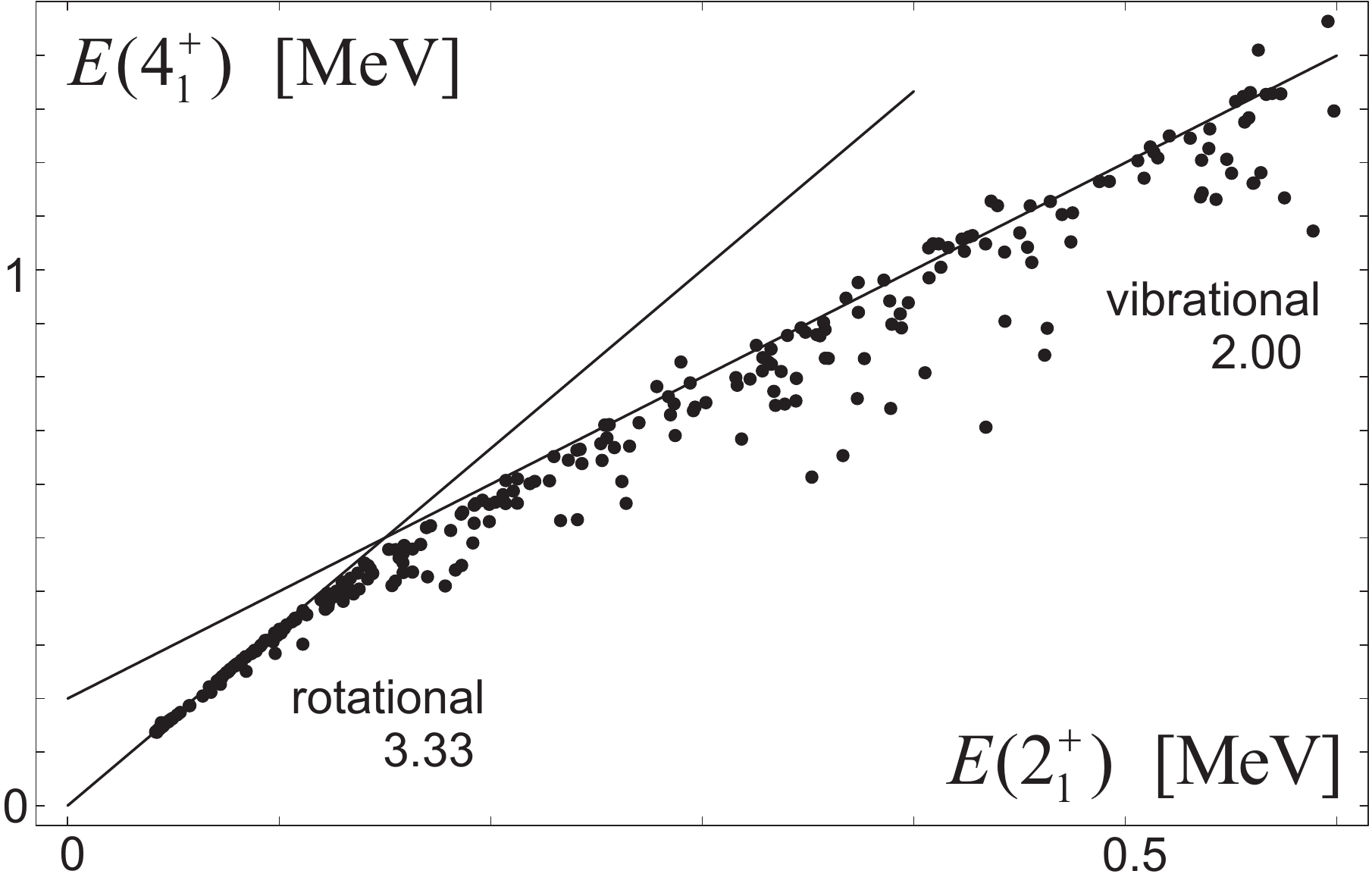}
\caption{\protect{\small The excitation energy of the $4_1^+$ state against one of the $2_1^+$ state for the nuclei with $E(2_1^+)<600$~keV.
The spherical-deformed transition is observed as the change of the slope between the indicated vibrational and rotational values at $E(2_1^+)\approx 150$~keV.}}
\label{fi:rick}
\end{figure}

The shape of the ground state is strongly present in the properties of the low-lying excited states. 
In order to unambiguously determine the shape types from the data, Casten {\it et al.} \cite{Casten85,Casten87,Casten93,Casten99} used some universal dependences obtained from spectroscopic observables for a large number of isotopes.
The best known example is shown in Fig.~\ref{fi:rick}, where the excitation energy of the first $4^+$ state is plotted against the one of the first $2^+$ state. 
All data together clearly yield a universal curve which shows two different slopes. 
At low excitation energies the slope is approximately 3.33, which indicates rotational motion, while at high energies the slope becomes approximately 2, indicating a spherical vibrator. 
The first-order shape-phase transition occurs where both slopes cross. 
One can thus deduce a typical excitation energy for the first excited state as a signature for the spherical-deformed phase transition.

Among the other indicators of shape-phase transitions are the electromagnetic properties, especially the quadrupole moment of the $2_1^+$ state and strengths of E2 transitions between excited states (cf. Subsec.~\ref{se:sig}).
For the prolate-oblate transition, the change of the sign of $\beta$ must induce the change of the sign of the $2_1^+$ quadrupole moment. 
A more difficult task is to observe and distinguish in data the second-order phase transition, as this relies on the determination of the critical exponent from rather scarce data points. 
In the following we focus on systematic studies of chains of nuclei along the U(5)-SU(3), U(5)-O(6), and  SU(3)-$\overline{\rm SU(3)}$ transitional paths.

\subsubsection{Spherical to prolate transition}

Early work on a systematic comparison with experimental data for the transitional region between U(5) and SU(3) was done by Scholten, Iachello and Arima \cite{Scholten78}. 
They studied in particular the Sm isotopes where the spherical-deformed transition is very well pronounced. 
These isotopes and in particular $^{152}$Sm, which is an almost ideal first-order transitional nucleus, formed the basis of the renewed interest in shape-phase transitions.
This was triggered again when a new study of $^{152}$Sm showed that the transition strength ratio $B({\rm E2};2^+_\gamma\to 0^+_2)/B({\rm E2};2^+_\gamma\to 0^+_1)$ is extremely small, but could be reproduced by a simple IBM calculation \cite{Casten98}. 
It was then shown that the sharp minimum in $B({\rm E2};2^+_\gamma\to 0^+_2)$ is located at the first-order phase transition, in the region where
spherical and deformed phases coexist \cite{Iachello98}. 
This interpretation launched the revival of interest in shape-phase transitions, especially after the introduction of the X(5) critical point symmetry \cite{Iachello01} and the subsequent interpretation of $^{152}$Sm as an example of a X(5) nucleus \cite{Casten01}. 
The interpretation of shape-phase transition caused an intense debate which is still ongoing (for a review see Ref.~\cite{Casten07}). 
As an example, the moments of inertia in the bands corresponding to both phases are almost the same, contrary to the model prediction \cite{Jolie99}.

Further nuclei located at or very close to the first-order transition were studied experimentally. 
They are the ${\cal N}=90$ isotones $^{150}$Nd \cite{Kruecken02} and  $^{154}$Gd \cite{Tonev04}. 
They also follow the X(5) pattern in energies and electromagnetic transition rates for the ground state band. 
The next candidate, $^{156}$Dy, follows still the X(5) energy trend but in its electromagnetic decay it already behaves as a rotor \cite{Tonev04}. 
Besides the ${\cal N}=90$ nuclei, which are situated directly in the region with $P=5$ (see Fig \ref{fi:chart}), the neutron deficient Os nuclei were recently proposed as other candidates. 
Lifetime measurements indicate that indeed $^{176,178}$Os exhibit the X(5) symmetry and are well described in the IBM as nuclei at the first-order critical point \cite{Dewald05,Dewald06}.

\begin{table}
\begin{center}
\begin{tabular}{|l|r|r|r|r|r|r|r|r|}
\hline
Nucleus &$^{148}$Nd & $^{150}$Nd &  $^{150}$Sm & $^{152}$Sm & $^{154}$Gd & $^{156}$Dy &$^{176}$Os & $^{178}$Os \\
\hline
$P$ &3.75 & 4.4 & 4.4  & 4.8 & 5.1    & 5.3 &4.5 & 4.6\\
\hline
\end{tabular}
\end{center}
\caption{\protect\small $P$ factors from Eq.~(\ref{promis}) for the isotopes situated close to the first-order transition.}
\label{ta:P}
\end{table}

A systematic comparison of rare earth nuclei was done by Garc{\'\i}a-Ramos, Arias, Barea, and Frank \cite{Ramos03}. 
Chains of Nd, Sm, Gd and Dy isotopes were fitted using the simple consistent-$Q$ hamiltonian as well as a complete IBM-1 hamiltonian.
Good descriptions for the excitation energies and $B({\rm E2})$ values were obtained together with a good description of the two-neutron separation energies (see Fig.~\ref{fi:chart}). 
Based on the fitted parameters the isotope chains could be placed in the separatrix plane studied before within the framework of the catastrophe theory \cite{Moreno96}. 
The results showed that none of the Dy isotopes is close to the critical point. 
The closest ones are $^{148}$Nd, $^{150}$Sm and less clearly $^{152}$Gd, followed by $^{150}$Nd and $^{152}$Sm. 
The finding that the latter two isotopes are not so close as $^{148}$Nd and $^{150}$Sm is a bit surprising in view of the low $P$-factor for $^{148}$Nd, see Table \ref{ta:P}. 
From the values shown in the table one may deduce that the first-order phase transition occurs when $P$ is about 4.5 rather than 5.

\begin{figure}[t]
\centering
\includegraphics[width=9.3cm]{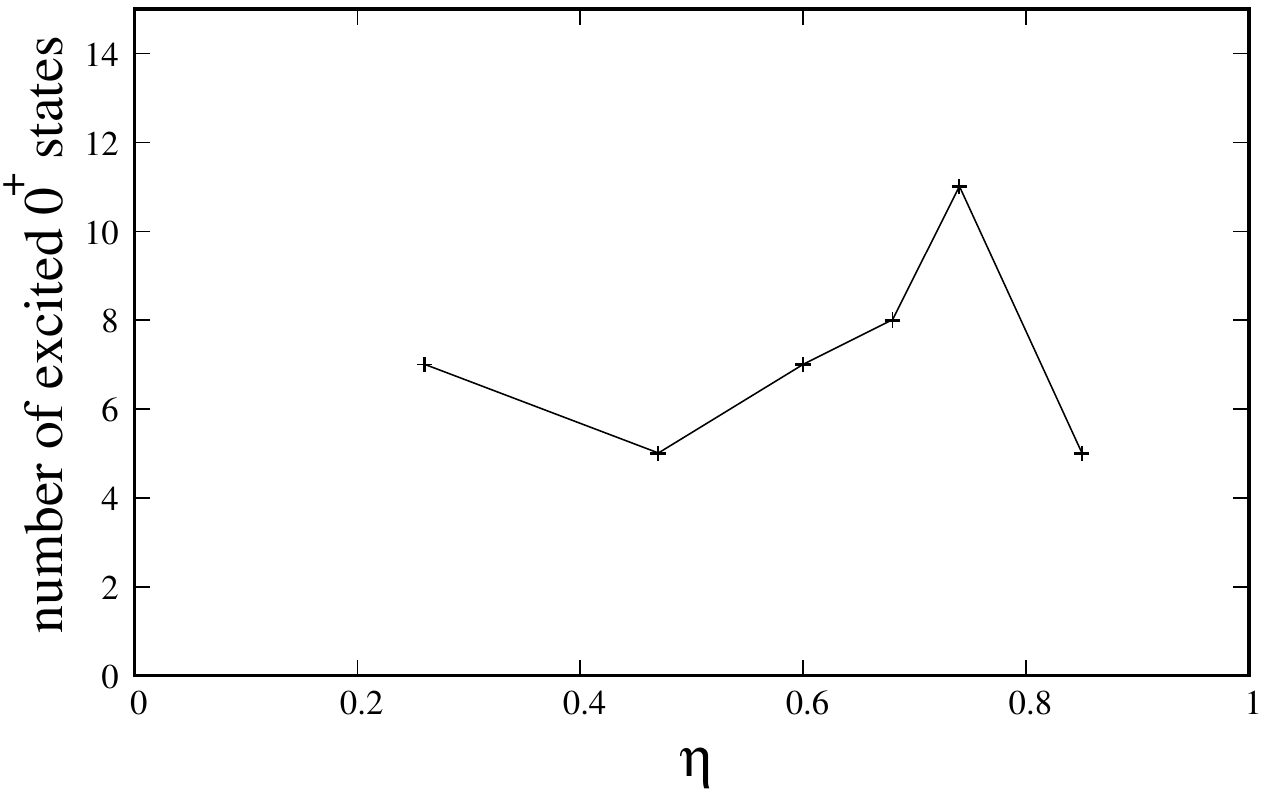}
\caption{\protect\small The number of $0^+$ states observed below 2.5~MeV as a function of $\eta$ (see text). The corresponding nuclei are from left to right:$^{162}$Dy, $^{168}$Er, $^{158}$Gd, $^{176}$Hf, $^{154}$Gd and $^{152}$Gd.}
\end{figure}

Using the simple IBM hamiltonian of the type (\ref{ising}), McCutchan {\it et al.} \cite{Libby04} performed systematic fits of Gd, Dy, Er, Yb and Hf isotopes, which allowed the location of these isotopes in the triangle. 
It was found that the phase transition is crossed inside the triangle for $^{154}$Dy and $^{156}$Er, while $^{152}$Gd lies on the U(5)-SU(3) leg before the phase transition and $^{154}$Gd lies close to it, but after the transition. 
The Yb and Hf isotopes do not cross the phase transition on the U(5)-SU(3) side. 

Another experimental observable was proposed by von Brentano {\it et al.} \cite{Brentano04} who observed that at the first-order transition from spherical to deformed shapes a rapid increase of the E0 strength for the $0^+_2\to 0^+_1$ transition takes places. 
A recent measurement in $^{154}$Sm confirmed this behavior nicely \cite{Wimmer08}:
$\rho^2(E0;0^+_2 \to 0^+_1)=0.020(2)$ for $^{150}$Sm, 0.051(5) for $^{152}$Sm, and 0.096(42) $^{154}$Sm.

Finally, as shown e.g. in Ref.~\cite{Cejnar00}, at the first-order phase transition the spectrum of excited states is maximally compressed.
Using the $(p,t)$ reaction at the highly sensitive Q3D spectrometer in Garching, an extended search for $0^+$ states in several even-even nuclei in the rare earth region was performed \cite{Meyer06}.
It was possible to identify the $0^+$ states up to very high energies.
Fig.~6 shows the numbers of observed $0^+$ states below 2.5~MeV for the nuclei studied in Ref.~\cite{Meyer06} against the fitted value \cite{Libby04b} of $\eta$ in the hamiltonian parametrization (\ref{ising}).
One clearly observes an increase of the $0^+$ state density at the critical value $\eta=0.8$ (corresponding in the fit to $^{154}$Gd). 
Note that here no corrections were applied for the changing boson number and the fitted $\chi$ parameter. 

\subsubsection{Spherical to $\gamma$-unstable and prolate to oblate transitions}

While there is ample evidence for atomic nuclei situated at the first-order phase transition between spherical and prolate deformed shapes, the situation is less clear for the second-order phase transition.
Despite the fact that the U(5)-O(6) transition in the IBM has been throughoutly studied from the theoretical side, only few examples of nuclei located along this leg of the symmetry triangle are known.
Early work \cite{Stachel82} indicated that the neutron-rich Ru and Pd isotopes are well described by this transitional class. 
More recently, the Ru isotopes were reanalysed by Frank, Alonso, and Arias \cite{Frank01} to locate the second-order critical point. 
Using two neutron separation energies, level energy systematics, and $B({\rm E2})$ values these authors concluded that the chain of Ru
isotopes in between $^{98}$Ru and $^{110}$Ru can be described as a U(5)-O(6) transition with the critical point situated at $^{104}$Ru. 

After the introduction of the E(5) critical point symmetry \cite{Iachello00}, the nucleus $^{134}$Ba was proposed as a candidate for the second-order critical point nucleus by Casten and Zamfir \cite{Casten00}. 
In this work, a comparison of $B({\rm E2})$ and energy ratios with an IBM calculation at the second-order phase transition was made, yielding a very good agreement. 
A subsequent systematic search in the nuclear database indicated that $^{128}$Xe might be another good candidate for the second-order critical nucleus \cite{Clark04}.
Very recently, the Zn isotopes were studied and $^{64}$Zn was proposed to be located close to the second-order transition \cite{Mihai07}.

\begin{figure}[t]
\centering
\includegraphics[width=18.5cm]{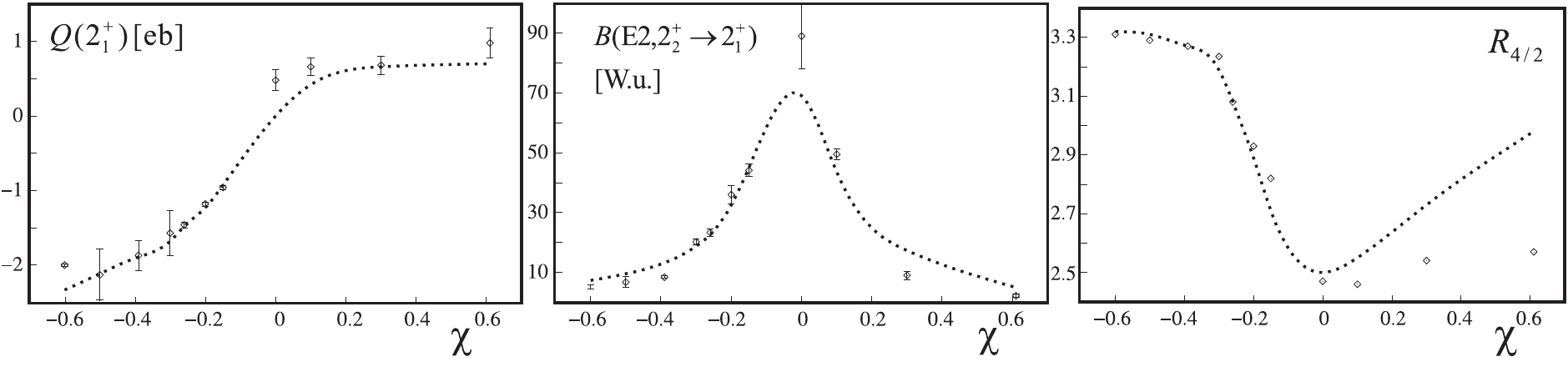}
\caption{\protect\small Experimental observables $Q(2^+_1)$, $B({\rm E2};2^+_2\to 2^+_1)$, and $R_{4/2}$,  for $^{180}$Hf, $^{182,184,186}$W, $^{188,190,192}$Os,  $^{194,196}$Pt, and $^{198,200}$Hg. The dashed lines represent theoretical values calculated for the appropriate $N$. Adapted from \cite{Jolie03}.}
\label{fi:Os-hg}
\end{figure}

For the SU(3)-$\overline{\rm SU(3)}$ first-order transition the critical point takes place at the O(6) dynamical symmetry.
Although oblate nuclei, or transitions from $\gamma$-soft to oblate shapes, are relatively rare in the known part of the nuclear chart, they do occur in the Pt and Hg isotopes close to the O(6) nucleus $^{196}$Pt. 
In Ref.~\cite{Jolie02}, three different signatures were investigated: the $4_1^+$ to $2_1^+$ energy ratio $R_{4/2}$, the $2_1^+$ quadrupole moment $Q(2^+_1)$, and the transition strength $B({\rm E2};2^+_2\to 2^+_1)$. 
In an attempt to span a larger part of the extended Casten triangle, these quantities are plotted in Fig.~\ref{fi:Os-hg} for nuclei ranging from the well deformed prolate rotor nucleus $^{180}$Hf up to $^{200}$Hg. 
For each nucleus the $\chi$ value from the quadrupole operator $Q_{\chi}$ was fitted using the hamiltonian (\ref{ising}) with $\eta=0$ and the respective value of $N$.
Clearly all experimental observables indicating the prolate-oblate phase transition are observed \cite{Jolie03}. 
The very rapid change of the quadrupole moment might explain why $^{196}$Pt has a nonvanishing quadrupole moment while still being a good O(6) nucleus.
Surprisingly, the Hg isotopes do not resemble vibrational or shell model like structures whom have the $R_{4/2}$ ratio around or below 2.


\section{Quantum phase transitions in IBM extensions}
\label{se:ext}

As mentioned in Sec.~\ref{se:iba}, the interacting boson model with just two types of bosons represents the simplest member of the IBM family.
Extensions are possible in several directions:
The distinction of proton-proton and neutron-neutron pairs leads to the two-component IBM-2, while mixed proton-neutron pairs are included in more sophisticated versions of the model (IBM-3 and -4).
The description of odd nuclei can be achieved when valence states of a single fermion are added into the Hilbert space within the IBFM.
To describe other types of excitations, one may consider non-scalar bosons of different angular momenta ($p,f,g,\dots$) together with (or instead of) the $d$ boson; this is also relevant if the IBM approach is applied in molecular or other types of physics.
The inclusion of higher order boson interactions, external rotation, or different particle-hole configurations represent some other possibilities to refine the model.
Each of these possibilities implies particular modifications of the model phase structure.
Some of these modifications will be outlined in this section, although the research in this field is still in progress.
The study of phase transitions in the IBM extensions is interesting not only from the viewpoint of concrete applications, but also because these models provide case examples of the influence of additional degrees of freedom on the QPT phenomena.


\subsection{Proton-neutron degrees of freedom (IBM-2)}
\label{se:ibm2}

In the version of the interacting boson model discussed up to now (IBM-1), no distinction was made between proton-proton and neutron-neutron pairs. 
In order to make the model more realistic, it is essential to introduce bosons of the proton and neutron type.
It is done in the two-component version of the model, the IBM-2 \cite{IA87}.
Hereby it is implicitly assumed that the shell model orbits which the protons and neutrons occupy are different, as is the case for medium mass and heavy nuclei.
The IBM-2 gives a successful phenomenological description of low-energy collective properties of virtually all such (even-even) nuclei.
The IBM-3 and 4, which are applicable in lighter nuclei, are not discussed here.

In the IBM-2, the total number of bosons $N$ is the sum of the neutron and proton boson numbers, $N_\nu$ and $N_\pi$, which are both conserved separately. 
The spectrum generating algebra of the IBM-2 is a product ${\rm U}_\nu(6)\otimes{\rm U}_\pi(6)$, consisting of neutron and proton generators $b^\dag_{\nu lm}b_{\nu l'm'}$ and $b^\dag_{\pi lm}b_{\pi l'm'}$, respectively. 
The model space is the product $[N_\nu]\times[N_\pi]$ of symmetric representations of ${\rm U}_\nu(6)$ and ${\rm U}_\pi(6)$. 
The most general $(N_\nu,N_\pi)$ conserving rotationally invariant IBM-2 hamiltonian will not be presented here, as it has 21 parameters. 
Instead, a strongly simplified hamiltonian is used. 
A comprehensive review of the model and its implications for nuclear structure can be found e.g. in Ref.~\cite{Lipas90}. 
Here we highlight three main features. 

The first is that the existence of two kinds of bosons offers the possibility to introduce a so called {\em $F$-spin\/} quantum number \cite{Otsuka78}, which is rather similar to isospin.
Since $F=\frac{1}{2}$, the bosons can be in two possible charge states: $M_F=-\frac{1}{2}$ for the neutron type boson and $M_F=+\frac{1}{2}$ for the proton type.
The $F$ spin is formally defined by the algebraic reduction
\begin{equation}
\begin{array}{ccccc}
{\rm U}(12)&\supset&{\rm U}_{\nu\pi}(6)&\otimes&{\rm U}(2)\\
\downarrow&&\downarrow&&\downarrow\\[0mm]
[N]&&[N-f,f]&&[N-f,f]
\end{array}\,,
\label{ibm2lab}
\end{equation}
\noindent
with $2F=N-2f$ the difference between the Young tableau labels that characterize both ${\rm U}_{\nu\pi}{\rm (6)}$ and U(2).
The algebra U(12) consists of the generators $b^\dag_{\rho lm}b_{\rho'l'm'}$, with $\rho,\rho'=\nu$ or $\pi$, which also includes operators
that change a neutron boson into a proton boson or {\it vice versa}.
Under the U(12) algebra, all bosons behave symmetrically, hence the representations of ${\rm U}_{\nu\pi}{\rm (6)}$ and U(2) are identical.

The SU(2) algebra associated with the $F$-spin, which is a subalgebra of U(2) in~(\ref{ibm2lab}), consists of the diagonal operator $F_z=\frac{1}{2}(N_\pi-N_\nu)$ and the raising and lowering operators $F_\pm$ that connect neutron and proton bosons.
It is clear that all IBM-2 hamiltonians must commute with $F_z$.
The hamiltonians which also commute with $F^2$ have the $F$-spin as a good quantum number. 
The classification and analysis of the $F$-spin conserving dynamical symmetries of the IBM-2 show that the three symmetry cases of the IBM-1 are recovered. 
These are the ${\rm U}_{\nu\pi}{\rm (5)}$, ${\rm O}_{\nu\pi}{\rm (6)}$ and ${\rm SU}_{\nu\pi}{\rm (3)}$ limits of the IBM-2 \cite{Isacker86}.

The second important aspect of the model is that it predicts states which are additional to those found in IBM-1.
Their structure can be understood in terms of the $F$-spin classification~(\ref{ibm2lab}) for $F$-symmetric hamiltonians. 
The states with the maximal $F$-spin value, $F=\frac{1}{2}N$, are symmetric in ${\rm U}_{\nu\pi}{\rm (6)}$ and represent exact analogues of the IBM-1 states. 
The next class of states has $F=\frac{1}{2}N-1$, and these are no longer symmetric in ${\rm U}_{\nu\pi}{\rm (6)}$ but belong to its representation $[N-1,1]$.
Such {\em mixed-symmetry states\/} were studied theoretically in 1984 by Iachello \cite{Iachello84} and since then they have been observed in many nuclei (see e.g. Ref.~\cite{Pietralla08}).
Of particular relevance are the $1^+$ states, since these are allowed in IBM-2 but not in IBM-1.
From the geometrical analysis performed in the limit of large boson numbers \cite{Ginocchio92} it emerges that the mixed-symmetry states correspond to linear or angular displacement oscillations, in which the neutrons and protons are out of phase, in contrast to the symmetric IBM-2 states for which such oscillations are in phase. 
The occurrence of mixed-symmetry states was first predicted in the context of geometric two-fluid models in vibrational \cite{Faessler66} and deformed \cite{Iudice78} nuclei, in which they appear as neutron-proton counter oscillations, pictorially referred to as scissors modes.

The third important feature offered by the IBM-2 is the possibility to get {\em triaxial shapes\/} besides the axially symmetric ones. 
This happens when proton and neutron fluids exhibit different types of deformation---prolate and oblate.
Hereby a new dynamical symmetry can be constructed, denoted by ${\rm SU}^*_{\nu\pi}{\rm (3)}$, for which the respective $\chi$ parameters are in absolute value equal to $\tfrac{\sqrt{7}}{2}$ but have opposite signs \cite{Dieperink82,Sevrin87}.
This possibility gives rise to the Dieperink tetrahedron, which has an extra dimension compared to the Casten triangle, and to a new, triaxial shape phase of the model.
Note that if all neutron-proton deformation combinations are taken into account, there are in fact four and not just two SU(3) limits: prolate-prolate, ${\rm SU(3)}$, oblate-oblate, ${\overline{\rm SU(3)}}$, prolate-oblate, ${\rm SU(3)}^*$, and oblate-prolate, ${\overline{\rm SU(3)}}\,^*$. 
Among these combinations, however, only two are inequivalent.

The analysis of shape-phase transitions in the IBM-2 is much more complex than the one in IBM-1. 
It was done independently and at the same time by Arias, Garc{\'\i}a-Ramos, and Dukelsky \cite{Arias04}, and by Caprio and Iachello \cite{Caprio04,Caprio05}.
The classical limit of the IBM-2 can be obtained using a two-component coherent state,
\begin{equation}
\ket{N_\pi,\alpha^{(2)}_\pi,N_\nu,\alpha^{(2)}_\nu}\propto
\left[s^\dag_\pi +\sum_m\alpha^{(2)}_{\pi m}d^\dag_{\pi m}\right]^{N_{\pi}}
\left[s^\dag_\nu+\sum_m\alpha^{(2)}_{\nu m}d^\dag_{\nu m}\right]^{N_\nu}\ket{0}
\,,
\label{ibm2coh}
\end{equation}
which is used to construct the energy surface for a given hamiltonian as a function of parameters $\alpha^{(2)}_{\pi m}$ and $\alpha^{(2)}_{\nu m}$.
Those can then be interpreted as 4 deformation parameters $\beta_\pi,\gamma_\pi,\beta_\nu,\gamma_\nu$, and $2\times 3$ Euler angles. 
In contrast to the IBM-1, the Euler angles do partly affect the energy surface through the relative orientation $(\theta_1,\theta_2,\theta_3)$ of the proton and neutron ellipsoids. 
Therefore, the IBM-2 energy functional needs to be minimized in 7 parameters.
 
As mentioned before, a tractable study of phase transitions using the complete IBM-2 hamiltonian (with 21 parameters) is hardly feasible. 
Hence only some simplified hamiltonians, resembling those used in the IBM-1, were studied. 
The following hamiltonian was used in both Refs.~\cite{Arias04,Caprio04}
\begin{equation}
H^{\chi_\pi,\chi_\nu}(\zeta)=(1-\zeta)\tfrac{1}{N}(n_{d\pi}+n_{d\nu})+\zeta\tfrac{1}{N^2}
\left[- (Q^{\chi_\pi}_{\pi}+Q^{\chi_\nu}_{\nu})\cdot(Q^{\chi_\pi}_{\pi}+Q^{\chi_\nu}_{\nu})\right]
\label{ibm2caprio}
\,,
\end{equation}
with $N$ the total boson number.
The form (\ref{ibm2caprio}) is a direct generalization of the IBM-1 hamiltonian (\ref{ising}). 
It contains (apart from an overall scaling factor that we do not account for) 3 control parameters $\zeta$, $\chi_\pi$, and $\chi_\nu$. 
The hamiltonian belongs to the type for which it was shown \cite{Ginocchio92} that the global minimum only occurs for aligned proton and neutron deformations, reducing the number of order parameters from 7 to 4, i.e. the $\beta$ and $\gamma$ shape variables of both fluids.

\begin{figure}[t]
\centering
\includegraphics[width=13.9cm]{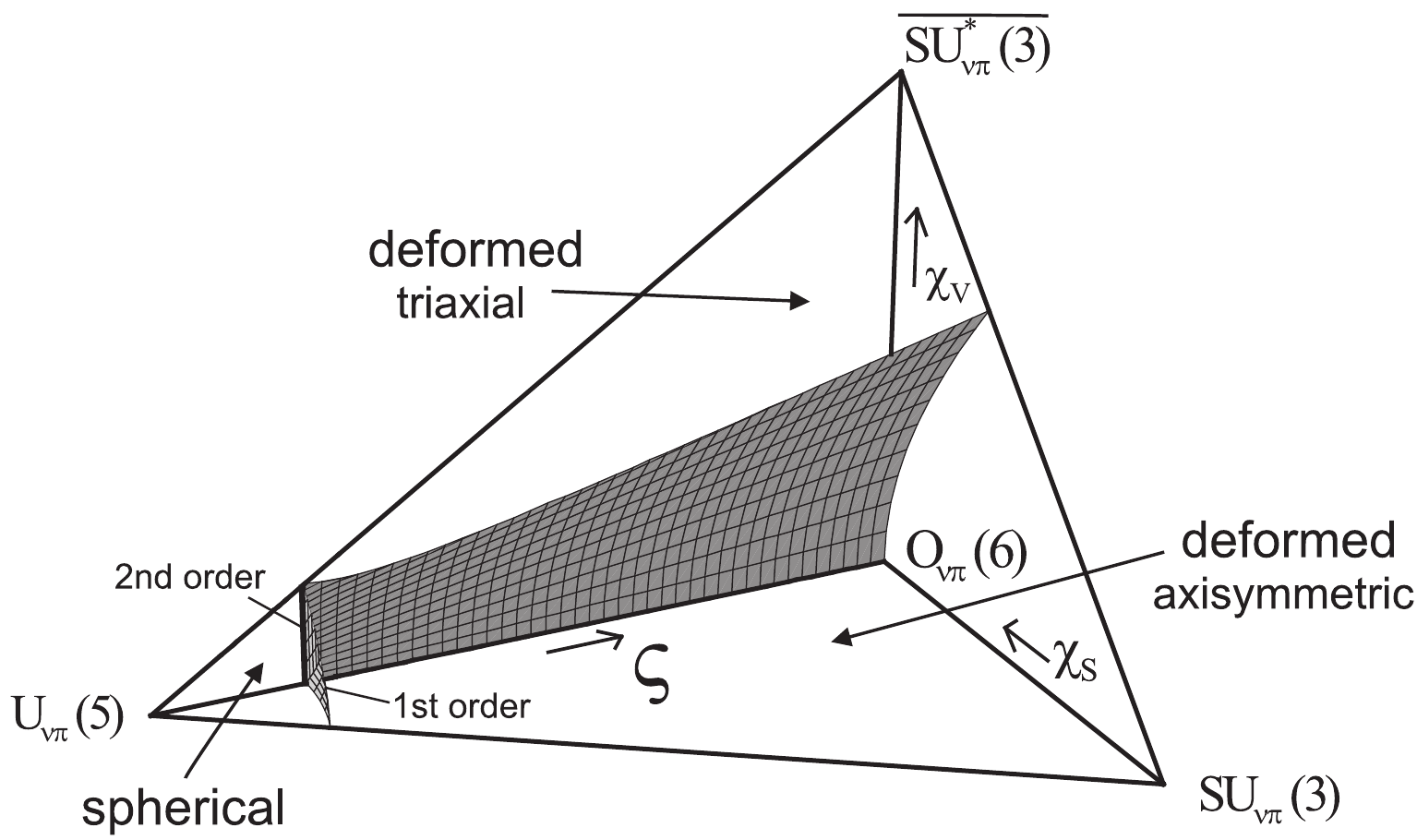}
\caption{\protect\small The phase structure of the IBM-2 hamiltonian (\ref{ibm2caprio}) in the space of parameters $\zeta$, $\chi_{{\rm S}}=\tfrac{1}{2}(\chi_{\pi}+\chi_{\nu})$, and $\chi_{{\rm V}}=\tfrac{1}{2}(\chi_{\pi}-\chi_{\nu})$. The surfaces represent first- and second-order phase transitions between spherical and deformed, and axially symmetric and triaxial shapes. Adapted from Caprio and Iachello \cite{Caprio05}.}
\label{fi:ibm2}
\end{figure}

The phase structure of hamiltonian (\ref{ibm2caprio}) can be studied analytically for $\zeta=1$, where the ${\rm O}_{\nu\pi}{\rm (6)}$, ${\rm SU}_{\nu\pi}{\rm (3)}$, ${\rm SU}_{\nu\pi}^*{\rm (3)}$, and the remaining two equivalent SU(3) dynamical symmetries are located.
The ground state at ${\rm O}_{\nu\pi}{\rm (6)}$ represents a $\gamma$-unstable deformed configuration, at ${\rm SU}_{\nu\pi}{\rm (3)}$ a $\gamma$-rigid deformed configuration, prolate or oblate for both fluids, whereas at ${\rm SU}^*_{\nu\pi}{\rm (3)}$ the ground state is triaxial since both fluids have opposite types of deformation (with the symmetry axes orthogonal to each other) \cite{Ginocchio92}.
A second-order phase transition between axially symmetric and triaxial deformed shapes occurs at the boundary determined by the following equations \cite{Caprio04},
\begin{eqnarray}
\frac{9\frac{N_\pi}{N_\nu}\beta_\pi(\beta^2_\pi-1)+\beta_\nu(2\beta^2_\pi-1)(\beta^2_\pi+1)}{2\beta_\pi(\beta^2_\pi-2)^2}
\cdot
\frac{9\frac{N_\nu}{N_\pi}\beta_\nu(\beta^2_\nu-1)+\beta_\pi(2\beta^2_\nu-1)(\beta^2_\nu+1)}{2\beta_\nu(\beta^2_\nu-2)^2}
=1\,,
\\
\beta_\rho=\sqrt{\tfrac{1}{14}\chi^2_\rho+1}-\tfrac{1}{\sqrt{14}}\chi_\rho\,,
\qquad\qquad\qquad\qquad\qquad\qquad\qquad
\nonumber
\end{eqnarray}
with $\beta_\rho$ expressing the deformation parameter of the neutron ($\rho=\nu$) and proton ($\rho=\pi$) ellipsoids. 
The location is strongly dependent on the ratio $N_\pi/N_\nu$ except at the ${\rm O}_{\nu\pi}{\rm (6)}$ limit which is always located at the second-order transition between $\gamma$-soft and triaxial deformation. 

A numerical analysis of the full parameter space showed that the above phase separatrix extends from $\zeta=1$ to $\zeta<1$, where it forms a surface of second-order phase transitions between axially symmetric and triaxial deformed phases. 
This plane is terminated at the spherical-deformed phase transition, which also forms a surface close to the ${\rm U}_{\nu\pi}{\rm (5)}$ dynamical symmetry.
The spherical-deformed phase transition is of the first order except at the intersection with the axisymmetric-triaxial phase transition, which happens at $\chi_\nu =\chi_\pi =0$.
The whole phase diagram in the symmetry tetrahedron is depicted in Fig.~\ref{fi:ibm2}.
A very detailed account, including an analysis of more complex hamiltonians can be found in Ref.~\cite{Caprio05}.

A closely related problem was recently analyzed by Iachello and P{\' e}rez-Bernal \cite{Iachello08} in the context of molecular physics, within the so called vibron model for coupled benders \cite{Iachello96}.
The model represents a generalization of the 2D vibron model, which is a lower-dimensional analog of the $sd$-IBM with the $d$ boson replaced by a two-component boson $\tau$ \cite{Iachello03b}. 
The spectrum generating algebra of the 2D vibron model coincides with U(3), while the extension investigated in Ref.~\cite{Iachello08} concerns  coupled systems (such as molecules of the ABBA type) with the spectrum generating algebra U(3)$\otimes$U(3).
A simple ground-state phase diagram of three basic molecular configurations was drawn in the plane of the hamiltonian control parameters.

\subsection{Systems with bosons and fermions (IBFM)}

One particularly important extension of the simple interacting boson model concerns odd-mass nuclei.
It is achieved by considering, in addition to the $s$ and $d$ bosons, a fermion coupled to the core with an appropriate boson-fermion interaction.
The resulting interacting boson-fermion model (IBFM) \cite{II91} is thus a specific version of the particle-core coupling model, which has been widely used in nuclear physics to describe odd-mass nuclei~\cite{BM75}.
The characteristic feature of the IBFM is that it lends itself very well to a study based on symmetry considerations whereby certain classes of boson--fermion hamiltonians can be solved analytically.
The basic building blocks of the IBFM are $N$ bosons, in terms of which the even-even core states are modeled, and fermions occupying a set of single-particle orbits with various angular momenta $j$.
Low-lying collective states of an odd-mass nucleus with $2N+1$ valence nucleons are approximated as a single fermion coupled to $N$-boson states.
A particularly attractive feature is the similarity in the description of even-even and odd-mass nuclei.
This has given rise to the development of a supersymmetric model \cite{Iachello80}.

The IBFM hamiltonian contains bosonic and fermionic parts and a boson-fermion interaction: $H_{\rm BF}=H_{\rm B}+H_{\rm F}+V_{\rm BF}$.
The bosonic hamiltonian can be taken directly from the IBM.
Because only one fermion is coupled to the boson core, the fermion hamiltonian can be written as $H_{\rm F}=\sum_j\epsilon_jn_j$,
where $n_j$ is the number of fermions in orbit $j$ and $\epsilon_j$ its single-particle energy.
The boson-fermion interaction is assumed to have a two-body character:
\begin{equation}
V_{\rm BF}=\sum_{ljl'j'J}v^{(J)}_{ljl'j'}\left[(b^\dag_l\times a^\dag_j)^{(J)}\times({\tilde b}_{l'}\times{\tilde a}_{j'})^{(J)}\right]^{(0)}_0. 
\label{4_hambf} 
\end{equation} 
This form, however, is too general for phenomenological applications.
A simplification is given by:
\begin{eqnarray} 
V_{\rm BF}^{\rm (m)}=\sum_j\kappa_j\left[(d^\dag\times{\tilde d})^{(0)}\times(a_j^\dag\times{\tilde a}_j)^{(0)}\right]^{(0)}_0
\,,\qquad
V_{\rm BF}^{\rm (q)}=\sum_{jj'}\kappa_{jj'}\left[Q_{\rm B}\times(a_j^\dag\times{\tilde a}_{j'})^{(2)}\right]^{(0)}_0
\,,\nonumber\\
V_{\rm BF}^{\rm (e)}=\sum_{jj'j''}\kappa_{jj'}^{j''}{\bf :}\left[(d^\dag\times{\tilde a}_j)^{(j'')}\times({\tilde d}\times a_{j'}^\dag)^{(j'')}\right]^{(0)}_0{\bf :}
\label{4_hamphen}\,.
\end{eqnarray}
Here ${\bf :}\bullet{\bf :}$ indicates normal ordering and $Q_{\rm B}$ denotes the boson quadrupole operator, Eq.~(\ref{Q}). 
The exchange interaction $V_{\rm BF}^{\rm (e)}$ takes account of the fact that the IBM bosons have an internal structure, and this leads to the exchange effects. The coefficients $\kappa$ can be related to the occupation probabilities and quasi particle energies derived from a BCS calculation (see \cite{II91}).

The spectrum generating Lie algebra associated with the IBFM is $[{\rm U}_{\rm B}(6)\otimes{\rm U}_{\rm F}(\Omega)]$.
The usual boson algebra ${\rm U}_{\rm B}(6)$ describes the collective core excitations, while ${\rm U}_{\rm F}(\Omega)$, with $\Omega=\sum_j(2j+1)$ denoting the size of the fermionic single-particle space, corresponds to the fermion degrees of freedom.
The algebra contains two sets of generators: in terms of bosons $b^\dag_{lm}\tilde b_{l'm'}$, and in terms of fermions $a^\dag_{jm}\tilde a_{j'm'}$ (both types written most favorably in the angular momentum coupled form). 
The existence of analytically solvable IBFM hamiltonians relies on isomorphisms between some boson and fermion algebras.
A simple example is provided by the angular momentum algebras defined separately for bosons and for fermions. 
The former consists of the angular momentum operators $L_{\rm B}$ which generate the boson algebra ${\rm O}_{\rm B}(3)$ and occurs in the lattice~(\ref{3_ibmlat}). 
The fermion angular momentum operators $J_{\rm F}$ are generators of an equivalent Lie algebra denoted as ${\rm SU}_{\rm F}(2)$.
An essential difference between these algebras is that while the bosons couple only to integer angular momenta, the fermions can form also states with half-integer momenta.
The summed operators $J_{\rm BF}=L_{\rm B}+J_{\rm F}$ generate the boson-fermion algebra ${\rm SU}_{\rm BF}(2)$ of the total angular momentum, which can be integer or half-integer depending on whether the number of fermions is even or odd.
Since ${\rm O}_{\rm B}(3)$ has to be a subalgebra of ${\rm U}_{\rm B}(6)$ and ${\rm SU}_{\rm F}(2)$ one of ${\rm U}_{\rm F}(\Omega)$, it follows that ${\rm SU}_{\rm BF}(2)$ is a subalgebra of the IBFM spectrum generating algebra.
The required form of all IBFM dynamical symmetries is the following:
\begin{equation} 
\left[{\rm U}_{\rm B}(6)\otimes{\rm U}_{\rm F}(\Omega)\right]\supset \cdots \supset{\rm SU}_{\rm BF}(2)
\,. 
\label{4_clas} 
\end{equation}
The subscript BF is usually omitted and the total angular momentum algebra is denoted as Spin(3).
A comprehensive review of various dynamical symmetries of the IBFM can be found in Ref.~\cite{II91}.
Numerous experimental examples are given in Ref.~\cite{Vervier87}.

It is clear that in absence of fermions the IBFM reduces to the standard IBM of even-even nuclei. 
This naturally leads to a {\em supersymmetric extension\/} of the IBFM that allows a simultaneous description of even-even and odd-mass nuclei.
Let us stress that a necessary condition for such an approach to be phenomenologically successful is that the energy scales for bosonic and fermionic excitations are comparable.
Nuclear supersymmetry treats spectral properties of different nuclei as arising from a single boson-fermion hamiltonian and a single set of transition operators.
This unification is achieved by embedding the IBFM spectrum generating algebra into a superalgebra ${\rm U}(6/\Omega)$:
\begin{equation} 
\begin{array}{ccccc}
{\rm U}(6/\Omega)&\supset&\bigl[{\rm U}_{\rm B}(6)&\otimes&{\rm U}_{\rm F}(\Omega)\bigr]\,.\\ \downarrow&&\downarrow&&\downarrow\\[0mm] [{\cal N}\}&&[N]&&[1^M]
\end{array}
\label{4_latsusy} 
\end{equation}
Here, the supersymmetric representation $[{\cal N}\}$ of ${\rm U}(6/\Omega)$ imposes symmetry in the bosons and anti-symmetry in the fermions, and contains the $[{\rm U}^{\rm B}(6)\otimes{\rm U}^{\rm F}(\Omega)]$ representations $[N]\times[1^M]$ with ${\cal N}=N+M$.
Thus, a single supersymmetric representation contains states in even-even ($M=0$) as well as odd-mass ($M=1$) nuclei.
Nuclear supersymmetry was originally postulated by Iachello \cite{Iachello80} as a symmetry among pairs of nuclei. 
Subsequently, it was extended to quartets of nuclei, where the neighboring odd-odd nuclei are also incorporated \cite{Isacker85b}.

To probe the ground-state phase transitions in the IBFM, we consider below a fermion in orbits with angular momenta $j=\tfrac{1}{2}$, $\tfrac{3}{2}$, and $\tfrac{5}{2}$.
The relevant dynamical algebra is $[{\rm U}_{\rm B}(6)\otimes{\rm U}_{\rm F}(12)]$ and an isomorphism between the boson and fermion algebras is established by introducing pseudo-spin and the pseudo-orbital angular momenta for the fermion: $S_{\rm F}=\tfrac{1}{2}$ and $L_{\rm F}=0,2$, respectively. 
This leads to a reduction of the spectrum generating algebra to $[{\rm U}_{\rm B}(6)\otimes{\rm U}_{\rm F}(6)\otimes{\rm U}_{\rm F}(2)]$, which allows to combine both ${\rm U}(6)$ algebras and to construct the three dynamical symmetries of the standard IBM \cite{Isacker84}. 
The embedding (\ref{4_latsusy}) results in the U(6/12) supersymmetric scheme. 
While this is sufficient for a theoretical study, the practical application is limited to those nuclei where the above fermionic shells are important, e.g. the ${\cal A}=$70--80, the Rh-Ag and the Pt-Os-Hg mass regions. 

\begin{figure}[t]
\centering
\includegraphics[width=18.5cm]{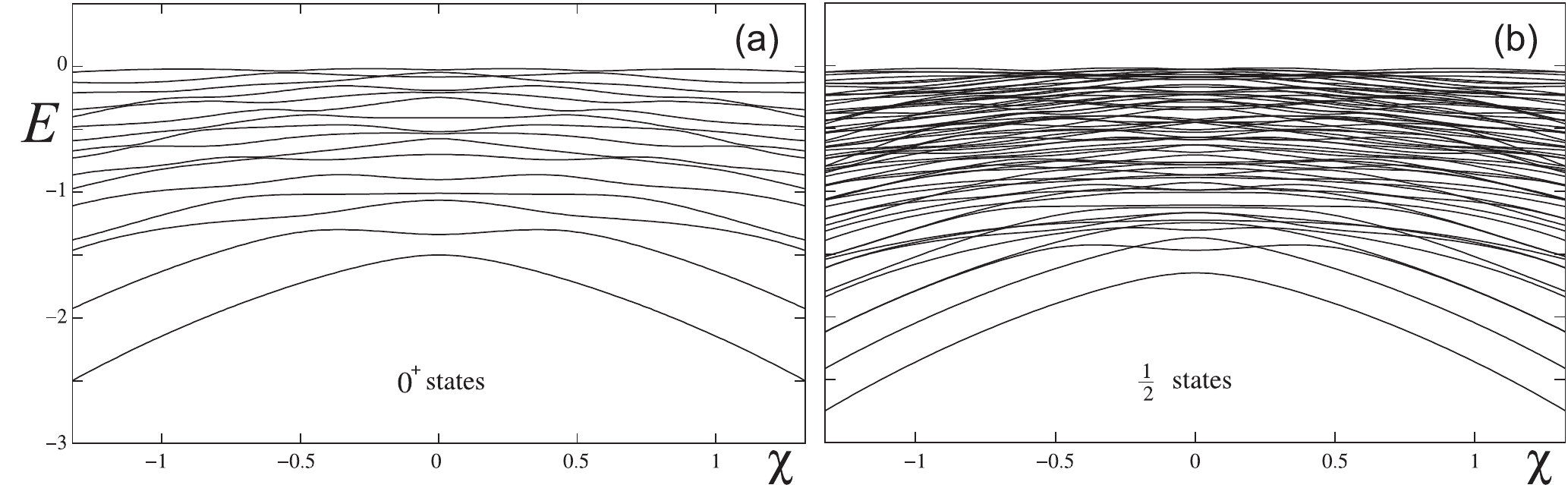}
\caption{\protect{\small  Absolute energies of the $0^+$ states in even-even nuclei (a) and of the corresponding $\tfrac{1}{2}$ states in odd-$\cal A$ nuclei (b) along the prolate-oblate shape transition. The calculation was done using the hamiltonian (\ref{Qhamiltoniansusy}) with $N=10$, $\eta=0$, and $\chi$ variable. Adapted from \cite{Jolie04}.}}
\label{fi:su3su3b}
\end{figure}

The first study of quantum phase transitions in odd-${\cal A}$ nuclei was due to Jolie, Heinze, Van Isacker, and Casten \cite{Jolie04}.
It relied on a supersymmetric extension of the results obtained for even-even nuclei. 
The simplified hamiltonian (\ref{ising}) was generalized to
\begin{equation}
H_{\rm BF}^{\chi}(\eta)\propto
(1-\eta)\tfrac{1}{N}\left(-Q^{\chi}_{\rm BF}\cdot Q^{\chi}_{\rm BF}\right)
+\eta C_1[{\rm U}_{\rm BF}(5)]
\label{Qhamiltoniansusy}
\end{equation}
with $C_1[{\rm U}_{\rm BF}(5)]$ the first-order Casimir operator of U$_{\rm BF}(5)$ and $Q^\chi_{\rm BF}$ the quadrupole operator of the form (\ref{Q}) but using generators of U$_{\rm BF}(6)$.
The use of U(6/12) supersymmetric scheme allowed to extend the Casten triangle to odd-mass nuclei.
In Ref.~\cite{Jolie04}, the application of Eq.~(\ref{Qhamiltoniansusy}) was discussed to the prolate-oblate phase transition on the ${\rm SU}_{\rm BF}(3)$--${\rm O}_{\rm BF}(6)$--$\overline{{\rm SU}_{\rm BF}(3)}$ line ($\eta=0$).

Figure~\ref{fi:su3su3b} compares the absolute energies of the $0^+$ states of the $N=10,M=0$ system to the ones of $J_{\rm BF}=\tfrac{1}{2}$ states in the $N=10,M=1$ system.
In the even-even system ($M=0$), the prolate-oblate phase transition is associated with the kink of the lowest energy at O(6) and the sequence of energies is symmetric around O(6), although the wave functions are different \cite{Jolie01}. 
A similar behavior is found for the odd-${\cal A}$ case ($M=1$).
Nevertheless, a closer inspection reveals some interesting differences.
In the even-even case, except at the dynamical symmetries with $\chi=0$ or $\pm\tfrac{\sqrt{7}}{2}$, all states undergo avoided level crossings indicating the absence of conserved quantum numbers (besides the total angular momentum).
This is not so in the odd-$\cal A$ calculation where one observes real crossings of several states.
The origin of these crossings follows from the conservation of the angular momentum $L_{\rm BF}=L_{\rm B}+L_{\rm F}$ by hamiltonian (\ref{Qhamiltoniansusy}) which results from the coupling of the total boson angular momentum to the pseudo-orbital angular momentum of the fermion.
$L_{\rm BF}$ is conserved in classifications contained in the lattices starting and ending with
\begin{eqnarray}
\begin{array}{cccccccccccccccccccccc}
{\rm U}(6/12)&\supset&
\bigl[{\rm U}_{\rm B}(6)&\otimes&{\rm U}_{\rm F}(12)\bigr]&\supset&
\bigl[{\rm U}_{\rm B}(6)&\otimes&{\rm U}_{\rm F}(6)&\otimes&{\rm U}_{\rm F}(2)\bigr]
&\supset&\cdots\\
\downarrow&&\downarrow&&\downarrow&&\downarrow&&\downarrow&&\downarrow&\\[0mm]
[{\mathcal N}\}&&[N]&&[1^M]&&[N]&&[1^M]&&[1^m]&&
\\ \\
&&&&&&
\cdots &\supset&
\bigl[{\rm O}_{\rm BF}(3)&\otimes&{\rm SU}_{\rm F}(2)\bigr]&\supset&{\rm Spin}(3)\,.\\
&&&&&& &&\downarrow&&\downarrow&&\downarrow\\
&&&&&& &&L_{\rm BF}&&S_{\rm F}&&J_{\rm BF}
\end{array}
\label{lattice}
\end{eqnarray}
Whatever embeddings are taken in the dotted part of (\ref{lattice}), $L_{\rm BF}$ remains a good quantum number and 
the same holds also for the pseudo-spin $S_{\rm F}$. 
A detailed analysis shows that the essential ingredient is the proper choice of the single-particle energies of the pseudo-spin doublet $j=\tfrac{3}{2}$ and $\tfrac{5}{2}$ \cite{Jolie04}.

\begin{figure}[t]
\centering
\includegraphics[width=15.8cm]{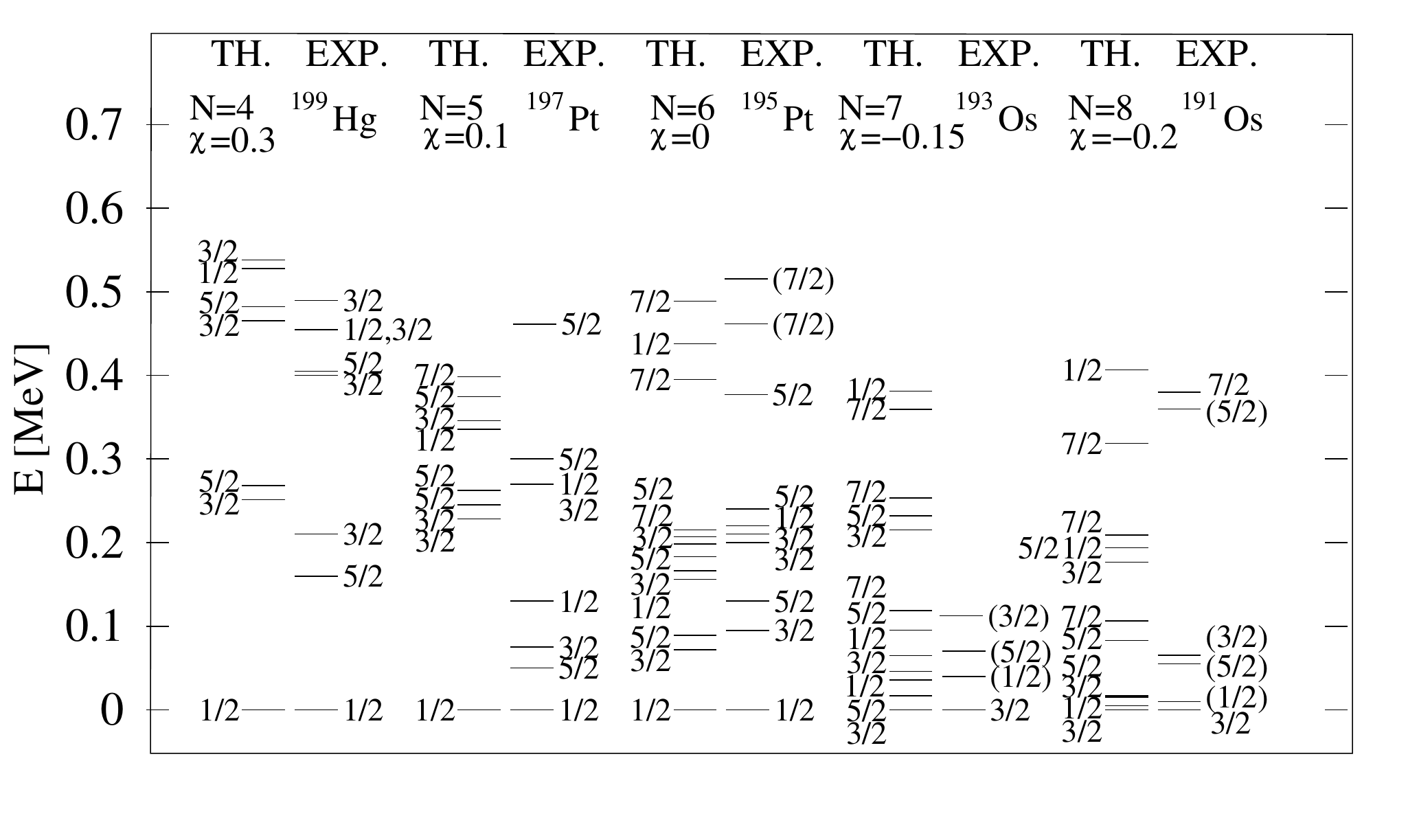}
\caption{\protect{\small Comparison between experimental and theoretical negative-parity states of odd-neutron isotopes between $^{191}$Os and $^{199}$Hg. Adopted from~\cite{Jolie04}.}}
\label{fi:susyex}
\end{figure}

In Ref.~\cite{Jolie04}, the hamiltonian (\ref{Qhamiltoniansusy}) was applied to odd-$\cal A$ nuclei between $^{191}$Os and $^{199}$Hg as a logical extension of the fit done for the even-even cores (see Fig.~\ref{fi:Os-hg}) \cite{Jolie03}. 
It was shown that the hamiltonian needs to be extended with $C_2[{\rm U}_{\rm BF}(6)]$ and $C_2[{\rm Spin}(3)]$, which do not affect the phase transitional behavior but allow a fine tuning of energies.
The result is shown in Fig.~\ref{fi:susyex}.
Although the hamiltonian is very simple, and although the parameters were kept constant across the region (after having adopted the values for $\eta$ and $\chi$ from the fit to the even-even partner), the overall structural changes, in particular the decreasing trend and compression of the low-lying levels with decreasing mass, were reproduced rather well.

Critical behaviors of odd-$\cal A$ nuclei have been recently studied for the ${\rm U}_{\rm BF}(5)$--${\rm O}_{\rm BF}(6)$ second-order transition by Alonso, Arias, Vitturi, and Fortunato \cite{Alonso05,Alonso06,Alonso07,Alonso07b}. 
These analyses were motivated by the fermion critical point symmetry E(5/4) introduced by Iachello \cite{Iachello05} as an extension to odd masses of the E(5) critical point symmetry \cite{Iachello00}. 
The E(5/4) symmetry is obtained in the geometric model when coupling a $j=\tfrac{3}{2}$ fermion to the E(5) core, but a similar system can be obtained also in the IBFM \cite{Alonso05}.
While the applicability of critical point symmetries is restricted only to critical points of phase transitions, the IBFM makes it possible to describe the whole transition from vibrational to $\gamma$-unstable nuclei. 
In Ref.~\cite{Alonso05}, a specific form of the interaction (\ref{4_hambf}) was used with only the quadrupole term in the boson-fermion interaction (\ref{4_hamphen}). 
Imposing $\chi=0$ for the boson quadrupole operator in both the boson hamiltonian and the boson-fermion interaction, the total hamiltonian conserves in addition to the O(5) quantum number also the ${\rm Spin}(5)$ quantum number along the whole transitional region. 
From the comparison between the E(5/4) and IBFM approaches at the critical point a moderate agreement was obtained. 

In a subsequent work \cite{Alonso06}, the boson-fermion interaction was extended with more terms conserving the ${\rm Spin}(5)$ symmetry. 
In addition to the electromagnetic transitions also single-particle transfer intensities were calculated. 
However, even the extended hamiltonian did not remove the discrepancies between both approaches.
The differences can be traced back to the square well form of the potential of the geometric model assumed in critical point symmetries, in contrast to the $\beta^4$ dependence of the critical IBM potential \cite{Ramos05}. 

Because the fermion space with only a $j=\tfrac{3}{2}$ orbit is rather restricted, Alonso, Arias, and Vitturi have extended recently both models to allow the fermion to occupy also orbits with $j=\tfrac{1}{2}$, $\tfrac{3}{2}$, and $\tfrac{5}{2}$ \cite{Alonso07,Alonso07b}. 
For the critical point symmetry this leads to the E(5/12) structure and for the IBFM it allows to use the U(6/12) supersymmetric approach as discussed before.
The hamiltonian (\ref{Qhamiltoniansusy}) was used to study the second-order phase transition in odd-$\cal A$ nuclei, posing now as a constraint $\chi=0$. 
The variable parameter $\eta$ controls the transition from ${\rm U}_{\rm BF}(5)$, at $\eta=1$, to ${\rm O}_{\rm BF}(6)$, at $\eta=0$. 
The energy spectrum and electromagnetic transition rates were compared with the critical point symmetry E(5/12) \cite{Alonso07b}. 
In this case, a better agreement between both approaches is obtained and signatures of the critical behavior are found in 
energies and $B({\rm E2})$ ratios of low-lying states. 
However, no comparison with experimental data is made in these works.

An extension of the coherent state formalism to odd-mass systems with a single fermion orbit was recently presented by Liu \cite{Liu07}. 
Under the neglect of the $\gamma$ degree of freedom, the energy functional was derived and studied for a hamiltonian with only a quadrupole term in the boson-fermion interaction. 
In this case, a somewhat surprising dependence of the positions of critical points $\eta_{\rm c}$ and $\chi_{\rm c}$ on the interaction strengths $\kappa$ from Eq.~(\ref{4_hamphen}) was obtained. 
Further studies including a proper treatment of the $\gamma$ degree of freedom are needed to verify these results. 

The study of shape-phase transitions in odd-$\cal A$ nuclei is more complicated than the one in even-even nuclei, also because the level densities are much higher in the former case. 
To investigate mixed systems of bosons and fermions is nevertheless of great interest as an increasing number of Bose-Fermi mesoscopic systems becomes available for experimental studies.

\subsection{Other extensions}

\subsubsection{Configuration mixing (IBM-CM)}

Besides the shape-phase transitions atomic nuclei may also exhibit shape coexistence.
This phenomenon, invoking multiple shapes present in the low-energy spectrum, was predicted and observed in many spherical nuclei near magic shells.
Particular configurations observed in these nuclei can be linked to the occupation of specific up- and/or down-sloping orbitals, coined \lq\lq intruder orbitals\rq\rq, which allowed for a simple understanding of the shape coexistence in the works of Heyde, Wood {\it et al.} \cite{Heyde83,Wood92}.
The method put forward by these authors has been used to predict shape coexistence e.g. in the Sn nuclei around mass number ${\cal A}=116$, and in the light Pb nuclei below ${\cal A}=196$.

The difference between shape coexistence and shape-phase transitions is not always very clear, especially from an experimental point of view \cite{Heyde04}.
In a shape-phase transition, the number of basis states is preserved through the transition as progressing from one limiting case into the other.
The eigenstates of one limit are spread out among the actual, intermediate eigenstates, which finally become the eigenstates of the other limit.
At any instant, the eigenstates form a complete basis of the same Hilbert space.
On the other hand, typical situations implying shape coexistence appear when a set of extra states (outside of the model space that is regularly considered for low-lying configurations) drops in excitation energy and produces new (intruder) low-lying states.
The Hilbert space is extended as ${\bf H}\equiv{\bf H}_{\rm nor}\oplus{\bf H}_{\rm int}$, where the two subspaces correspond to normal and intruder states.
Any hamiltonian acting in this space decomposes into three parts: the one of normal states, the one of intruder states, and the one describing their interaction.
In practical IBM applications, this division is implemented by coupling the $sd$ boson subspaces with total numbers of bosons $N$ and $N+2$.
These are used to describe shape coexistence between normal (valence) excitations and intruder excitations that include 2p-2h configurations.
Both sets of states can interact with each other via a mixing hamiltonian that does not conserve the boson number.


Both these seemingly different pictures were connected in the approach recently proposed by Frank, Van Isacker, and Iachello \cite{Frank06}.
These authors have introduced a method for studying phase transitions in systems with mixed configurations, which they call type-II phase transitions.
As explained above, a configuration mixing (CM) hamiltonian consists of parts corresponding to normal and intruder states (these form diagonal blocks of the hamiltonian) and their interaction (which appears off the diagonal).
To cope this structure with the QPT paradigm expressed by the hamiltonian in Eq.~(\ref{Hlin01}), a general phase transitional CM hamiltonian was considered in the form:
\begin{equation}
H(\eta,\omega,\Delta)=\left[\begin{array}{cc} (1-\eta) H_1 & \omega W \\  \omega W & \eta H_2+\Delta\end{array}\right]
\,.
\label{CM}
\end{equation}
Here, $H_1$ and $H_2$ are operators acting in the space of normal and intruder states, respectively, while $W$ represents the interaction hamiltonian scaled by the mixing strength $\omega$.
The normal and intruder parts of the hamiltonian are driven by an external parameter $\eta$ between the $\eta=0$ and $\eta=1$ limits, in which only one of the parts is active. 
An offset $\Delta$ energetically separates intruder and normal configurations.

To determine the phase structure of the CM system, the concept of an eigenpotential \cite{Frank04} has been employed.
Hamiltonian (\ref{CM}) is evaluated in the appropriate group coherent states of the respective subspaces.
These, in the IBM case, may be associated with condensate states of $N$ and $N+2$ bosons (for $H_1$ and $H_2$, respectively).
The interaction $W$ couples both types of coherent states, so one obtains a $2\times 2$ hamiltonian matrix expressed in terms of the coherent state parameters that capture the geometry of the problem.
The lower eigenvalue of this matrix (the lower eigenpotential) represents the coherent state expectation of the ground state energy.
The minimization of this eigenvalue in the coherent state parameters yields the ground state energy as a function of the hamiltonian control parameters, which can be subject to the phase transitional analysis.

In Ref.~\cite{Frank06}, an example has been analyzed when the normal and intruder configurations are described by the U(5) and O(6) dynamical symmetries \cite{Lehmann95}.
The coherent state averages of the hamiltonian blocks in Eq.~(\ref{CM}) took forms $\ave{H_1}=\beta^2/(1+\beta^2)$, $\ave{H_2}=\tfrac{1}{4}[(1-\beta^2)/(1+\beta^2)]^2$, and $\ave{W}=\omega$ (since $W=s^{\dagger}s^{\dagger}+d^{\dagger}\cdot d^{\dagger}+{\rm h.c.}$).  
If $\omega=\Delta=0$, the hamiltonian (\ref{CM}) describes two noninteracting configurations with a pair of degenerate $\beta=0$ and $\beta>0$ minima, and the entire line $\eta\in[0,1]$ is critical.
For $\omega>0$, the phase structure becomes more complicated and strongly dependent on the offset $\Delta$.
If $\Delta$ is small, the first-order critical curve in the plane $\eta\times\omega$ is surrounded by spinodal and antispinodal lines that converge to a single curve with increasing $\omega$. 
For $\omega\to\infty$, the transition becomes of the second order; this limit is equivalent to the normal U(5)-O(6) case.
For larger values of $\Delta$, however, a region with a single spherical minimum develops near $\eta=1$ which keeps growing until it eventually extincts the phase coexistence regions.
For $\Delta>\frac{3}{4}$ the ground state eigenpotential has only one minimum, either spherical or deformed.

It should be stressed that these structures are essentially different from those obtained within the Landau approach to phase transitions.
This is because the above procedure, involving the diagonalization of the potential matrix, does not lead to the ground state energy in the form of a polynomial in the order parameter.
General consequences and other examples of this generalization are still to be investigated.
A characteristic signature of the type-II first-order phase transition seems to be a large phase coexistence region.
In the original paper \cite{Frank06}, the Sr and Zr nuclei at ${\cal N}=58$ were proposed as candidates for this type of transition. 
The above approach was recently applied by Hellemans, Van Isacker, De Baerdemacker, and Heyde \cite{Hellemans07} to the case of so called U(5)$-Q^{\chi}\cdot Q^{\chi}$ mixing in nuclei, which includes both U(5)-O(6) and U(5)-SU(3) types of phase coexistence.
Let us note that similar models can be applied also in molecular physics.

\subsubsection{External rotation (IBM cranked)}
\label{se:crank}

Interesting results have been obtained by analyzing the phase transitional properties of the system of IBM bosons in a rotating frame.
The so called cranking approach, which imposes an external rotation onto an ensemble of interacting particles, is a standard tool to describe moments of inertia and other collective aspects of rotational motions.
The cranking analysis of nuclear quadrupole shapes performed on the fermionic level by Alhassid {\it et al.} \cite{Levit84,Alhassid86,Alhassid87} in the framework of the Landau theory led to the determination of a shape-phase diagram for hot rotating nuclei.
The cranking approach within the IBM was pioneered by Schaaser and Brink \cite{Schaaser84,Schaaser86} and by Dukelsky {\it et. al.} \cite{Dukelsky83,Cambiaggio85}.
The mean field phase transitional analysis of the cranked IBM was performed by Cejnar \cite{Cejnar02,Cejnar03}.

\begin{figure}[t]
\centering
\includegraphics[width=11.1cm]{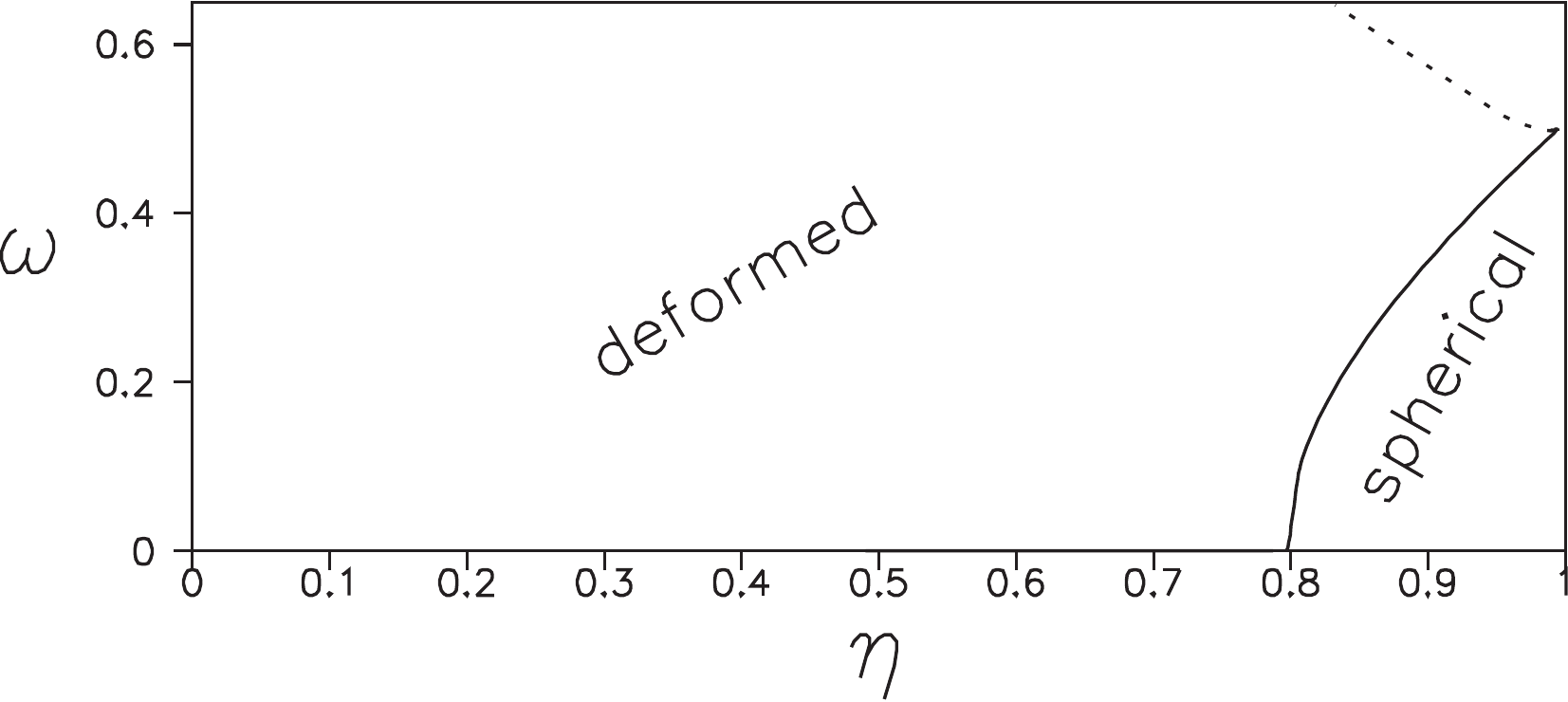}
\caption{\protect\small The phase diagram of the cranked hamiltonian (\ref{ising}) in the plane $\eta\times\omega$. Cranking frequency $\omega$ is expressed relative to $\Omega$ from the energy scaling factor $a=\hbar\Omega$. The upper frequency limit is indicated by the dashed line. Adapted from \cite{Cejnar02}.}
\label{fi:crank}
\end{figure}

The procedure is similar to that discussed in Subsecs.~\ref{se:fin} and \ref{se:geo}.
It consist of evaluating the expectation value of the cranked hamiltonian $H'=H-{\vec\omega}\cdot{\vec L}$ (where ${\vec\omega}$ is a vector of the cranking angular frequency and ${\vec L}$ the angular momentum operator) in a condensate state~(\ref{condenIBM}).
An important difference, however, concerns the form of the condensate: its coefficients cannot satisfy the usual condition $\alpha_{+m}=(-)^m\alpha^*_{-m}$, resulting from the hermicity of the Cartesian matrix $\alpha_{ij}$, since in such a case the contribution of the cranking term ${\vec\omega}\cdot{\vec L}$ vanishes.
Instead, one has to assume some modified conditions \cite{Schaaser86} that for a concrete choice of rotation along the $x$ axis can be represented by the following parametrization \cite{Cejnar03}:
\begin{equation}
\alpha_0=\beta\cos\gamma\,,\quad
\alpha_{\pm 1}=\tfrac{1}{\sqrt{2}}\beta\sin\gamma\sin\delta\,,\quad
\alpha_{\pm 2}=\tfrac{1}{\sqrt{2}}\beta\sin\gamma\cos\delta\,.
\end{equation} 
The generalized condensate states for $\omega\neq 0$ then carry angular momentum while the standard results for $\omega=0$ are recovered by setting $\delta=0$.
One has to be careful since $\beta$ and $\gamma$ do not any more carry the standard shape interpretation. 
The geometry involved in the generalized coherent state must be read out from the expectation value of the quadrupole operator in these states.
Minimization of the cranked energy functional with respect to $\beta$, $\gamma$, and $\delta$ for the simple IBM hamiltonian (\ref{ising}) 
leads to the phase diagram depicted in Fig.~\ref{fi:crank}.
Note that the finiteness of angular momentum within the IBM ($L\leq 2N$) puts an upper bound on the physical domain of the cranking frequency $\omega$. 
In the U(5) limit, e.g., the limiting frequency coincides with the boundary of the spherical phase, see Ref.~\cite{Cejnar03}.

The effect of external rotation may be related to the effect of an external magnetic field.
Indeed, Fig.~\ref{fi:crank} resembles a magnetic phase diagram of a superconductor.
The same type of phase diagram is expected to hold for superfluid and normal phases in nuclei \cite{Mottelson60,Goodman81}, but in the context of quadrupole shape phases it represents a modification of the earlier shell model based result of Ref.~\cite{Alhassid86}.
In particular, the present approach predicts a certain critical rotational frequency that induces a transition from the spherical to deformed phase (similarly as there exists a critical magnetic field for a transition from the superconducting to normal phase), which is in contrast to the previous result.
This becomes interesting in connection with the recent proposal of Regan {\it et al.} \cite{Regan03} to interpret experimental data on the yrast backbending effect in spherical nuclei as signatures of a rotation induced spherical-to-deformed shape phase transition.
A discussion of experimental yrast band data in $^{108-114}$Cd, $^{104-108}$Pd, $^{100,102}$Ru, and $^{100}$Mo within the cranked IBM framework can be found in Ref.~\cite{Cejnar04}.

\subsubsection{Other types of bosons and higher-order interactions}
\label{se:other}

Other IBM extensions, whose phase structure has so far been explicitly analyzed only to a limited extent, include in particular (a) the incorporation of bosons of other types (these can either replace, or supplement the existing $d$ boson) and/or (b) the addition of higher-order interactions (to supplement the present one- and two-body interactions).

The case (a) with the $d$ boson {\em replaced\/} by another type of boson, let us denote it as $b$, is well studied for lower dimensions.
In fact, the Lipkin model corresponds to $b\equiv t$ (a scalar or pseudoscalar boson).
Such modifications of the standard IBM-1 are experimentally relevant in molecular physics, where $b\equiv p$ (a boson with spin $l=1$, hence with 3 angular momentum projections) corresponds to the 3D vibron model \cite{FI94,IL95}, while $b\equiv\tau$ (a two-component boson with no angular momentum content) to the 2D vibron model \cite{Iachello96,Iachello03b}.
Related approaches have been also applied in condensate state physics \cite{Iachello02,Isacker07}.
The phase structure of such models in general has been considered by Cejnar and Iachello \cite{Cejnar07}.
It turns out that all two-level boson models of this type, with ${\rm U}(n)$ as a spectrum generating algebra, exhibit a second-order QPT between \lq\lq spherical\rq\rq\ and \lq\lq deformed\rq\rq\ phases, characterized by dynamical symmetries ${\rm U}(n-1)$ and ${\rm O}(n)$ (the spherical phase can be unambiguously characterized by the ground state having the form of $s$-boson condensate).
The first-order transitions, however, are a much scarcer spice.
If the angular momentum conservation is required [for models with $b$ corresponding to a boson with an even number of components we consider rotations generated by the group O(2) instead of O(3)], the occurrence of a first-order transition of the above type is limited to spectrum generating group dimensions $n=2,6,10,\dots$ (corresponding to cases when boson $b$ has even angular momenta).

Phase transitions in the 2D vibron model ($n=3$) were discussed by Pan {\it et al.} \cite{Pan05b}.
A detailed study was recently presented by P{\' e}rez-Bernal and Iachello \cite{Bernal08}.
As already mentioned in Subsec.~\ref{se:ibm2}, the latter calculations were even extended to a two-component version of the model (analogous to the IBM-2) \cite{Iachello08}.
Phase transitions in the 3D vibron model ($n=4$) were investigated by Van Roosmalen \cite{Roosmalen82} and by Arias, Dukelsky, Dusuel, Garc{\'\i}a-Ramos, and Vidal \cite{Dusuel05,Arias07}, using techniques beyond the mean field (cf. Subsec.~\ref{se:bey}).
The case $n=5$ was briefly discussed in Ref.~\cite{Cejnar07}, but so far this model has no application.
From the QPT viewpoint, all these results are essentially equivalent to those of the standard IBM ($n=6$) between U(5) and O(6) dynamical symmetries (except that the IBM yields a richer system of excited states).

The problem becomes increasingly difficult starting from the cases $b\equiv f$ and especially $b\equiv g$, i.e. bosons with angular momenta $l=3$ and 4.
For $l=4$, both first- and second-order transitions between spherical and deformed phases are supposed to take place.
To determine the phase structure seems to be complicated already on the mean field level since the energy functional depends in general on $2l+1$ variables (radius $\beta$ plus $2l$ hyperspherical angles) while the elimination of Euler angles (which is essential in the $l=2$ case) is not feasible for $l>2$.
Moreover, in nuclear physics the bosons with higher angular momenta are useful mostly as supplements of the dominant quadrupole degrees of freedom, which are carried by the system of $s$ and $d$ boson.
One should therefore deal with situations in which $p,f,g,\dots$ bosons are considered {\em together\/} with $s$ and $d$.
The phase structure then depends on a very large number of parameters and severe simplifications are necessary.
Although analytical and phenomenological studies of additional non-quadrupole degrees of freedom within the IBM exist (see e.g. the coherent state analysis in Ref.~\cite{Devi90} and the $spdf$ calculations in Ref.~\cite{Babilon05}), no comprehensive QPT analysis is available for such systems so far.

The situation is similar also for the case (b) of the above classification: the higher-order interactions.
Inclusion of such interactions within otherwise standard $sd$-IBM-1 may lead to non-trivial effects (cf. Ref.~\cite{Isacker99}).
The most general energy functional can be fabricated just from various powers of expressions $\beta^2$ and $\beta^3\cos3\gamma$ [since for $l=2$ there exist only two independent scalar couplings of shape coordinates, see Eq.~(\ref{scalcomb})].
For instance, the energy functional up to three-body terms looks as follows
\begin{equation}
V(\beta,\gamma)=\frac{A\beta^2+B\beta^3\cos 3\gamma+C\beta^4+D\beta^5\cos 3\gamma+E\beta^6\cos^2 3\gamma+F\beta^6}{(1+\beta^2)^3}
\,,
\label{V3body}
\end{equation}
where the coefficients $A,\dots,F$ depend on the concrete hamiltonian.

The form (\ref{V3body}) generates a new type of phase transition, namely the one connected with the angle variable $\gamma$, as recently discussed by Iachello \cite{Iachello03}.
To make it clear, let us rewrite the energy functional in the form $V(\beta,\gamma)=f(\beta)-g(\beta)\cos 3\gamma+h(\beta)\cos^2 3\gamma$, where the functions $f$ and $g$ can be easily read off from Eq.~(\ref{V3body}), and denote the point of global minimum as $(\beta_{\rm m},\gamma_{\rm m})$.
For $h(\beta_{\rm m})=0$ the minimum in angle variable is apparently at $\gamma_{\rm m}=0$ [for $g(\beta_{\rm m})>0$] or at $\gamma_{\rm m}=\frac{2}{3}\pi$ [for $g(\beta_{\rm m})<0$], both cases corresponding to axially symmetric (deformed) shapes.
An increase of $h(\beta_{\rm m})$ above the critical value $=\frac{1}{2}g(\beta_{\rm m})$ eventually turns $\gamma=0$ (or $\frac{2}{3}\pi$) into a saddle point, creating a new minimum $0<\gamma_{\rm m}<\frac{2}{3}\pi$.
This corresponds to a second-order phase transition from axially-symmetric to triaxial shapes \cite{Iachello03}.
Such a situation was discussed for a concrete choice of three-body IBM-1 hamiltonian by Jolos \cite{Jolos04,Jolos04b}.

A special type of a three-body hamiltonian was considered by Thiamov{\' a} and Cejnar \cite{Thiamova06}.
It was chosen to preserve the O(6) dynamical symmetry in the whole parameter space, although the underlying O(5) symmetry could be broken \cite{Isacker99}.
The three-body interaction generated by the term $[Q_0Q_0Q_0]^{(0)}$, where $Q_0$ represents the O(6) quadrupole operator, induces the occurrence of the $\beta^3\cos 3\gamma$ term in the energy functional and hence a crossover from $\gamma$-soft to $\gamma$-rigid shapes.
There is no shape-phase transition in angle variable in this model (since the term $\propto\cos^2 3\gamma$ is absent), but the phase structure is very much reminiscent of the extended Casten triangle of hamiltonian (\ref{ising}) with its three phases and a triple point in between.

Let us conclude this section by noting that the general phase structure of $k$-body IBM hamiltonians was not yet determined, and this not even for $k=3$.


\section{Excited-state quantum phase transitions}
\label{se:es}

So far we have been probing the system of interacting bosons at about zero temperature.
However, phase transitions driven by interaction parameters can also appear at finite temperatures.
In the last part of this review, we turn our attention to possible extensions of the QPT description to states with an arbitrarily high degree of excitation.
In contrast to the traditional thermodynamical treatment, we will mostly deal with individual states instead of ensembles. 
We will show that in the infinite size limit excited states throughout the whole spectrum may exhibit nonanalytic evolutions with the control parameter at a certain sequence of \lq\lq critical\rq\rq\ points. 
For this phenomenon we use the name \lq\lq excited state quantum phase transition\rq\rq\ (ESQPT) \cite{Cejnar06,Caprio08}.
Concrete investigations in this part are mostly restricted to the IBM between U(5) and O(6) dynamical symmetry limits, but we believe that continuation of work in the directions outlined here may bring new results also in more general situations. 

\subsection{Finite temperatures}
\label{se:te}

We start with some remarks on the thermodynamical extension of the ground-state QPT.
A system interacting with a heat bath at fixed temperature $T$ is naturally described \cite{Reichl} by the Helmholtz free energy $F=\ave{E}-TS$, where $\ave{E}={\rm Tr}(H\varrho)$ is the energy average and $S={\rm Tr}(-\varrho\ln\varrho)$ the entropy associated with the quantum density operator $\varrho$ (we set the Boltzmann constant to unity, so the temperature is measured in units of energy and the entropy is dimensionless).
Maximizing the entropy while keeping the energy average fixed leads to the familiar thermal population of states $p_i(T)=\frac{1}{Z(T)}\exp(-T^{-1}E_i)$, hence $\varrho=\varrho_0(T)=\tfrac{1}{Z(T)}\exp(-T^{-1}H)$, with $Z(T)=\sum_i\exp(-T^{-1}E_i)={\rm Tr}\exp(-T^{-1}H)$ being the canonical partition function.
The equilibrium value of the free energy is obtained by substituting $\varrho=\varrho_0(T)$ in the above formula, which yields
\begin{equation} 
F_0(T)=\ave{E}_T-TS_0(T)=-T\ln Z(T)
\,.
\label{Helm0}
\end{equation}
The following well-known relations can be derived for the derivatives of the equilibrium free energy with respect to temperature,
\begin{equation}
\tfrac{\partial}{\partial T}F_0(T)=-S_0(T)\,,\qquad
\tfrac{\partial^2}{\partial T^2}F_0(T)=-\tfrac{1}{T^3}\underbrace{\bigl[\ave{E^2}_T-\ave{E}_T^2\bigr]}_{T^2{\cal C}(T)}
\label{HelmdT}
\,,
\end{equation}
where ${\cal C}(T)$ is the specific heat.
Since the energy dispersion in the second equation is never negative, the equilibrium entropy $S_0(T)$ is a nondecreasing function of temperature.
The thermodynamical phase transition of $n$th order ($n=1,2,\dots$) corresponds to a situations when the value $S_0(T)$ itself (for $n=1$) or its $(n-1)$th derivative (for $n>1$) jumps discontinuously (in the thermodynamical limit) at a certain transitional temperature $T_{\rm c}$.
If the $(n-1)$th derivative ($n>1$) gets infinite at $T_{\rm c}$, the phase transition is called just continuous (with no order in the Ehrenfest sense).

At zero temperature, the equilibrium free energy coincides with the ground-state energy: $F_0(0)=\ave{E}_0=E_0$.
The ground-state QPT, expressed through the dependence of $E_0$ on control parameters, may therefore be a special case of a more general phenomenon that resides in the space of the model control parameters {\em and\/} temperature.
In infinite lattice models with finite range of interactions, the extension of QPTs observed at $T=0$ to $T>0$ is of general interest and numerous results can be found in the literature \cite{Sachdev99,Vojta03}.
In finite many-body models, results on $T>0$ phase transitions are not abundant.
This is in spite of the fact that an analytic approach to describe thermodynamical phase transitions in pseudospin models as well as in some non-spin algebraic models was formulated by Gilmore \cite{Gilmore78,Gilmore79}.

Basic steps of Gilmore procedure can be briefly outlined using the hamiltonian based on the U(2) spectrum generating algebra, i.e. the Lipkin model of Subsec.~\ref{se:lip}.
We may use the representation in terms of a spin-$\tfrac{1}{2}$ array (counting $N$ constituents) with infinite-range interactions.
First, one makes clear that in the mean field approximation (which for the present class of models becomes exact in the $N\to\infty$ limit) the thermal $N$-body density operator can be expressed as a product of 1-body density operators, i.e. in the form 
$\varrho^{(N)}=\varrho_1\otimes\varrho_2\otimes\dots\varrho_N\equiv(\varrho)^{\otimes N}$,
where $\varrho_k=\varrho$ are identical operators acting in the 2-dimensional Hilbert space of $k$th particle.
Each $\varrho_k$ can be written in the usual form $\varrho=\tfrac{1}{2}{\bf 1}+{\vec a}\cdot{\vec\sigma}$, with
${\bf 1}$ standing for the unity matrix and $(\sigma_x,\sigma_y,\sigma_z)$ for Pauli matrices. 
Polar coordinates in the decomposition $(a_x,a_y,a_z)\equiv(r\sin\theta\sin\phi,r\sin\theta\cos\phi,r\cos\theta)$ represent a convenient set of parameters to determine the state $\varrho^{(N)}$.
The minimization of the free energy $F^{(N)}={\rm Tr}[H\varrho^{(N)}]-T{\rm Tr}[-\varrho^{(N)}\ln\varrho^{(N)}]$ with respect to these parameters yields the equilibrium value $F^{(N)}_0$ for each set of external parameters and temperature.
The free energy for large values of $N$ can be written as $F^{(N)}=N f(r,\theta,\phi)$.
For $T=0$ the minimization gives $r=\tfrac{1}{2}$; then $\varrho^{(N)}$ represents a pure state and the method reduces to the condensate state procedure used for the ground state. 
For $T>0$, however, one gets $r<\tfrac{1}{2}$.
Phase transitions (if any) can be detected as nonanalytic evolutions of $F^{(N)}_0=N f_0$ with the model control parameters.
In this way, a finite-$T$ counterpart of the ground-state phase transition in the Lipkin model was detected \cite{Gilmore78,Gilmore81}.

With a hamiltonian satisfying the linear ansatz (\ref{Hlin}) the equilibrium free energy $F_0$ depends just on two parameters: $T$ and $\eta$.
As may be anticipated from the Gilmore method, for finite models the description of phase transitions in the $\eta\times T$ plane will not be essentially different from the Landau theory of thermodynamical phase transitions in the $p\times T$ plane, where $p$ stands for another intensive thermodynamical variable like pressure.
One therefore expects that phase transitions will form some curves (or parts of curves) in $\eta\times T$.
Consider a point $P$ on such a curve and at this point mark the normal and tangent directions $n$ and $t$, respectively.
For the classification of the phase transition at $P$, discontinuous partial derivatives of $F_0$ along the normal direction $n$ are substantial (partial derivatives along $t$ are continuous).
However, since partial derivatives along an arbitrary direction can be expressed through partial derivatives of the respective order along $n$ and $t$, the same classification is likely to be found if probing the transition along any of the two parameters $\eta$ and $T$. 

Let us explicitly evaluate the first two partial derivatives of $F_0$ with respect to $\eta$ and $T$.
The derivatives $\frac{\partial}{\partial T}F_0$ and $\frac{\partial^2}{\partial T^2}F_0$ at constant $\eta$ obviously yield Eq.~(\ref{HelmdT}).
For the derivatives containing $\eta$ one obtains,
\begin{equation}
\tfrac{\partial}{\partial\eta}F_0(\eta,T)=\ave{V}_T
\,,\quad
\tfrac{\partial^2}{\partial\eta^2}F_0(\eta,T)=
-\tfrac{1}{T}\left\langle\left[V-\ave{V}_T\right]^2\right\rangle_T+\left\langle\tfrac{d^2}{d\eta^2}E_i\right\rangle_T
\,,\label{Helmde}
\end{equation}
where $\ave{X}_T=\sum_ip_i(T)X_i$ with $X_i=\matr{\Psi_i}{X}{\Psi_i}$ represents the thermal average of a quantity $X$.
An analogous expression can be derived for $\frac{\partial^2}{\partial T\partial\eta}F_0(\eta,T)$.
Note that the $T=0$ limit of these equations coincides with Eq.~(\ref{derivatives}), discussed above in connection with the ground state.

The first equations of both pairs (\ref{HelmdT}) and (\ref{Helmde}) show that a jump of entropy (in the vertical, i.e. temperature direction) is likely to be connected with a jump of the interaction average $\ave{V}_T$ (in the horizontal, i.e. interaction parameter direction).
These are equivalent expressions of a generic first-order phase transition at such places in $\eta\times T$ where the phase separating curve is not parallel with one of the axes.
Similarly, as seen from the second equations in both pairs (\ref{HelmdT}) and (\ref{Helmde}), a nonanalyticity of the specific heat is expected to have an equivalent counterpart in the expression of the thermal dispersion of $V$ and/or the thermal average of the level curvatures $\frac{d^2}{d\eta^2}E_i$.

In view of the above discussion, one may tend to anticipate that any QPT at $\eta=\eta_{\rm c}$ and $T=0$ smoothly continues to the $T>0$ domain, where it can be detected as a thermodynamical phase transition at $T=T_{\rm c}(\eta)$.
Such a statement is the content of a so called crossover theorem, formulated by Gilmore for the specific systems he analyzed \cite{Gilmore78,Gilmore81}.
It should be stressed, however, that this rule cannot be applied in general.
The simplest counterexample arises if symmetry constrains disable repulsion of some levels and the ground state undergoes an {\em unavoided\/} crossing.
Then a $T=0$ phase transition (typically a first-order one) takes place at the crossing point, with no continuation to $T>0$.

To illustrate this mechanism on a trivial example, let us take a specific IBM hamiltonian with the U(5) dynamical symmetry:
\begin{equation}
H(\eta)=(1-\eta)n_d+\eta n_s=\eta N+(1-2\eta)n_d
\quad\Rightarrow\quad
V(\eta;\beta)=\eta+(1-2\eta)\frac{\beta^2}{1+\beta^2}
\,.
\label{transfiN}
\end{equation}
Here, $n_s$ and $n_d$ stand for the numbers of $s$ and $d$ bosons, and $V$ represents the classical potential derived from the hamiltonian.
The eigenstates of (\ref{transfiN}) for any value of the control parameter $\eta$ can be classified by the U(5) quantum numbers and the spectrum becomes fully degenerate at $\eta=\eta_{\rm c}=\tfrac{1}{2}$.
For $\eta<\eta_{\rm c}$ the energy of the ground state, which is an $s$-boson condensate, linearly increases with $\eta$.
At $\eta=\eta_{\rm c}$ the ground state flips ito the $d$-boson condensate and its energy starts to linearly decrease.
From the viewpoint of the above nomenclature, such a situation would be classified as the first-order ground-state quantum phase transition; this being so for an arbitrary (even finite) number of bosons $N$.
However, since all levels vary with the control parameter in an analytic (linear) way, the introduction of a nonzero temperature makes the free energy an {\em analytic\/} function of both $\eta$ and $T$.
The phase-transitional behavior detected at $T=0$ disappears for an arbitrarily small temperature $T>0$.
Other examples of finite-$N$ ground-state crossings on the O(6)-U(5) side of the Casten triangle can be found in Refs.~\cite{Arias03b,Heinze06}.

In spite of its success in pseudospin models, the approach based on canonical thermodynamics has some drawbacks in general finite systems.
In particular, we know that quantal spectra of such systems (for finite numbers of particles) have finite numbers of states, so the level density in the upper part of the spectrum is a decreasing function of energy.
Under such circumstances, the introduction of temperature can work as in usual thermodynamical systems only at some low temperatures. 
For higher temperatures, the thermal averaging smoothenes all properties over the whole spectrum and disables a direct access to states with higher energies. 
(The only way to bypass this problem would be to consider negative absolute temperatures as in \cite{Cejnar00}.)
For this reason, it is usually more convenient to use directly the excitation energy instead of temperature.
Quantum phase transitions for excited states then show up as singularities of the level density and also as nonanalyticities in the evolution of individual level energies and wave functions. 
Since the logarithm of the level density is proportional to the microcanonical entropy, the present approach is related to the microcanonical thermodynamics, see e.g. Ref.~\cite{Gross01}.
As canonical and microcanonical results should converge in the proper thermodynamical limit, the present approach does not imply any loss of thermodynamical information.

\subsection{Excited level dynamics}
\label{se:dy}

It is clear that evolution of the spectrum of the general linear hamiltonian (\ref{Hlin}) with parameter $\eta$ is fully determined by a \lq\lq snapshot\rq\rq\ containing all energies $E_i\equiv E_i(\eta)$ (with $i=0,\dots,n-1$) and all matrix elements $V_{ij}\equiv\matr{\psi_i(\eta)}{V}{\psi_j(\eta)}$ at an arbitrary {\em single\/} value of $\eta$.
This viewpoint was elaborated by Pechukas and Yukawa \cite{Pechukas83,Yukawa85}, who considered the level dynamics induced by hamiltonian (\ref{Hlin}) with variable $\eta$ as a classical problem of determination one-dimensional motions of $n$ interacting particles.
In this approach, parameter $\eta$ plays the role of time, while $E_i$ are positions and $\frac{d}{d\eta}E_i\equiv{\dot E_i}$ velocities of individual particles.
The dynamical equations
\begin{equation}
\tfrac{d}{d\eta}E_i=V_{ii}
\,,\quad
\tfrac{d^2}{d\eta^2}E_i=2\sum_{j(\neq i)}\frac{|V_{ij}|^2}{E_i-E_j}
\,,\quad
\tfrac{d}{d\eta}V_{ij}=\sum_{k(\neq i)}\frac{V_{ik}V_{kj}}{E_i-E_k}+\sum_{k(\neq j)}\frac{V_{ik}V_{kj}}{E_j-E_k}
\label{PY}
\end{equation}
(which immediately result from ordinary perturbation theory) are analogous to those describing dynamics of a gas of particles interacting through two-dimensional Coulomb force $F_{ij}\propto q_iq_j/(x_i-x_j)$.
The essential difference, however, is the fact that the product of charges $q_iq_j$ is replaced by $|V_{ij}|^2$, which cannot be factorized and is also subject to variation due to the third equation.
The knowledge of $E_i$, ${\dot E_i}$ and $V_{ij}$ at some $\eta=\eta_{\rm ini}$ allows the determination of level dynamics for any $\eta$.

From the computational viewpoint, there is certainly no advantage in using the system of coupled differential equations (\ref{PY}) instead of diagonalizing the hamiltonian (\ref{Hlin}) for each particular value of $\eta$.
However, Pechukas-Yukawa theory is fruitful in gaining deeper insight into the mechanisms leading to QPT behaviors.
It shows that whether a given model will exhibit a QPT or not is entirely determined by the \lq\lq initial conditions\rq\rq, i.e. by values of $E_i$, ${\dot E_i}$ and $V_{ij}$ at an arbitrary $\eta_{\rm ini}$.
In case when the two phases of the model coincide with some dynamical symmetries, the initial point is naturally put to one of these symmetries, which is therefore made responsible for the phase-transitional behavior at some $\eta_{\rm c}\neq\eta_{\rm ini}$.
Moreover, the evolution of quantum energies and interaction matrix elements with $\eta$ is pictured as an intuitively imaginable process {\it {\` a} la\/} classical mechanics.
This makes it easier to guess which types of initial conditions are \lq\lq QPT friendly\rq\rq\ and which are not.
In particular, a plausible condition for a QPT to occur is a fast initial compression of the whole spectrum, which leads to quickly increasing level repulsion and may eventually produce a sharp scattering of levels.
In the following we show that this type of mechanism indeed applies in the IBM case.

\begin{figure}[t]
\centering
\includegraphics[width=13cm]{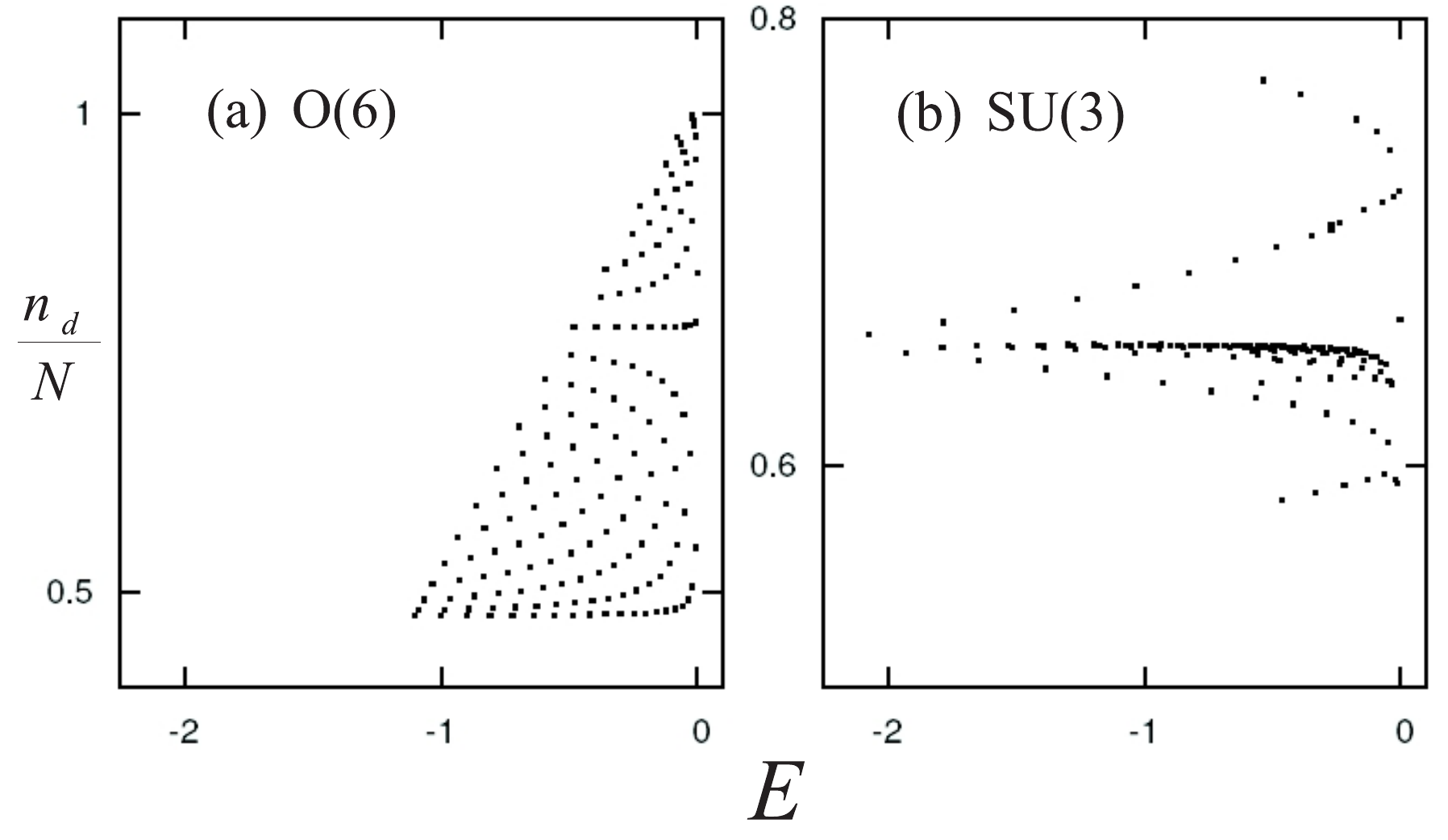}
\caption{\protect{\small The grid of IBM-1 states with $L=0$ in the plane $n_{\rm d}$ versus energy per boson for $N=40$. Panels (a) and (b) correspond to O(6) and SU(3) limits, respectively.}}
\label{fi:gri}
\end{figure}

Let us have a closer look on the initial conditions for the IBM simplified hamiltonian (\ref{ising}).
A natural initial point is $\eta_{\rm ini}=0$, which coincides with the O(6) dynamical symmetry for $\chi=0$, with the SU(3) or ${\overline{\rm SU(3)}}$ dynamical symmetries for $\chi=\pm\tfrac{\sqrt{7}}{2}$, or is intermediate between these cases for other values of $\chi\in[-\tfrac{\sqrt{7}}{2},+\tfrac{\sqrt{7}}{2}]$.
As noticed in Ref.~\cite{Heinze06}, a large fraction of the spectrum with $L=0$ in the O(6) and SU(3), or ${\overline{\rm SU(3)}}$, limits exhibits a fast initial compression, having almost a sharp point of focus on the U(5) side of the transition. 
This is true on the whole $\eta=0$ side of the extended Casten triangle.

From the first equation in (\ref{PY}) we can easily derive the condition for a subset ${\cal S}$ of eigenstates of hamiltonian (\ref{Hlin}) at $\eta=\eta_{\rm ini}$ pointing to a single focus $(\eta_{\rm foc},E_{\rm foc})$; it reads as 
\begin{equation}
\matr{\psi_i(\eta_{\rm ini})}{H(\eta_{\rm foc})}{\psi_i(\eta_{\rm ini})}=E_{\rm foc} 
\quad {\rm for\ }i\in{\cal S}
\,.
\label{focus}
\end{equation}
In our case, with $\eta_{\rm ini}=0$ and the focal point at $(\eta_{\rm foc},E_{\rm foc})=(1,f)$, the condition (\ref{focus}) can be written as $\matr{\psi_i(0)}{n_d}{\psi_i(0)}=f\,N$.
The value of $\ave{n_d}_i$ is easy to calculate for the ground state with large $N$, using the coherent state relation $\ave{n_d}_0=\beta_{\rm m}^2/(1+\beta_{\rm m}^2)$.
In the O(6) limit we have $\ave{n_d}_0=\frac{1}{2}N$, while in the SU(3) and ${\overline{\rm SU(3)}}$ limits $\ave{n_d}_0=\frac{2}{3}N$.
In the O(6) limit, the condition $\ave{n_d}_i\approx\frac{1}{2}N$ is valid for a great majority of excited states with $\tau=0$ and $L=0$, but it is increasingly violated for growing values of seniority and angular momentum, see Fig.~\ref{fi:gri}(a).
In the SU(3) and ${\overline{\rm SU(3)}}$ limits, on the other hand, the condition $\ave{n_d}_i\approx\frac{2}{3}N$ holds for a large portion of the whole $L=0$ spectrum, see Fig.~\ref{fi:gri}(b).
This means that the initial compression in the latter case is faster and involves more states, which is consistent with the fact that the subsequent QPT is of the first order, i.e. sharper that the one observed in the O(6)-U(5) transition.

\begin{figure}[tp]
\centering
\includegraphics[width=14.8cm]{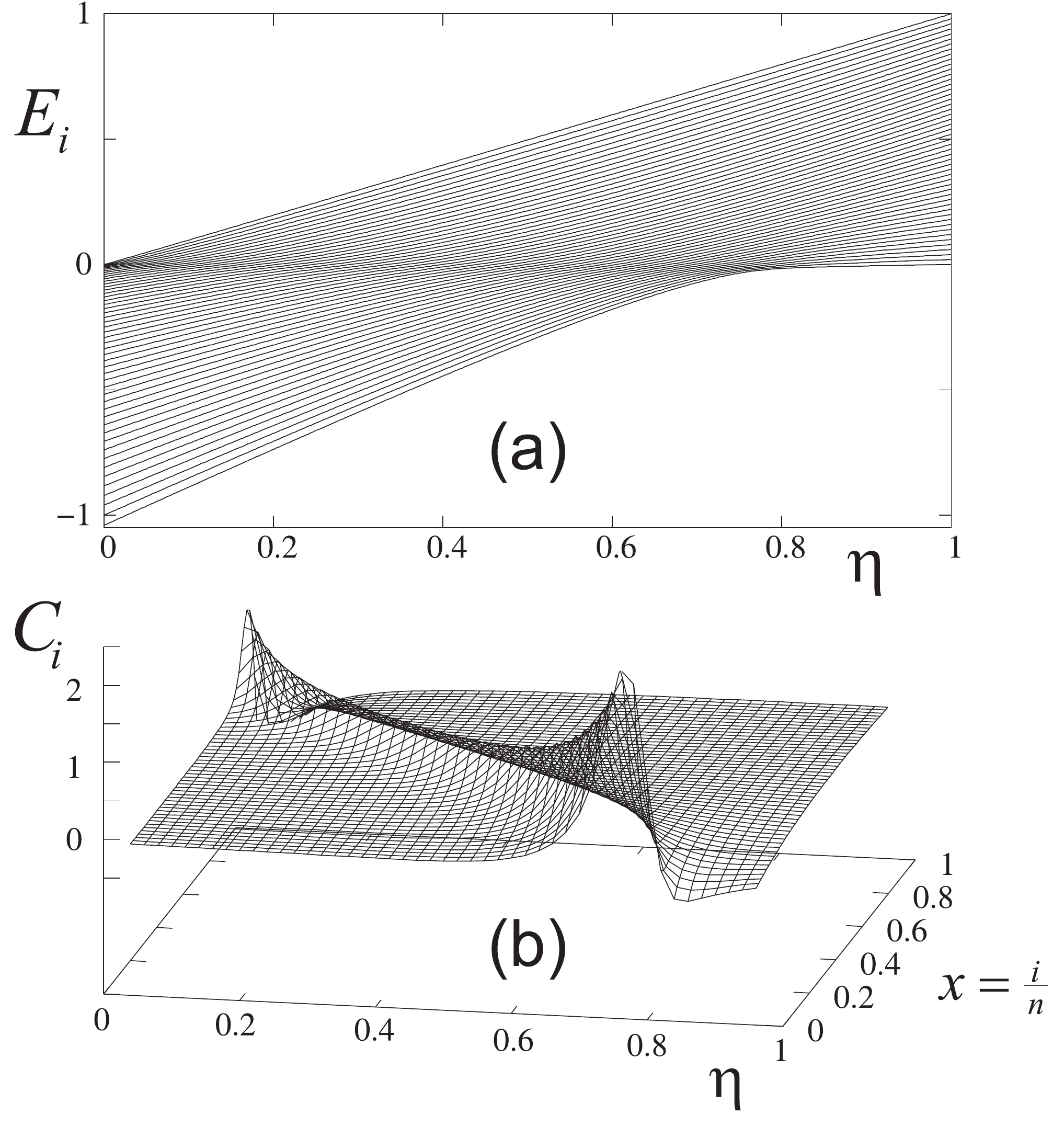}
\caption{\protect{\small Panel (a) shows a \lq\lq shock wave\rq\rq\ formed in the $N=100$ spectrum of IBM states with $L=\tau=0$ along the transition from O(6) to U(5), described in the parametrization (\ref{ising}). The energy is expressed per boson. Panel (b) shows the \lq\lq specific heat\rq\rq\ from Eq.~(\ref{Cbr}) evaluated across the spectrum, with $x$ denoting the relative excitation ratio ($x=0$ for the ground state and 1 for the highest state).}}
\label{fi:shock}
\end{figure}

An interesting type of solution of the system in Eq.~(\ref{PY}) which may apply under certain initial conditions is a formation of a wave propagating through the spectrum.
A spectacular example of such a phenomenon is shown in panel (a) of Fig.~\ref{fi:shock}.
It displays the spectrum of the IBM states with zero spin and seniority between the O(6) and U(5) dynamical symmetries [hamiltonian (\ref{ising}) with $\chi=0$].
The \lq\lq shock wave\rq\rq, which is initiated in the uppermost part of the spectrum at the O(6) side of the transition, keeps moving due to the initial compression of the spectrum and eventually reaches the ground state at the QPT critical point $\eta_{\rm c}=\tfrac{4}{5}$.
This effect was for the first time investigated by Heinze, Cejnar, Jolie, and Macek in Ref.~\cite{Heinze06}.
Examples from a broader class of models of the same type have been subsequently analyzed in detail by Caprio, Cejnar, and Iachello \cite{Caprio08}.

As discussed in Subsec.~\ref{se:se}, the curvatures of individual level trajectories as they pass through the wave in Fig.~\ref{fi:shock}(a), i.e. the deceleration and acceleration due to the repulsive force in (\ref{PY}), becomes infinite in the asymptotic size regime.
Therefore, the wave can be viewed as a manifestation of quantum phase transitions for individual excited states (ESQPT) at a running value of the control parameter.
This interpretation is supported by the analysis of the behavior of branch points in complex extended control parameter (cf. Subsec.~\ref{se:ther}).
Indeed, the branch points corresponding to individual avoided crossings of levels along the wave propagation get closer to the real $\eta$ axis as the number of bosons increases.  
Panel (b) of Fig.~\ref{fi:shock} shows the \lq\lq specific heat\rq\rq\ from Eq.~(\ref{Cbr}), which measures the proximity of branch points on the $i$th level Riemann sheet to the real axis for a given $N$. 
The moving peak coincides with the trajectory of the wave in panel (a).
The analysis by Cejnar, Heinze, and Macek \cite{Cejnar07b} showed that the behavior of the peak for very large $N$ is consistent with the assumption of a continuous ESQPT (the \lq\lq latent heat\rq\rq\ $Q$ is zero).
This conclusion is confirmed by semiclassical arguments \cite{Cejnar06,Caprio08} that will be outlined in Subsec.~\ref{se:se}.
Thermodynamical consequences may be anticipated from Eq.~(\ref{Helmde}) and are probably in agreement with the early analysis by Gilmore \cite{Gilmore78} (cf. Subsec.~\ref{se:te}).

The transition from SU(3) or ${\overline{\rm SU(3)}}$ to U(5) does not seem to show any ESQPT effect, at least not at such degree of coherence as in the previous case.
The exceptional behavior of the [O(6)-U(5)]$\supset$O(5) transition is probably connected with the fact that the corresponding hamiltonian is integrable but, at the same, creates repulsion of levels with the same quantum numbers $L$ and $\tau$.
Note that the level repulsion is not a generic feature of integrable systems since the underlying integrals of motions [unlike the O(5) Casimir invariant in the present case] usually depend on the control parameter \cite{Yuzbashyan02}.
The \lq\lq laminar flow\rq\rq\ of levels (due to integrability) and the level repulsion [due to persistent O(5) symmetry] constitute ideal conditions for the formation of a coherent wave.
In nonintegrable transitional regimes, the chaotic (\lq\lq turbulent\rq\rq) flow of levels (with numerous collisions of neighboring levels) obscures the observation of any collective aspect in level dynamics.
So far, the existence ESQPTs was not systematically investigated in these regimes.

\subsection{Monodromy}
\label{se:mo}

It turns out that the sequence of excited state quantum phase transitions along the O(6)-U(5) path is closely related to so called monodromy. 
In classical mechanics, monodromy is a topological property of some integrable systems that usually indicates the presence of an anomalous (pinched) torus of trajectories in the phase space \cite{Cushman97}.
This anomaly prevents a full analytic description of the system (in spite of its integrability) and creates a dichotomic distinction for the types of classical orbits. 
The pinched torus shows up as a point defect in the lattice of quantum states, in which the states are organized with respect to their quantum numbers.
If a closed loop around the defect is followed, the form of an elementary cell in the lattice varies such that after the return to the initial point the transformed cell does not fit with the one taken at start. 
Hence the term monodromy, i.e. \lq\lq once around\rq\rq.

\begin{figure}[tp]
\centering
\includegraphics[width=18.5cm]{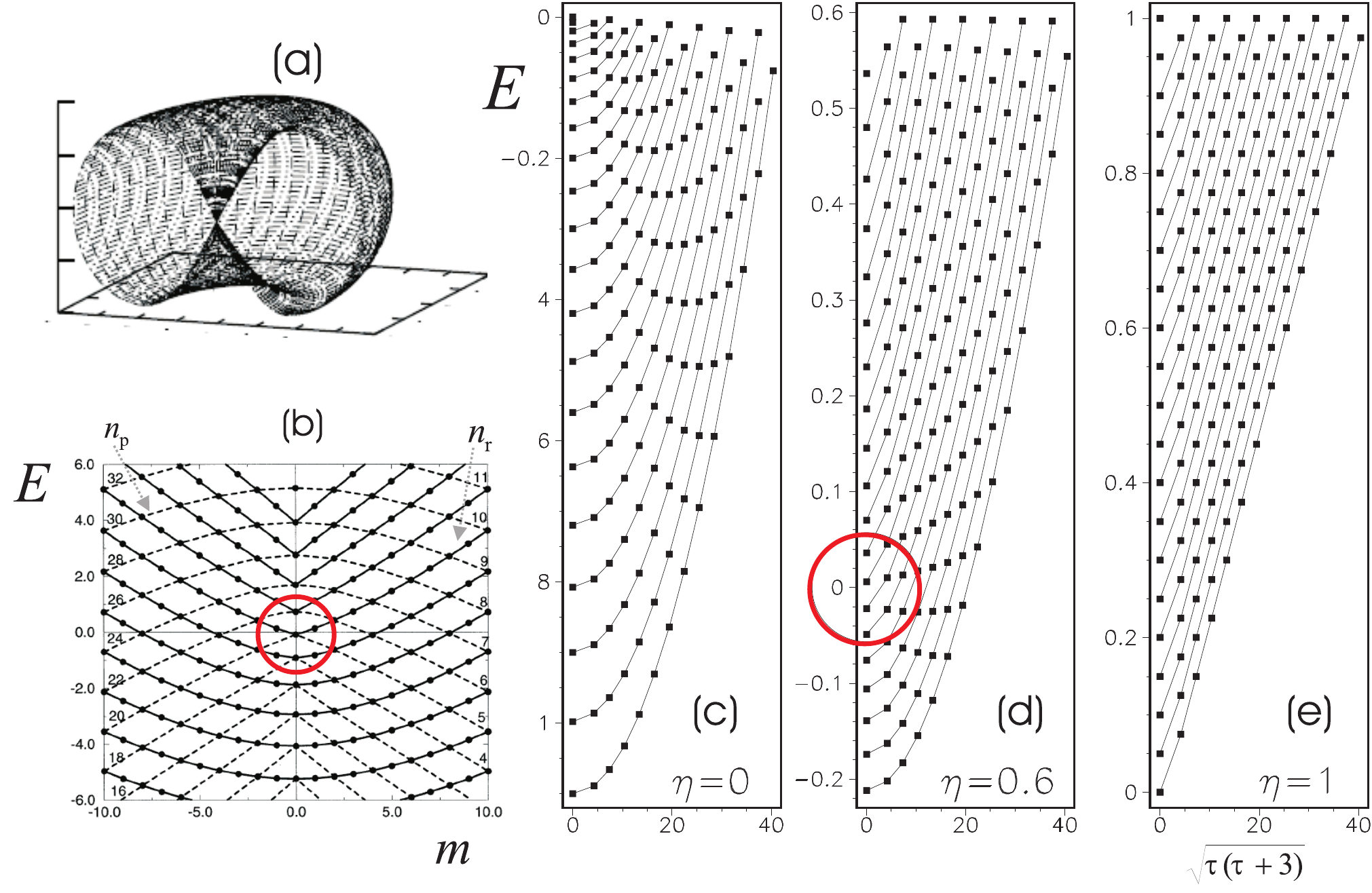}
\caption{\protect{\small Panel (a): a schematic view of the pinched torus in the phase space $x\times y\times p_r$ of the sombrero potential. Panel (b): the lattice of quantum states in the sombrero potential (adapted from Child~\cite{Child98}). Full (broken) lines join points with the same radial (principal) quantum numbers; the circle indicates a crystal defect associated with the monodromy. Panels (c)--(e): quantum lattices of IBM states with $L=0$ for the transition between the O(6) ($\eta=0$) and U(5) ($\eta=1$) limits for $N=40$ (adapted from Ref.~\cite{Macek06}).}}
\label{fi:mon}
\end{figure}

The simplest systems that exhibit the monodromy are (a) the spherical pendulum \cite{Cushman97,Efstathiou04} and (b) particle moving in the sombrero potential \cite{Child98}.
In both these cases, the pinched torus---see Fig.~\ref{fi:mon}(a)---is formed by orbits passing the points of unstable equilibrium (at the north pole of the pendulum sphere or at the top of the sombrero potential) with just the critical energy $E_{\rm c}$ at which to reach that point takes infinite time.
Since this happens at zero angular momentum $J_z$, the corresponding motions can be reduced to an effectively two-dimensional phase space, whose intersection with the pinched torus defines a classical {\em separatrix}.
The dynamics on both sides of the separatrix shows qualitatively different features: while orbits below the critical energy (inner side of the separatrix) are essentially vibrations around the stable equilibrium, those above the critical energy (outer side of the separatrix) extend over a larger part of the phase space.
In the quantum case, if all angular momenta $J_z=m\hbar$ are considered, the states below and above the critical energy form different types of patterns in the plane $E\times J_z$, as can be seen---for the sombrero potential---in Fig.~\ref{fi:mon}(b).


Monodromy in the two-dimensional pendulum and sombrero systems affects bending vibrations of some triatomic molecules, such as H$_2$O \cite{Child98}, where the concept was also confirmed on the experimental level \cite{Cushman04}.
The algebraic description of these molecules is possible within the 2D vibron model, which is a two-level bosonic model with a scalar boson and a two-component boson \cite{Iachello96,Iachello03b,Bernal08}, cf. Subsec.~\ref{se:other}.
A detailed study (including the analysis of excited state quantum phase transitions, monodromy, scaling laws and thermodynamical quantities) was recently presented by Perez-Bernal and Iachello \cite{Bernal08}.
The phase transitional properties of the 2D vibron and related models are very close to those of the $sd$-IBM along its O(6)-U(5) transition.
It turns out that in absence of O(3) rotations ($L=0$) the seniority quantum number $\tau$ can be regarded as an analog of the O(2) angular momentum quantum number $m$ considered in the above 2D examples.

Recall that if the IBM hamiltonian (\ref{3_ibmhamc}) contains no admixture of the SU(3) invariant, the cubic term in the classical potential (\ref{Vfunc}) is missing: $B=0$.
In this case, the critical point at $A=0$ separates the sombrero and oscillator types of potential (located on the deformed and spherical sides, respectively).
The full classical hamiltonian in the parametrization (\ref{ising}) with $\chi=0$ is given by
\begin{equation}
{\cal H}_{\rm cl}(\eta)=T_{\rm rot}+
\underbrace{\left[\tfrac{\eta}{4}+\left(1-\tfrac{3}{4}\eta\right)\beta^2\right]\left[(1+\beta^2)^2p_{\beta}^2+\left(\tfrac{p_{\gamma}}{\beta}\right)^2\right]}_{T_{\rm vib}}+
\underbrace{\frac{(5\eta-4){\beta}^2+\eta{\beta}^4}{(1+{\beta}^2)^2}}_{V}
\,,
\label{clasa}
\end{equation}
where $T_{\rm rot}$ and $T_{\rm vib}$ stand for the rotational and vibrational kinetic energies; $p_{\beta}\in[0,\infty]$ and $p_{\gamma}\in[0,1]$ are momenta associated with the shape coordinates $\beta$ and $\gamma$, respectively.
The radius $\beta\in[0,\infty]$ can be compressed to ${\tilde\beta}\in[0,\sqrt{2}]$ through the relation $\beta={\tilde\beta}/\sqrt{2-{\tilde\beta}^2}$ \cite{Klein81}, while the radial momentum is transformed accordingly:
$p_{\beta}\mapsto{\tilde p}_{\beta}\in[0,\sqrt{2}]$.
Eq.~(\ref{clasa}) then takes the form
\begin{equation}
{\tilde{\cal H}}_{\rm cl}(\eta)={\tilde T}_{\rm rot}+
\left[\tfrac{\eta}{2}+(1-\eta){\tilde\beta}^2\right]\left[{\tilde p}_{\beta}^2+\left(\tfrac{p_{\gamma}}{{\tilde\beta}}\right)^2\right]+
(\tfrac{5}{2}\eta-2)\,{\tilde\beta}^2+(1-\eta)\,{\tilde\beta}^4
\,,
\label{clasb}
\end{equation}
which is more convenient in the present case since the transformed potential is no longer distorted by the $\beta$-dependent denominator.
If considering motions with $L=0$, the classical phase space is generated solely by the vibrational degrees of freedom, described either by polar coordinates ${\tilde\beta}$ and $\gamma$, or by the corresponding Cartesian coordinates $x={\tilde\beta}\cos\gamma$ and $y={\tilde\beta}\sin\gamma$.
The angular momentum $p_{\gamma}=xp_y-yp_x$ is then related to the classical limit of the O(5) Casimir invariant for $L=0$ through $\frac{1}{N^2}C_2[{\rm O(5)}]_{\rm cl}=p_{\gamma}^2$.

As shown by Cejnar, Macek, Jolie, Heinze, and Dobe{\v s} \cite{Macek06,Cejnar06}, the sombrero potential from Eq.~(\ref{clasb}) for $\eta<\eta_{\rm c}=\frac{4}{5}$ generates both classical and quantum signatures of monodromy.
It was recognized that the energy $E_{\rm c}=0$ corresponding to the pinched torus separates two different dynamical regimes that can be linked, on both quantum and classical levels, to the respective limiting dynamical symmetries O(6) and U(5).
In particular, the following observations were made: 
(a) Arrangements of states in the $E\times \tau$ quantum lattice below and above $E_{\rm c}$ resemble those associated with dynamical symmetry limits O(6) and U(5), respectively \cite{Macek06}. In the O(6) type of lattice states form well distinguished chains with an approximately quadratic dependence on $\tau$, while in the U(5) type of lattice the dependence on $\tau$ is essentially linear, seen Fig.~\ref{fi:mon}(c)-(e).
(b) The form of wave functions changes from O(6)-like to U(5)-like at $E_{\rm c}$ \cite{Caprio08}. The crossover is most clearly seen for $\tau=0$ states. Below $E_{\rm c}$ the average $\ave{n_d}_i$ gradually decreases with energy and the spread of the $n_d$ distribution increases, while above $E_{\rm c}$ these trends are reverted.
(c) Classical trajectories change their typical forms at $E_{\rm c}$ \cite{Macek06}. This can shown by expressing the ratio $R_{\gamma/\beta}=T_{\gamma}/T_{\beta}$ of periods associated with $\gamma$ and $\beta$ degrees of freedom. Trajectories below $E_{\rm c}$ exhibit a broad distribution of $R_{\gamma/\beta}$ with the lower bound at $\approx 3$ and a strong preference for large values. This corresponds to flower-like orbits in the sombrero potential. The trajectories above $E_{\rm c}$ show a narrow distribution concentrated at values $<3$, which with an increasing energy converges to the lower limit $R_{\gamma/\beta}=2$. This corresponds to the bouncing-ball orbits traversing across the central maximum.

\subsection{Semiclassical considerations}
\label{se:se}

The relation of monodromy to nonanalytic evolution of individual $\tau=0$ excited states with the control parameter was for the first time noticed \cite{Cejnar06} in application of the \lq\lq displaced oscillator\rq\rq\ approximation of Rowe \cite{Rowe04,Rowe04b} to hamiltonian (\ref{ising}).
The approximation is valid on the O(6) side of the transitional path for asymptotic boson numbers, when $x=2\tfrac{n_d}{N}-1$ can be treated as a continuous variable.
Eigenstates of the hamiltonian are expressed as conventional wave functions $\psi_i(x)=\scal{n_d}{\psi_i}$ and the hamiltonian itself turns into a differential operator. 
This is due to the fact that two-body interactions conserving the O(5) invariant connect states only with $\Delta n_d=0,\pm 2$, which is an infinitesimal change of $x$.
For the simplified hamiltonian (\ref{ising}) with $\chi=0$ this procedure yields the following result:
\begin{equation}
H_{\rm osc}(\eta)=-\tfrac{4}{N^2}(1-\eta)\tfrac{d}{dx}(1-x^2)\tfrac{d}{dx}
+\underbrace{(1-\eta)}_{A(\eta)}\biggl[x-\underbrace{\tfrac{\eta}{4(\eta-1)}}_{x_0(\eta)}\biggr]^2
+\underbrace{\tfrac{(5\eta-4)^2}{16(\eta-1)}}_{E_0(\eta)}
\,.
\label{sha}
\end{equation}
The above hamiltonian is hermitian and can be regarded as a possible quantization of a shifted oscillator with a specific $x$-dependent mass parameter.

\begin{figure}[tp]
\centering
\includegraphics[width=11.1cm]{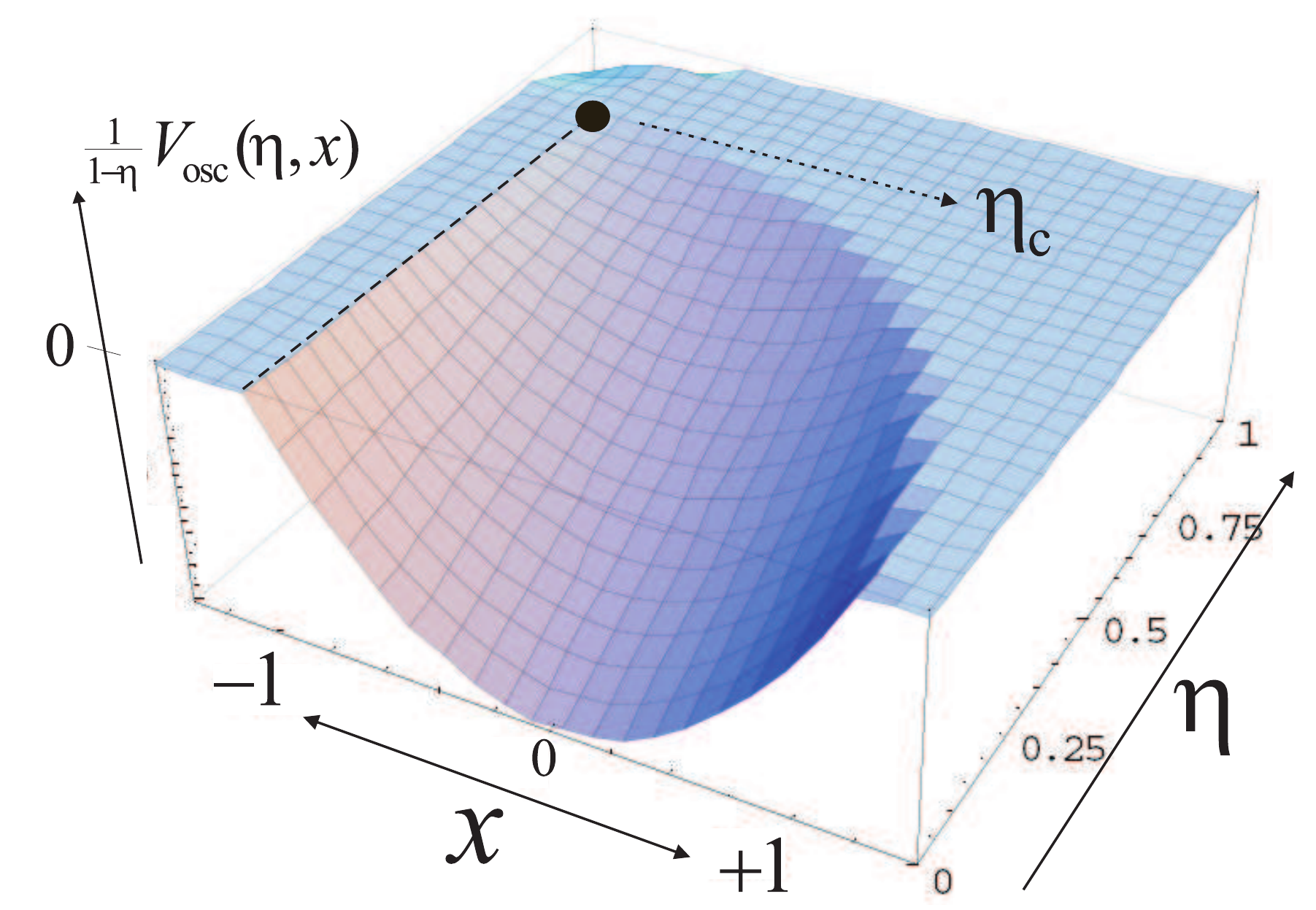}
\caption{\protect{\small The oscillator potential from Eq.~(\ref{sha}) scaled by $(1-\eta)^{-1}$. The physical domain is limited by the upper energy $E_{\rm c}=0$ where the potential reaches $x=-1$.}}
\label{fi:osc}
\end{figure}

The oscillator potential from Eq.~(\ref{sha}) is shown in Fig.~\ref{fi:osc}.
The energy $E_0$ and position $x_0$ of the oscillator minimum depend smoothly on the control parameter, as well as the width $A^{-1/2}$.
The departure from the O(6) limit ($\eta=0$) is therefore expressed just as a smooth shift and rescaling of wave functions and a gradual change of the corresponding energies.
However, the solutions loose sense if the energy exceeds a certain limit $E_{\rm c}$, where the semiclassical wave function overflows the physical domain $x\in[-1,+1]$.
Fig.~\ref{fi:osc} shows that $E_{\rm c}=0$ (independent of $\eta$), i.e. it coincides with the top of the central maximum of the sombrero potential in the present parametrization.
At this point, the solution becomes singular since the mass diverges and the analytic continuation from the O(6) limit cannot be maintained any more.
Therefore, the energy $E_{\rm c}=0$ is the upper limit where for each given level with $\tau=0$ the concept of quasi-O(6) dynamical symmetry can be applied.
Above this energy, the level converts to the judicature of another type of analytic continuation that can be named quasi-U(5) (cf. Subsec.~\ref{se:sy}).
As seen in Fig.~\ref{fi:osc}, the quasi-O(6) energy region (below $E_{\rm c}$) shrinks with increasing $\eta$ and disappears at $\eta_{\rm c}$, where the ground state reaches the critical energy.

In fact, the same conclusions can be obtained also directly from the classical hamiltonian (\ref{clasb}).
For ${\tilde T}_{\rm rot}=0$ and the relative seniority $\tfrac{\tau}{N}\to 0$ (hence $p_{\gamma}=0$), the sombrero potential creates a phase space separatrix at total absolute energy $E_{\rm c}=0$.
The action along a periodic orbit at energy $E$ can be expressed as
\begin{equation}
S(\eta,E)=2\int_{\beta_0(\eta,E)}^{\beta_1(\eta,E)}\sqrt{2M(\eta;\beta)[E-{\tilde V}(\eta;\beta)]}\,d\beta
\,,
\label{act}
\end{equation}
where $M(\eta;\beta)$ is a position-dependent \lq\lq mass\rq\rq\ that can be extracted from Eq.~(\ref{clasb}), while $\beta_0$ and $\beta_1$ represent classical turning points of the radial motion.
It can be shown that the action as a function of energy has a singular tangent at $E=E_{\rm c}$, when the separatrix is crossed.
At this critical energy, the inner turning point $\beta_0(\eta,E)$, which for $E<E_{\rm c}$ is a decreasing function of energy, reaches the value $\beta_0(\eta,E)=0$ and gets fixed \cite{Cejnar06}.

As discussed by Caprio, Cejnar, and Iachello \cite{Caprio08}, the singularity at $E=0$ has two crucial consequences: (a) it implies an infinite local growth of the $\tfrac{\tau}{N}=0$ level density, and (b) creates a singularity in the dependence of individual quantum energies on $\eta$.
Both these conclusions can be obtained from the Bohr-Sommerfeld quantization condition 
\begin{equation}
S(\eta,E_i)=2\pi\hbar\left(i+\tfrac{1}{2}\right)={\rm const}
\label{ebk}
\,,
\end{equation}
where $\hbar\propto N^{-1}$.
The gap between adjacent levels is given by $\Delta\propto\tfrac{1}{N}(\tfrac{\partial}{\partial E}S)^{-1}$, with $\tfrac{\partial}{\partial E}S$ equal to the classical orbit period $t_{\rm p}$.
For $E\to E_{\rm c}$ the period grows asymptotically (it takes an infinite time to get over the central maximum of the potential), thus the level density has an anomaly. 
The variation of the $i$th level energy with $\eta$ can be associated with a contour curve of the function $S(\eta,E)$ satisfying Eq.~(\ref{ebk}).
Exploiting the condition $dS=\tfrac{\partial S}{\partial\eta}d\eta+\tfrac{\partial S}{\partial E}dE=0$ and evaluating both partial derivatives of the action in Eq.~(\ref{act}), one obtains
\begin{equation}
\tfrac{d}{d\eta}E_i=-\frac{\tfrac{\partial}{\partial\eta}S(\eta,E)}{\tfrac{\partial}{\partial E}S(\eta,E)}\biggr|_{E=E_i(\eta)}
=\biggl\langle\tfrac{\partial}{\partial\eta}{\tilde{\cal H}}_{\rm cl}(\eta)\biggr\rangle
=\biggl\langle\left(\tfrac{1}{2}-{\tilde\beta}^2\right)\left[{\tilde p}_{\beta}^2+\left(\tfrac{p_{\gamma}}{{\tilde\beta}}\right)^2\right]+
\tfrac{5}{2}{\tilde\beta}^2-{\tilde\beta}^4\biggl\rangle
\,,
\label{flow}
\end{equation}
where $\langle f\rangle=\tfrac{1}{t_{\rm p}}\int_0^{t_{\rm p}} f(p,q)\,dt$ stands for the time average over the specific periodic orbit.
Note that Eq.~(\ref{flow}) is a semiclassical analog of the first Eq.~(\ref{PY}).
As the level energy crosses the separatrix value $E_{\rm c}=0$ at some value of the control parameter, $\eta=\eta_{\rm c}^{(i)}$, the average on the right hand side of Eq.~(\ref{flow}) drops to zero since the motion gets temporarily frozen at the top of the sombrero potential.
A more detailed analysis \cite{Caprio08} shows that this is a logarithmic type of singularity, with the second derivative $\tfrac{d^2}{d\eta^2}E_i$ approaching to $\mp\infty$ as $\eta$ converges to $\eta_{\rm c}^{(i)}$ from the left and right, respectively.
We therefore encounter a continuous quantum phase transition (with no Ehrenfest classification) of excited level $i$.

It should be stressed that anomalous properties of wave functions at the classical phase space separatrix for a double well potential were studied already in 1987 by Cary, Rusu, and Skodje \cite{Cary87,Cary92,Cary93}.
The relation of quantum phase transitions to classical instability points was anticipated by Heiss and M{\"u}ller \cite{Heiss02} and a singular evolution of the quantum spectrum in the above sense has been first investigated within the Lipkin model by Leyvraz, Heiss, Scholtz, and Geyer \cite{Heiss05,Leyvraz05}.
The classical limit of the Lipkin model represents a 1D projection of the IBM classical hamiltonian for $L=0$ and $\tfrac{\tau}{N}=0$.
If the above analysis within the IBM is extended to states with $\tfrac{\tau}{N}\neq 0$ (hence $p_{\gamma}\neq 0$), one finds out that the centrifugal barrier in Eq.~(\ref{clasb}) removes the phase space separatrix and the nonanalytic behavior of $E_i$ is washed out.
Therefore, the ESQPT behavior in the $N\to\infty$ limit applies only to states with $\tfrac{\tau}{N}=0$ \cite{Cejnar06}.

\section{Concluding remarks}

In this review, we attempted to summarize the present state of knowledge on quantum phase transitions in the interacting boson model.
This subject is important from both experimental and theoretical viewpoints.
It has direct applications in nuclear structure physics, where shape phases and critical regions in the chart of nuclides have been subject to intense research in the last decade.
Closely related models, with analogous types of quantum phase transitions, are used also in molecular physics.

With respect to recent reviews \cite{Casten07,Casten08} of the experimental aspects of the structural evolution in nuclei, here a stronger accent was put to theoretical problems.
Our aim was to present the family of interacting boson models as a useful framework for probing the origins and fundamental properties of quantum phase transitions in many-body systems, which at the same time holds a close relation to experimental data.
Rather than going into deep details within a restricted area, we decided to show a vast variety of diverse approaches and related problems.
The reader interested in a specific issue is encouraged to follow the relevant references.

Basic properties of the interacting boson model connected with phase transitions are contained in its simplest version, the IBM-1, which has been mostly discussed in the QPT context so far.
One of the main advantages of this model is a simultaneous occurrence of first- and second-order quantum phase transitions.
We showed the roots of the ground-state critical behavior in the mean-field dynamics and outlined the related approaches going beyond the mean field and to the direction of statistical physics.
These concepts are essential for the description and fundamental understanding of the ground-state evolution.

Various generalizations of symmetry were presented as important guides for the QPT physics of many-body systems.
The symmetries emerge (somewhat surprisingly) in transitions between different dynamical regimes of the system, giving rise to specific signatures of quantum critical behaviors in the spectra of low-lying states.
The symmetry related aspects of QPT's in many-body systems constitute an important subject of ongoing research.

Extensions of the IBM-1 to various sides, also reported in this review, provide enriched phase structures and modified realizations of the QPT phenomena.
Phase transitions in these generalized models represent an interesting subject on its own, but they can also be seen as concrete cases of a general branch of problems concerning the QPTs influenced by a coupling of the system to additional degrees of freedom.
In the IBM extensions, the main effort has been spent so far to identify the basic form of the relevant phase diagrams, while various sophisticated approaches developed in the simpler cases still wait for application.

The last part of the review was devoted to excited-state quantum phase transitions, which represent a promising topic for future analyses.
In these transitions, global features of the whole quantum spectrum are subject to nonanalytic changes of the phase transitional type.
New hints for traditional physics of phase transitions may follow from the study of these issues.
An open problem remains the extension of the concept of excited-state QPTs to first-order transitions.
The interacting boson model, which itself seems to be at the first-order phase transition between simplicity and complexity (with both aspects coexisting within one framework), may serve as a firm ground for such studies.

\section*{Acknowledgments}

The authors would like to acknowledge
M.~Caprio, R.F.~Casten, J.~Dobe{\v s}, S.~Heinze, F.~Iachello, M.~Macek, P.~Str{\' a}nsk{\' y}, G.~Thiamov{\' a}, P.~Van~Isacker, P.~von~Brentano, and V.~Werner 
for fruitful collaboration in the domain of quantum phase transitions, and
Y.~Alhassid, J.M.~Arias, D.~Bonatsos, O.~Casta{\~ n}os, A.~Dewald, J.~Dukelsky, A.~Frank, J.E.~Garc{\'\i}a-Ramos, K.L.G.~Heyde, V.~Hellemans, R.V.~Jolos, A.~Leviatan, E.A.~McCutchan, P.~Regan, D.~Rowe, D.D.~Warner, and V.~Zamfir
for many interesting discussions over the last years.

This work was supported by the Czech Science Foundation (grant 202/06/0363), Czech Ministry of Education (contracts 0021620859 and LA~314), Grant Agency of Charles University (grant 222/2006/B-FYZ/MFF), and by Deutsche Forschungsgemeinschaft (grant JO391/5-1).

\def\Journal#1#2#3#4{{#1} {#2} (#4) #3}
\def\ANNP{\em Ann. Phys. (N.Y.)}
\def\ANNPL{\em Ann. Phys. (Leipzig)}
\def\MATH{{\em J. Math. Phys.}}
\def\CHEM{{\em J. Chem. Phys.}}
\def\CHEML{{\em Chem. Phys. Lett.}}
\def\MOPH{\em Mol. Phys.}
\def\INT{{\em Int. J. Mod. Phys.} E}
\def\JPA{{\em J. Phys. A: Math. Theor.}} 
\def\JPAold{{\em J. Phys. A: Math. Gen.}} 
\def\JPG{{\em J. Phys. G: Nucl. Part. Phys.}} 
\def\NCA{{\em Nuovo Cimento} A }
\def\RNC{\em Rivista Nuovo Cimento}
\def\NP{{\em Nucl. Phys.}}
\def\NPA{{\em Nucl. Phys.} A}
\def\NPB{{\em Nucl. Phys.} B\ }
\def\PHYS{{\em Physica}}
\def\PRO{{\em Prog. Theor. Phys.}}
\def\PLA{{\em Phys. Lett.} A\ }
\def\PLB{{\em Phys. Lett.} B}
\def\PLD{{\em Phys. Lett.} D}
\def\PL{{\em Phys. Lett.}}
\def\PRL{{\em Phys. Rev. Lett.\/}\ }
\def\PREV{\em Phys. Rev.}
\def\PREP{\em Phys. Rep.}
\def\PRA{{\em Phys. Rev.} A}
\def\PRD{{\em Phys. Rev.} D}
\def\PRC{{\em Phys. Rev.} C}
\def\PRE{{\em Phys. Rev.} E}
\def\PRB{{\em Phys. Rev.} B}
\def\PS{\em Phys. Scr.}
\def\PhilML{{\em Phil. Mag. Lett.}}
\def\RMF{{\em Rev. Mex. Fis.}}
\def\RMP{{\em Rev. Mod. Phys.}}
\def\RPP{{\em Rep. Prog. Phys.}}
\def\ZPC{{\em Z. Phys.} C}
\def\ZPA{{\em Z. Phys.} A}
\def\Nat{\em Nature}
\def\PRSL{{\em Proc. Roy. Soc. Lond.} A}
\def\PPNP{\em Prog. Part. Nucl. Phys.}

\end{document}